\documentclass[usenatbib,a4paper,times,fleqn]{mnras}
\usepackage{graphicx,natbib,times,amssymb,amsmath,pstricks}
\usepackage{aas_macros}
\usepackage{bm,url}
\usepackage{textcomp}
\usepackage{breakurl}
\usepackage{subfig}
\usepackage{booktabs,times}
\usepackage{siunitx}
\usepackage{media9}
\usepackage{animate}

\newcommand{\newtext}[1]{{\black #1}}

\title[Gas and multi-species dust dynamics]{Gas and multi-species dust dynamics in viscous protoplanetary discs: the importance of the dust back-reaction}

\author[Dipierro et al.]{Giovanni Dipierro$^{1}$\thanks{giovanni.dipierro@leicester.ac.uk}, Guillaume Laibe$^2$, Richard Alexander$^{1}$ and Mark Hutchison$^{3,4}$   \\
$^{1}$Department of Physics and Astronomy, University of Leicester, Leicester, LE1 7RH, United Kingdom\\ 
$^{2}$Univ Lyon, Univ Lyon1, Ens de Lyon, CNRS, Centre de Recherche Astrophysique de Lyon UMR5574, F-69230, Saint-Genis-Laval, France\\
$^{3}$Physikalisches Institut, Universit{\"a}t Bern, Gesellschaftstrasse 6, 3012 Bern, Switzerland \\
$^{4}$Institute for Computational Science, University of Zurich, Winterthurerstrasse 190, CH-8057 Z{\"u}rich, Switzerland}
\pagerange{\pageref{firstpage}--\pageref{lastpage}} \pubyear{2018}
\defcitealias{nakagawa86a}{NSH86}
\defcitealias{dipierro16a}{DLPL16}
\date{}
\begin{document}
\label{firstpage}
\bibliographystyle{mnras}
\maketitle

\begin{abstract}
We study the dynamics of a viscous protoplanetary disc hosting a population of dust grains with a range of sizes. We compute steady-state solutions, and show that the radial motion of both the gas and the dust can deviate substantially from those for a single-size dust population. Although the aerodynamic drag from the dust on the gas is weaker than in the case where all grains are optimally coupled to the gas, the cumulative ``back-reaction'' of the dust particles can still alter the gas dynamics significantly. In typical protoplanetary discs, the net effect of the dust back-reaction decreases the gas accretion flow compared to the dust-free (viscous) case, even for dust-to-gas ratios of order $1\%$.
In the outer disc, where dust grains are typically less strongly coupled to the gas and settle towards the midplane, the dust back-reaction can even drive outward gas flow. Moreover, the radial inward drift of large grains is reduced below the gas  motion in the inner disc regions, while small dust grains follow the gas dynamics over all the disc extent.
The resulting dust and gas dynamics can give rise to observable structures, such as gas and dust cavities. Our results show that the dust back-reaction can play a major role in both the dynamics and observational appearance of protoplanetary discs, and cannot be ignored in models of protoplanetary disc evolution. 
\end{abstract}

\section{Introduction}

Solids in protoplanetary discs represent the primary building blocks of both rocky planets and cores of gas giants and play an essential role in regulating most of the thermal features of discs \citep{testi14a}. 
The birth place of protoplanetary discs is the diffuse interstellar medium (ISM), which on average is composed in mass of 1\% dust and 99\% gas \citep{mathis77a,weingartner01a}. At the beginning of the star formation process, dust particles are well mixed with the gas and the dust-to-gas ratio is expected to be uniform. However, even during the early phases of protostellar collapse, before protoplanetary discs are formed, the solid component  is expected to evolve significantly from typical ISM dust to a population of solids characterized by a wider range of sizes \citep{weidenschilling80a,kim94a}. 
Large dust grains with size $ \gtrsim 1\,\mathrm{\mu}\text{m}$ couple more loosely to the turbulent fluctuations of the gas. As a result, they segregate spatially, concentrate locally and increase the local dust-to-gas ratio by up to an order of magnitude \citep{liseau15a,larsson17a,tricco17a}. 
Observations of molecular clouds and young protostellar objects provide robust evidence of the presence of larger grains \citep[e.g.,][]{furlan06,hughes07,ricci10a,pagani10a,andersen15a,miotello14a} and of regions where the dust-to-gas ratio is up to an order of magnitude higher than the canonical ISM value \citep{liseau15a,ansdell16a,ansdell17a,larsson17a,long17a}. However, uncertainties on carbon abundances can affect the gas mass estimates  and detailed physico-chemical disk models need to be carried out to properly account for freeze-out and isotopologue-selective processes \citep{miotello16a,miotello17a}.
Similarly, solids concentrate during the formation of the disc itself, and the initial dust content of a disc is expected to be richer than the ISM. An intense differential concentration of solids occurs in the disc under the combined effect of the gravity of the star and gas drag \citep{bate17a}. Millimetre-sized grains settle close to the midplane, while micron-sized particles are scattered by turbulence over large scales \citep{dubrulle95a}. Additionally, the inward drag-induced drift of grains in typical protoplanetary discs makes grains concentrate radially \citep{weidenschilling77a}, an effect amplified by dust growth \citep{laibe14b}. In parallel, gas loss due to winds, either photoevaporative \citep[e.g.,][]{throop05,gorti16a} or magnetically-driven \citep[e.g.,][]{suzuki09,gressel15,bai16} may also enhance the dust-to-gas ratio. These low-density flows preferentially deplete the disc of gas, entraining only sub-micron grains \citep{takeuchi05b,hutchison16a,hutchison16b} and leaving most of the dust content behind \citep{alexander14a}. The combined action of dust dynamical concentration and gas dispersal would produce wide regions of high dust-to-gas ratio close to the central star in the late stages of protoplanetary disc evolution. At such ratios, the dust influences the disc dynamics as much as the gas (since gas drags dust as much as dust drags gas) and the disc is expected to enter a regime dominated by the back-reaction.

The dust back-reaction is known to play an important role in several aspects of planet formation.
It regulates the dust scale height \citep{weidenschilling80a,cuzzi93a,youdin02a} 
and drives streaming instabilities \citep{youdin05a,youdin07b,jacquet11a,auffinger18a}, which might concentrate pebbles locally up to a threshold sufficient to form planetesimals by gravitational instability \citep{johansen07a,johansen09b,bai10a}. 
Over long time-scales, the back-reaction can destroy vortices \citep{fu14b,rubsamen15a} or affect the morphology and location of pressure maxima \citep{taki16a,weber18a}. Indeed, the back-reaction is capable of affecting the evolution of the gas -- even for dust-to-gas ratios of just a few per cent -- since the flow powered by the back-reaction in the midplane is more intense than the viscous accretion flow, resulting in an inner region depleted of gas \citep{gonzalez17a,kanagawa17a}. When grain growth is additionally considered, the back-reaction may also produce a large-scale pressure maximum that traps the grains drifting from the outer disc \citep{gonzalez17a}. 

From an observational point of view, recent multi-wavelength, high-sensitivity and high-resolution dust continuum measurements revealed a difference between the gas and dust disc sizes and, importantly, an anti-correlation between the wavelength and the radial extent of the continuum emission: the disc emission observed at short wavelengths is more extended than the disc emission observed at long wavelengths \citep[e.g.][]{guilloteau11a,andrews12a,perez12a,tazzari15a}. These findings have been interpreted as evidence for a size-sorted spatial density distribution in the dust, induced by the combination of radial drift and grain growth processing (see the review by \citealt{testi14a} and references therein).  Moreover, scattered light observations confirm the presence of stratification, with smaller grains extending found higher up in the disc atmosphere while larger grains have settled about the disc midplane \citep{pinte08a,duchene10a,avenhaus18a}. The general picture emerging from these observational results is that large dust grains populate the inner disc regions around the midplane, while the gas and small dust grains have a larger spatial extent both radially and vertically \citep{ansdell18a}, although the larger size of the gas disc might be ascribed to optical depth effects \citep{facchini17a}. Since most of the solid mass is typically concentrated in larger grains, regions of higher dust-to-gas ratio are therefore expected to develop very efficiently around the midplane in the inner disc regions. 

Moreover, dust tracers -- from near-infrared (IR) emission from small grains in the inner disc, through mid-IR and far-IR emission, to mm and cm emission from larger solids at large radial distances -- are readily observed in hundreds of protoplanetary discs \citep{testi14a}. In the youngest clusters the incidence of these tracers is close to 100\%, but this fraction declines to $<$ 5--10\% by ages of 3--5 Myr \citep{haisch01a,hernandez07a}. Accretion signatures and molecular line emission vanish on similar time-scales \citep{fedele10a,mathews12a}, and upper limits on all of these tracers in disc-less stars show that disc dispersal is extremely efficient \citep{pascucci06a,ingleby09a}.  All these signatures disappear near-simultaneously, and relatively few objects (so-called ``transitional'' discs) show evidence of partial disc clearing. This suggests that disc dispersal occurs relatively quickly, after lifetimes of a few Myr \citep[e.g.][]{simon95a,andrews05a,koepferl13a}. However, it is also notable that the majority of transitional discs shows signs of {\it inner} disc clearing, suggesting that disc dispersal proceeds (rapidly) from the ``inside-out'' \citep[see, e.g.,][]{owen15a}. This is consistent with our theoretical understanding of disc evolution and dispersal \citep{alexander14a,gorti15a,gorti16a}, but to date most secular evolution models have considered dust only as a trace contaminant. Given that most disc observations trace only the dust component, the possibility that dust-gas dynamics may also drive inside-out disc evolution clearly warrants further investigation. All of these results suggest that the back-reaction of the dust on the gas is likely to be significant, and needs to be taken into account in disc evolution models. 

To determine precisely the extent to which the dust affects the gas dynamics, one should take into account the fact that the mass of solids is locally distributed (unequally) among grains with different size that contribute to the back-reaction with different intensities. Interestingly, as dust grains of different sizes are coupled through their respective interaction with the gas, the back-reaction from grains of a given size may trigger a gas flow which in turn affect the motions of grains of other sizes \citep{tanaka05a,kretke09a,bai10a,okuzumi12a}. Moreover, while previous studies have considered only inviscid discs, \textit{we aim to determine whether the dust-driven gas flow can dominate over the viscously-driven accretion flow when the cumulative back-reaction from a grain-size distribution is accounted for properly}. To this end, we use analytic calculations to determine the steady-state gas and dust velocities in a viscous disc hosting a population of grains with different sizes. 
The paper is organised as follows: in Sect.~\ref{sect:dynamics} we formalise the dynamics of viscous discs hosting grains of multiple sizes in 2D, and discuss its evolution. In Sect.~\ref{sect:cumeffback} we analyse the dust and gas dynamics at the midplane of typical protoplanetary discs.
In Sect.~\ref{sect:dynamicsvert}, we integrate the analysis over the vertical direction in typical disc models with realistic grain-size distributions. 
Finally, in Sect.~\ref{sect:discussion} we discuss the implications of our results in the context of disc evolution models and summarize our findings in Sect.~\ref{sec:conclusion}.

\section{Dynamics of multiple species of dust grains}
\label{sect:dynamics}


\subsection{Equations of motion}
We assume a thin, axisymmetric, non-magnetic, non-self-gravitating, viscous, and vertically isothermal dusty protoplanetary disc. We assume that the angular momentum is transported across the disc by an effective viscous-like mechanism \citep{shakura73a}, without detailing the origin of the transport (see Sect.~\ref{sec:limitation} for a discussion). The dust is modelled as a continuous pressure-less and inviscid fluid made of $n$ separate phases of compact homogeneous, collisionless and indestructible spheres of fixed radii \citep{garaud04a}. \newtext{Notations are described in Appendix~\ref{app:notations}}. The equations of motion for the gas and the $i^{\mathrm{th}}$ dust phase are \citep[e.g.][]{tanaka05a,laibe14d}
\begin{align}
\frac{\partial \mathbf{u}}{\partial t}+(\mathbf{u}\cdot \mathbf{\nabla})\,\mathbf{u}  = &  \frac{1}{\rho_{\mathrm{g}}}\sum_i K_i(\mathbf{v}_i-\mathbf{u})-\mathbf{\nabla}\Phi \nonumber \\
&-\frac{1}{\rho_{\mathrm{g}}}(\mathbf{\nabla} P  - \mathbf{\nabla} \cdot \sigma) , 
\label{eq:gene3}\\
 \frac{\partial \mathbf{v}_i}{\partial t}+(\mathbf{v}_i\cdot \mathbf{\nabla})\,\mathbf{v}_i   = &- \frac{K_i}{\rho_{\mathrm{d}i}}(\mathbf{v}_i-\mathbf{u})-\mathbf{\nabla}\Phi,
 \label{eq:gene4}
\end{align}
where $\mathbf{u}$ and $\mathbf{v}_i$ denote the velocity of the gas and the $i^{\mathrm{th}}$ dust phase respectively. $\rho_{\rm g}$ and $\rho_{\mathrm{d}i}$ denote the gas and the $i^{\mathrm{th}}$ dust phase densities, respectively. $\Phi$, $P$ and $\sigma$ refer to the gravitational potential of the star, the pressure, and the viscous stress tensor of the gas, respectively. $K_i$ is the drag coefficient between the gas and the $i^{\mathrm{th}}$ dust species. 
Its expression depends on the local gas properties and grain sizes (e.g. \citealt{birnstiel10a,laibe12a}), and is related to the stopping time $t_i^{\mathrm{s}}$ of the mixture of gas and the single $i^{\mathrm{th}}$ dust species according to 
\begin{equation}
K_i \equiv \frac{\rho_{\mathrm{d}i} \rho_{\mathrm{g}}}{t_i^{\mathrm{s}} \left(\rho_{\mathrm{g}}+\rho_{\mathrm{d}i}\right )} = \frac{\rho_{\mathrm{d}i}}{t_i^{\mathrm{s}} \left ( 1+\epsilon_i \right)},
\label{eq:tstopmixture}
\end{equation}
where $t_i^{\mathrm{s}}$ is the typical time for the drag to damp the local differential velocity between the gas and the $i^{\mathrm{th}}$ dust phase. 
Individual dust-to-gas ratios, $\epsilon_i \equiv \rho_{\mathrm{d}i}/\rho_{\mathrm{g}}$, add up to give the total dust-to-gas ratio $\epsilon \equiv \rho_{\mathrm{d}} / \rho_{\rm g}$, where $\rho_{\rm d} \equiv \sum_i \rho_{\mathrm{d}i}$ is the total dust density. 
We define the Stokes number for the $i^{\mathrm{th}}$ dust species as
\begin{equation}
\mathrm{St}_i \equiv \left(1+\epsilon_i \right) t_i^{\mathrm{s}} \Omega_{\mathrm{k}} ,
\label{eq:stokesimpl}
\end{equation}
i.e. proportional to the ratio of the stopping time of the $i^{\mathrm{th}}$ dust phase to the orbital time $ \Omega_{\mathrm{k}}^{-1}$, where $\Omega_{\mathrm{k}}$ is the Keplerian angular velocity. We want to caution that two different definitions of the Stokes number differing by a factor $\left(1+\epsilon_i \right)$ can be found in literature \newtext{(\citealt{dipierro17a,kanagawa17a}, see Sect. 3.2 of \citealt{laibe12b})}. In this paper, we adopt the definition of the stopping time in a gas-dust mixture defined in Eq.~\ref{eq:tstopmixture}, but for convenience we consider the Stokes number in the form expressed by Eq.~\ref{eq:stokesimpl} in the case of a mixture of multiple dust species. 
The stopping time $t_i^{\mathrm{s}}$ depends on the size of the particle $s_i$ relative to the mean free path of the gas molecules $\lambda_{\mathrm{mfp}}$ \citep{whipple72a}, defined as
\begin{equation}
\lambda_{\mathrm{mfp}}\equiv\frac{\mu m_{\mathrm{H}}}{\rho_{\mathrm{g}} \sigma_{\mathrm{coll}}} \approx 1.15 \,\mathrm{cm} \left(\frac{10^{-9} \mathrm{g\, cm^{-3}}}{\rho_{\mathrm{g}}} \right),
\end{equation}
where $m_{\mathrm{H}}\approx 10^{-24}$ g is the proton mass, $\sigma_{\mathrm{coll}}\approx 2\times 10^{-15}\, \mathrm{cm^2}$ is the molecular collisional cross section of the gas \citep{chapman70a}  and $\mu$ is the mean molecular weight (typically taken $\sim 2.3$ due to the abundances of molecular hydrogen and helium in protoplanetary discs).
When the grain size $s_i$ is smaller than the mean free path, i.e. $s_i<9\lambda_{\mathrm{mfp}}/4$, the drag force can be computed by considering the transfer of angular momentum between collisions of individual gas molecules at the grains surface. This case is typically referred to as the \emph{Epstein regime} \citep{epstein24a} and is the relevant drag law for the majority of grains with size $\lesssim \mathrm{cm}$. However, the gas in the very inner disc regions can be sufficiently dense for the dust particles to enter the Stokes drag regime. 
In typical protoplanetary discs, the Stokes number can therefore be expressed by  (see Table C.1 in \citealt{laibe12c})
\begin{equation}
\mathrm{St}_i=  \sqrt{\frac{\pi \gamma}{8}}\frac{ \rho_{\mathrm{grain}} s_i \Omega_{\mathrm{k}}}{ \rho_{\mathrm{g}} c_{\mathrm{s}}}  
  \begin{cases}
       1, & s_i< \frac{9}{4}\lambda_{\mathrm{mfp}},\\
        \frac{4}{9}\frac{s_i}{\lambda_{\mathrm{mfp}}}, & s_i \geq \frac{9}{4}\lambda_{\mathrm{mfp}}, 
  \end{cases}
 \label{eq:stoppingtimecase}
\end{equation}
where $\rho_{\mathrm{grain}}$ denotes the material grain density, while $c_{\mathrm s}$ and $\gamma$ are the sound speed and the adiabatic index of the gas respectively. 
%

Following \citet{nakagawa86a} and \citet{tanaka05a}, we perform a perturbative expansion of the velocities of each phase relative to the local Keplerian velocity $\mathbf{v_{\mathrm k}} \equiv r\Omega_{\mathrm k} \mathbf{e}_{\phi}$, since the deviations are of order $(H_{\mathrm{g}}/ r)^{2} \ll 1$, where $H_{\mathrm{g}}$ is the gas disc scale height (Sect.~\ref{sect:gasdensity}). In cylindrical coordinates, the equations of motion for the perturbed velocities are
\begin{align}
\frac{	\partial u_{r}}{\partial t} &=\sum_i\frac{K_i}{\rho_{\mathrm g}}(v_{i,r}-u_{r})  -\frac{1}{\rho_{\mathrm g}}\frac{\partial P}{\partial r}+2\Omega_{\mathrm k} u_{\phi},
\label{eq:1} \\
\frac{	\partial u_{\phi}}{\partial t}&=\sum_i\frac{K_i}{\rho_{\mathrm g}}(v_{i,\phi}-u_{\phi})  -\frac{\Omega_{\mathrm k}}{2}u_{r} +\frac{1}{\rho_{\mathrm g}}\nabla \cdot \sigma|_\phi, \\
\frac{	\partial v_{i,r}}{\partial t}&=-\frac{K_i}{\rho_{\mathrm di}}(v_{i,r}-u_{r})  +2\Omega_{\mathrm k} v_{i,\phi},
\label{eq:3} \\
\frac{	\partial v_{i,\phi}}{\partial t}&=-\frac{K_i}{\rho_{\mathrm di}}(v_{i,\phi}-u_{\phi})  -\frac{\Omega_{\mathrm k}}{2} v_{i,r} ,
\label{eq:4}
\end{align}
where for simplicity we use $u_{\phi}$ and $v_{i,\phi}$ to denote the relative azimuthal velocities with respect of the Keplerian velocity. 
Eqs.~\ref{eq:1}--\ref{eq:4} are a linear system of  $2n + 2$ differential equations of $2n + 2$ velocities, where $n$ is the number of particle bins in the grain-size distribution. 

\subsection{Steady-state velocities}
We look for steady-state circular orbits, assuming that the initial eccentricities are damped after a typical time $\sim \displaystyle \max \left (t_{i}^{\mathrm{s}} \right)$.
We follow the derivation of \citet{tanaka05a} to obtain the steady-state solution of Eqs.~\ref{eq:1}--\ref{eq:4}. The radial and azimuthal gas velocities relative to the Keplerian velocity  are given by
\begin{align}
u_{r}&=\frac{-\lambda_1 v_{\mathrm{P}} +\left(1+\lambda_0\right) v_{\mathrm{visc}}}{\left(1+\lambda_0\right)^2+\lambda_1^2}  , \label{eq:vradg} \\
u_{\phi}&=  \frac{1}{2}\left [\frac{ v_{\mathrm{P}}\left(1+\lambda_0\right)  + v_{\mathrm{visc}}\lambda_1}{\left(1+\lambda_0\right)^2+\lambda_1^2} \right ], \label{eq:vthetag}
\end{align}
while the relative velocities of the $i^{\mathrm{th}}$ dust phase are
\begin{align}
v_{i,r}&= \frac{v_{\mathrm{P}} \left[\left(1+\lambda_0\right) \mathrm{St}_i -\lambda_1 \right] +v_{\mathrm{visc}} \left(1+\lambda_0+\mathrm{St}_i \lambda_1 \right)}{\left[\left(1+\lambda_0\right)^2+\lambda_1^2 \right] \left (1+\mathrm{St}_i^2 \right)}, \label{eq:vraddi}\\
v_{i,\phi}&=  \frac{1}{2}\left \lbrace \frac{v_{\mathrm{P}} \left(1+\lambda_0+\mathrm{St}_i \lambda_1 \right)  -v_{\mathrm{visc}} \left[\left(1+\lambda_0\right) \mathrm{St}_i -\lambda_1 \right]}{\left[\left(1+\lambda_0\right)^2+\lambda_1^2 \right] \left (1+\mathrm{St}_i^2 \right)} \right \rbrace , \label{eq:vthetadi}
\end{align}
where
\begin{equation}
v_{\mathrm{P}} =  \frac{1}{\rho_{\mathrm g} \Omega_{\mathrm k}}\frac{\partial P}{\partial r} 
\label{eq:vp}
\end{equation}
is the typical particle drift velocity and 
\begin{eqnarray}
v_{\mathrm{visc}} & \equiv & \frac{2}{\Omega_{\mathrm k}\rho_{\mathrm g}}\nabla \cdot \sigma|_\phi \nonumber \\
& = &  \frac{1}{r \rho_{\mathrm g} \frac{ \partial }{ \partial r}\left(   r v_{\mathrm k} \right)  }\frac{\partial}{\partial r} \left(\eta r^3 \frac{\partial \Omega_{\mathrm k}}{\partial r}\right) 
\label{eq:vvisc}
\end{eqnarray}
is the radial velocity of the gas due to viscous torques, derived by \citet{lynden-bell74a}. Here $\eta= \nu \rho_{\mathrm g}$ is the shear viscosity coefficient, with $\nu$ denoting the kinematic viscosity. The relative contributions of the individual dust phases are encoded by the parameters $\lambda_{0}$ and $\lambda_{1}$ defined by
\begin{equation}
 \lambda_k \equiv \sum_i \frac{  \mathrm{St}_i^k}{1+\mathrm{St}_i^2 } \epsilon_i .
 \label{eq:lambda01sum}
\end{equation}
As shown by \citet{tanaka05a}, Eqs.~\ref{eq:vraddi} and \ref{eq:vthetadi} can be simplified and dust velocities can be expressed in terms of the gas velocities as
\begin{align}
v_{i,r}=&\frac{u_{r} +2 u_{\phi}\mathrm{St}_i }{1+\mathrm{St}_i^2} \label{eq:vdrtermsvg},\\
v_{i,\phi}=&  \frac{1}{2}\left (\frac{- u_{r}  \mathrm{St}_i + 2 u_{\phi} }{1+\mathrm{St}_i^2}\right ) . \label{eq:vdptermsvg}
\end{align}
Eqs.~\ref{eq:vdrtermsvg} and \ref{eq:vdptermsvg} have a similar form to Eqs.~14 and 15 in \citet{tanaka05a}, but the gas velocities in our Eqs.~\ref{eq:vradg} and \ref{eq:vthetag} include additional terms related to the viscous evolution (i.e. terms $\propto v_{\mathrm{visc}}$). In Appendix~\ref{app:numericaltests}, we validate our equations against 3D Smoothed Particle Hydrodynamics (SPH) simulations containing a mixture of gas and dust with multiple grain sizes \citep{price17a,hutchison18a}.

In the case of a population comprising a single equally-sized dust species with uniform dust-to-gas ratio $\epsilon$, adopting the definition of Stokes number $\mathrm{St}$ shown in Eq.~\ref{eq:stokesimpl}, Eqs.~\ref{eq:vradg}--\ref{eq:vthetadi} reduce to the steady-state velocities derived in \citet{dipierro17a} and \citet{kanagawa17a}, given by
\begin{align}
u_{r}&=- \frac{\epsilon \,v_{\mathrm{P}}}{\mathrm{St}+\mathrm{St}^{-1} \left(1+\epsilon\right)^2}   + \frac{v_{\mathrm{visc}}}{1+\epsilon} \left(1 + \epsilon \frac{ \mathrm{St}^{2}}{\left(1+\epsilon\right)^2+\mathrm{St}^{2} } \right)   , \label{eq:vradsimp} \\
u_{\phi}&=  \frac{1}{2}\left [ \frac{v_{\mathrm{P}}}{1+\epsilon}\left(1 + \epsilon \frac{ \mathrm{St}^{2}}{\left(1+\epsilon\right)^2+\mathrm{St}^{2} } \right)+\frac{\epsilon v_{\mathrm{visc}}}{\mathrm{St}+\mathrm{St}^{-1}\left(1+\epsilon\right)^2}\right ] , \label{eq:vthetagsimp}\\
v_{r}&=  \frac{ v_{\mathrm{P}}}{\mathrm{St}+\mathrm{St}^{-1}\left(1+\epsilon\right)^2} + \frac{1+\epsilon }{\left(1+\epsilon\right)^2+\mathrm{St}^{2}}v_{\mathrm{visc}}, \label{eq:vradustsimp} \\
v_{\phi}&= \frac{1}{2}\left [ \frac{1+\epsilon }{\left(1+\epsilon\right)^2+\mathrm{St}^{2}}v_{\mathrm{P}}   - \frac{ v_{\mathrm{visc}}}{\mathrm{St}+\mathrm{St}^{-1} \left(1+\epsilon\right)^2} \right] . \label{eq:vthetadsimp}
\end{align}
\citet{kanagawa17a} found that the back-reaction has a non-negligible effect on the gas dynamics for equally-sized grains with Stokes number $\sim 0.1-1$ close to the midplane, even for low dust-to-gas ratios. With Eqs.~\ref{eq:vradg}--\ref{eq:vthetadi}, we can generalise this analysis to the case where the solid mass is distributed over a large range of dust sizes.

\subsection{Physical interpretation}
\label{sect:physicalinterp}

Alone, the disc gas undergoes slow radial viscous accretion (if $v_{\mathrm{visc}}<0$) while orbiting at sub-Keplerian velocity (if $\partial P/ \partial r<0$), since gravity from the central star is weakly mitigated by the negative radial pressure gradient of the gas. Similarly, dust alone follows a Keplerian orbit. 
In a dust-gas mixture, drag tends to damp the differential velocities between the phases. In the radial direction, drag forces the dust to move along with the gas accretion flow, which in turn is slowed by the dust back-reaction. We will call this motion radial \emph{viscous drag}. In the azimuthal direction, dust experiences a continuous headwind from the gas, which makes it lose angular momentum and fall on to the star (if $\partial P/ \partial r<0$). The back-reaction pushes gas in the opposite direction, towards regions of low pressure. 
Therefore, the headwind (tailwind) causes the dust (gas) to lose (gain) angular momentum and spiral towards (away from) the star, assuming that the pressure maximum (minimum) is located close to (far from) the star.
This motion is usually called radial \emph{drift}. The viscous drag and drift components of the gas and dust motions depend on the gas pressure and viscosity, respectively, since those are the sources of the differential motions, as well as the Stokes number, the dust-to-gas ratio and the dust size distribution. An order-of-magnitude analysis of the different velocity sources (Eqs.~\ref{eq:vp} and \ref{eq:vvisc}) gives
\begin{align}
\left| v_{\mathrm{P}} \right|  \sim & \left( \frac{H_{\mathrm{g}}}{r} \right)^{2} v_{\rm k} , \label{eq:magnvp}\\
\left|  v_{\mathrm{visc}} \right|  \sim & \,  \alpha\left( \frac{H_{\mathrm{g}}}{r} \right)^{2} v_{\rm k} \label{eq:magnvvisc},
\end{align}
where $H_{\mathrm{g}}$ is the usual gas disc scale height in vertical hydrostatic equilibrium (see Sect.~\ref{sect:gasdensity}) and we use $\nu=\alpha c_{\rm s} H_{\mathrm{g}}$ \citep{shakura73a}. Thus, $\left| v_{\mathrm{P}} \right|/\left|  v_{\mathrm{visc}} \right|\sim \alpha^{-1}$. Since $\alpha^{-1}\sim \left[10^{2}, 10^{4} \right]$ in typical protoplanetary discs \citep{flaherty15a,flaherty17a}, radial drift is expected to dominate over radial viscous drag for a wide range of grain sizes. 

Typically, for dust made of equally-sized grains (see Eq.~\ref{eq:vradustsimp}) with low dust-to-gas ratio ($\epsilon \ll 1$), the direct viscous drag dominates the radial motion for smaller particles, i.e. when $\mathrm{St}\lesssim \alpha$. Larger particles ($\alpha \lesssim \mathrm{St}\sim 1$) are significantly influenced by the headwind from the gas, drifting towards pressure maxima \citep[e.g.][]{laibe12c}. Particles with $\mathrm{St}\gg 1$ do not dynamically interact with the gas and vice versa.
Eq.~\ref{eq:vradsimp} shows that for $\epsilon \lesssim \alpha/\left (\mathrm{St}-\alpha\right)$, the dust back-reaction by grains with $\alpha \ll \mathrm{St}\ll 1$ is negligible and the gas moves with the unperturbed radial velocity $v_{\mathrm{visc}}$, decreased by a factor $\left(1+\epsilon \right)$. However, when $\epsilon \gtrsim \alpha/\left (\mathrm{St}-\alpha\right)$, dust grains with $\alpha \ll \mathrm{St}\ll 1$ affect the gas motion, pushing it strongly back in the regions of lower pressure. In particular, grains with $\mathrm{St}=1$ affect the gas motion as soon as their dust-to-gas ratio $\epsilon\gg \alpha/\left(1-\alpha\right)\sim \alpha$. Since these grains represent a significant fraction of the dust mass in real discs \citep[e.g.][]{brauer08a,laibe08a}, their dynamical interaction with the gas cannot be neglected when investigating the gas disc evolution \citep{kanagawa17a}.

\subsection{The case of continuous dust distributions}
\label{sect:continudistrib}
In protoplanetary discs, the local dust size distribution results from the complex interplay between dynamical size sorting and grain growth processings \citep{birnstiel16a,johansen17a}. The resulting distribution in size is usually modelled as a truncated power-law \citep{draine06a}
\begin{equation}
\frac{ \mathrm{d} n}{ \mathrm{d} s}= A \left(\frac{s}{s_{\rm{max}}}\right)^{-q} \qquad \mathrm{for} \quad s_{\rm{min}}<s<s_{\rm{max}}  ,
\label{grainsizedistrib}
\end{equation}
where $s_{\rm{min}}$ and $s_{\rm{max}}$ denote the minimum and maximum sizes of the grains respectively and $A$ is a normalisation factor. The dust-to-gas ratio per unit grain size is therefore given by
\begin{equation}
\frac{ \mathrm{d} \epsilon}{\mathrm{d} s}=\frac{ m\left( s\right)}{\rho_{\mathrm{g}}} \frac{ \mathrm{d} n}{\mathrm{d}s},
\label{depsds}
\end{equation}
where $m\left( s\right)=4\pi \rho_{\mathrm{grain}} s^3/3$ is the mass of a spherical grain of radius $s$ and material density $\rho_{\mathrm{grain}}$. The factor $A$ is obtained by integrating Eq.~\ref{depsds} over the size distribution. Denoting $x \equiv s_{\rm max} / s_{\rm min}$, one obtains
\begin{equation}
A = a\left(x \right) \frac{ \rho_{\mathrm{d}}}{s_{\rm{max}} \,m(s_{\rm{max}}) },
\end{equation}
where
\begin{equation}
a\left(y \right) \equiv
\begin{cases}
\displaystyle  \left(q-4\right) \left[ y ^{q-4}-1 \right ]^{-1} , & \text{$q \neq 4$,} \\ 
\displaystyle \left[\log \left( y\right) \right]^{-1}, & \text{$q = 4$.}
\end{cases}
\label{eq:def_f}
\end{equation}
In this continuous limit, the coefficients $\lambda_0$ and $\lambda_1$ defined in Eq.~\ref{eq:lambda01sum} are expressed by
\begin{equation}
 \lambda_k=\int_{s_{\mathrm{min}}}^{s_{\mathrm{max}}}  \frac{\mathrm{St}^k}{1+\mathrm{St}^2 } \frac{ \mathrm{d} \epsilon}{\mathrm{d} s} \,\mathrm{d}s .
\label{eq:lambda01}
\end{equation}
Assuming that all dust grains interact with the gas in the Epstein regime (see the first form of Eq.~\ref{eq:stoppingtimecase}), it is convenient to rewrite Eq.~\ref{eq:lambda01} under the form
\begin{equation}
 \lambda_k= \epsilon \, \frac{a \left(x' \right)}{\mathrm{St}_{\rm max}^{4 - q}}   \int_{\mathrm{St}_{\mathrm{min}}}^{\mathrm{St}_{\mathrm{max}}} \frac{\mathrm{St}^{3+k-q}}{1 + \mathrm{St}^{2}} \mathrm{d} \mathrm{St} ,
 \label{eq:lambda_int}
\end{equation}
where $x' \equiv \mathrm{St}_{\rm max} / \mathrm{St}_{\rm min}$ and $\mathrm{St}_{\rm min}$ and $\mathrm{St}_{\rm max}$ correspond to the minimum and the maximum Stokes numbers of the distribution, respectively. The integral in Eq.~\ref{eq:lambda_int} can be evaluated by means of the hypergeometric function ${}_2\mathrm{F}_1$ \citep{abramowitz72a}, as
\begin{equation}
 \lambda_k= \epsilon \, \frac{a \left(x \right)}{\mathrm{St}_{\rm max}^{4 - q}}  \Delta W ,
 \label{eq:lambda_int2}
\end{equation}
where $\Delta W=\left[ W\!\left(\mathrm{St}_{\rm max} \right) - W\!\left(\mathrm{St}_{\rm min} \right) \right] $, with the function $W$ defined as
\begin{equation}
W(y) \equiv 
\begin{cases}
 \frac{y^{\xi}}{\xi} \left[1 -{}_2\mathrm{F}_1\left(1, \frac{ \xi}{2} ; 1+\frac{\xi}{2} ; -y^2 \right) \right]  , & \text{$q \neq k+2$,} \\[5pt]
\displaystyle \frac{1}{2}\log \left(1 + y^{2}   \right), & \text{$q = k+2$,}
\end{cases}
\end{equation}
where $\xi = k-q+2$.
Restricting our analysis to size distributions relevant for dust dynamics in typical protoplanetary discs, i.e. for which $\mathrm{St}_{\rm min} \ll 1$ and $\mathrm{St}_{\rm max} \gg 1$, we derive the expansion of $\Delta W$ accurate to a few percent, given by
\begin{equation}
\Delta W \simeq   
\begin{cases}
\frac{\mathrm{St}_{\rm max}^{\xi}}{\xi} + \frac{\pi}{2} \mathrm{cosec} \left[ \frac{\pi}{2} \left(k - q \right)\right]- \frac{\mathrm{St}_{\rm min}^{2+\xi}}{2+\xi}  , & \text{$q \neq k+2$,} \\[5pt] 
\log \left(\mathrm{St_{max}}\right)-\frac{\mathrm{St_{min}}^2}{2}  , & \text{$q = k+2$,}
\end{cases}
\label{eq:casedeltah}
\end{equation}
which is valid for $\lambda_{0}$ when $q\neq 4$ and for $\lambda_{1}$ when $q \neq5$. 
%
If we consider size distributions that are reasonably steep $(q<4)$,  $\lambda_{0}$  and $\lambda_{1}$ are mostly dominated by the value of $\mathrm{St}_{\rm max}$. One exception concerns the case $q > 4$, where $\lambda_{1}$ is dominated by the value of $\mathrm{St}_{\rm min}$. However, since we limit our study to size distributions with $q<5$, we find that value of $\mathrm{St}_{\rm min}$ does not strongly affect the dynamics of the mixture in our model, assuming $\mathrm{St}_{\rm min}\ll1$ (see also Sect.~\ref{sect:gasdyn2D}). 
\begin{figure*}
\begin{center}
\includegraphics[width=0.975\textwidth]{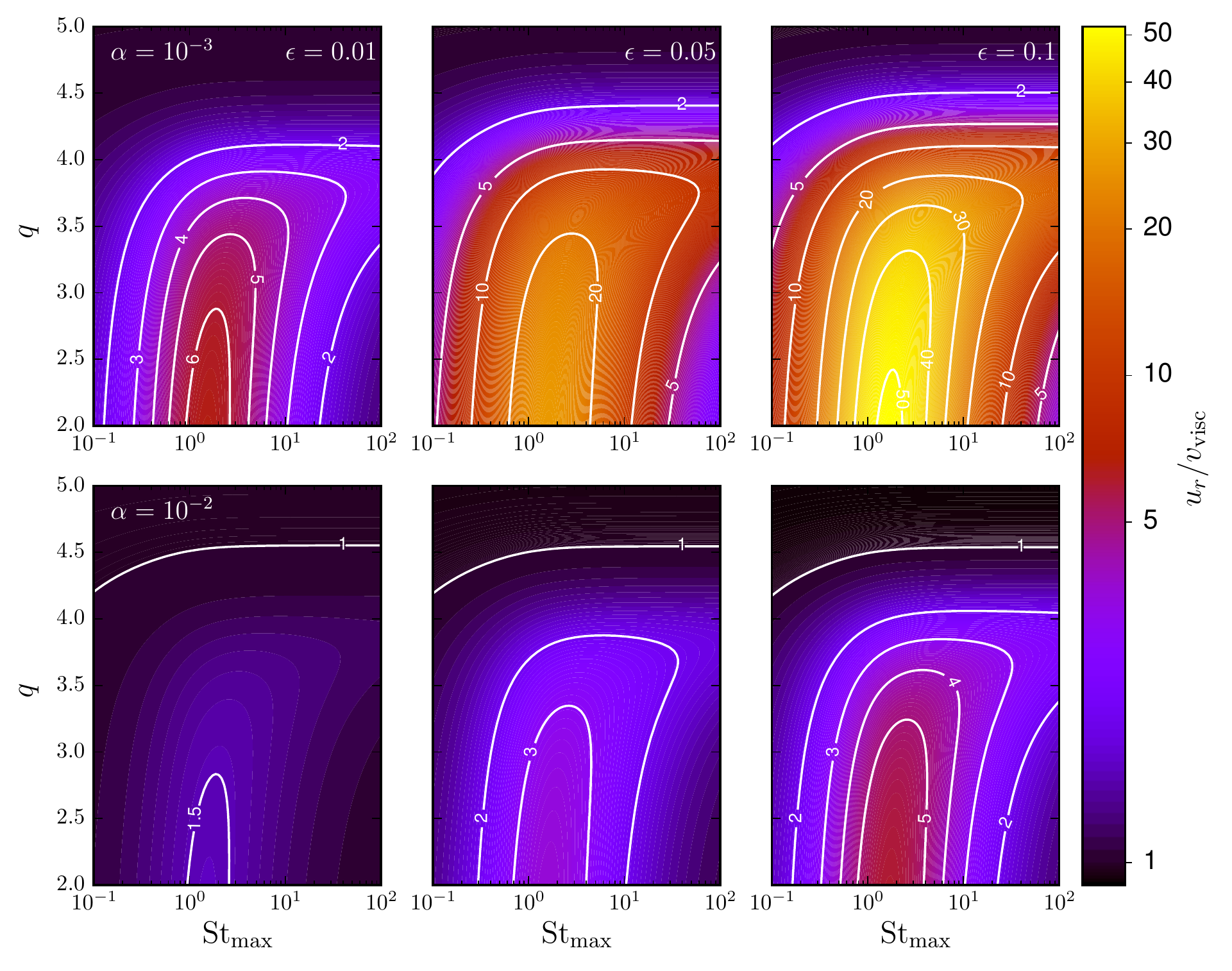}
\caption{Color and contour plot of the ratio of the gas radial velocity with respect to the viscous (i.e. dust-free) gas velocity for different grain-size distributions, described by two free parameters: $q$ and $\mathrm{St_{max}}$. The panels correspond to different values of the dust-to-gas ratio (from \textit{left} to \textit{right}): $[0.01,0.05,0.1]$ and different values of $\alpha$: (\textit{top}) $10^{-3}$ and (\textit{bottom}) $10^{-2}$. Significant deviations from the dust-free gas velocity are induced by the dust back-reaction, and are maximised for low viscosity and for shallower grain-size distribution (i.e. $q \leq 3$) with $\mathrm{St_{\mathrm{max}}}\sim [0.5,10]$.}
\label{fig:vrgsa}
\end{center}
\end{figure*}

\section{Gas and dust evolution at the disc midplane}
\label{sect:cumeffback}
In this Section we compute the steady-state gas and dust velocities at the midplane of typical protoplanetary discs with the aim  to infer the role of the back-reaction in the gas and dust dynamics adopting a continuous grain-size distribution (Sect.~\ref{sect:continudistrib}).

\subsection{Disc model}
\label{sect:discmodel1d}
We model a locally isothermal disc ($\gamma=1$) orbiting around a central star with mass $M_{\star}=1\,M_{\odot}$ with gas surface density $\Sigma_{\mathrm{g}} \propto r^{-1}$ and temperature $T\propto r^{-0.5}$. 
The dust component is modelled assuming truncated power-law grain-size distribution (Eq.~\ref{grainsizedistrib}), adopting different values of the power law index, $q$, and the minimum and maximum values of the grain-size distribution $s_{\mathrm{min}}$ and $s_{\mathrm{max}}$, corresponding to the minimum and maximum values of Stokes number $\mathrm{St_{min}}$ and $\mathrm{St_{max}}$.
Experimental studies of fragmentation over a single target found that $q$ can range from $\sim 1.9$ for low-velocity collisions to $\sim 4$ for catastrophic impacts \citep{davis90a,deckers14a}. Importantly, there is no specific reason to assume that in protoplanetary discs, $q$ should take the same value as in the ISM ($q_{\rm ISM}=3.5$). Moreover, since small grains with $\mathrm{St}\ll1$ do not affect strongly the gas motion, we set the minimum grain size of our distribution such as $\mathrm{St_{min}}=10^{-4}$ (no significant effect is found when decreasing this value).  The grain-size distribution is therefore described by just two free parameters: $q$ and $\mathrm{St_{max}}$.
In this paper, we focus our study on values of $q$ in the range $[2,5]$ and $\mathrm{St_{max}}$ in the range $[0.1,100]$, in order to explore the resulting dust and gas dynamics taking into account the different outcome of the grain growth processing. We assume a material grain density $\rho_{\mathrm{grain}}=1\, \mathrm{g\, cm^{-3}}$.  Moreover, in order to investigate the contributions from the viscous and pressure terms on to the dust and gas motions, we explore different values of the dust-to-gas ratio: $\epsilon =[0.01,0.05,0.1]$ and viscosity: $\alpha = [10^{-3},10^{-2}]$.
We discretise the continuous particle size distribution into $N_{\mathrm{bins}}=5000$ logarithmically spaced bins.
A numerical convergence test shows that a $N_{\mathrm{bins}} \simeq 50$ size bins is sufficient to discretise Eq.~\ref{eq:lambda01}. 

\subsection{Gas dynamics}
\label{sect:gasdyn2D}
Adding up the contributions of the different dust populations on to the gas motion, we find that the back-reaction has two effects (see Eq.~\ref{eq:vradg}). First, the drag from the tightly-coupled dust grains reduces the radial velocity of the gas (compared to a gas-only disc) as follows, 
\begin{equation}
u_{\mathrm{drag}}=\frac{1+\lambda_0}{\left(1+\lambda_0\right)^2+\lambda_1^2} v_{\mathrm{visc}} < v_{\mathrm{visc}},
\label{eq:vgrvisc}
\end{equation}
where $v_{\mathrm{visc}}$ at the midplane (see Eq.~\ref{eq:vvisc}) is given by 
\begin{equation}
v_{\mathrm{visc}}(r,z=0)=-\frac{3 \nu}{r} \frac{\partial \log \left( \nu \rho_{\mathrm{g}0} r^{-1/2}\right)}{\partial \log r},
\label{eq:vviscz0}
\end{equation}
where $\rho_{\mathrm{g}0 } \propto\Sigma_{\mathrm{g}}/H_{\mathrm{g}} \propto \Sigma_{\mathrm{g}} \Omega_{\mathrm{k}}/c_{\mathrm{s}} $ is the gas density at the midplane. In our disc model, the viscous velocity in the midplane is positive (see also Sect.~\ref{sect:gasvel3D}).
Second, the back-reaction produces a strong radial drift of gas (and, indirectly, small tightly-coupled grains) toward the outer disc regions, where the pressure decreases ($\partial P/ \partial r<0$ in our model). From Eq.~\ref{eq:vradg}, this effect occurs when the back-reaction from large grains is sufficient to push the gas outwards and carry small grains away. Quantitatively,
\begin{equation}
u_{\mathrm{drift}}=-\frac{\lambda_1}{\left(1+\lambda_0\right)^2+\lambda_1^2} v_{\mathrm{P}},
\label{eq:udrift}
\end{equation}
where
\begin{equation}
\frac{v_{\mathrm{P}}}{v_{\mathrm{k}}}  \left(r,z=0 \right) =- \left( \frac{H_{\mathrm{g}}}{r} \right)^2 \left | \frac{\partial \log P_0}{\partial \log r}\right| ,
\label{eq:vpvkz}
\end{equation}
where $P_0=\rho_{\mathrm{g}0} c_{\mathrm{s}}^2$ is the pressure at the disc midplane. Since in our model (see also Sect.~\ref{sect:dustvelmodel}) $v_{\mathrm{P}}$ is negative, the drift component $u_{\mathrm{drift}}$ is therefore positive. Furthermore, $u_{\mathrm{drift}}$ varies with the local grain-size distribution, since its value is mostly determined by the mass carried by the grains with $\mathrm{St}\sim 1$ (see Eq.~\ref{eq:lambda_int}). By comparing the order of magnitude of $v_{\mathrm{visc}}$ and $v_{\mathrm{P}}$ (Eqs.~\ref{eq:magnvp} and \ref{eq:magnvvisc}), we expect $u_{\mathrm{drift}}$ to dominate the gas motion when enough mass is embodied in large grains. However, since $u_{\mathrm{drag}}$ and $u_{\mathrm{drift}}$ both depend on $\lambda_{k}$, the relative importance of these components depends on the grain-size distribution. 
\begin{figure}
\begin{center}
\includegraphics[height=0.36\textwidth]{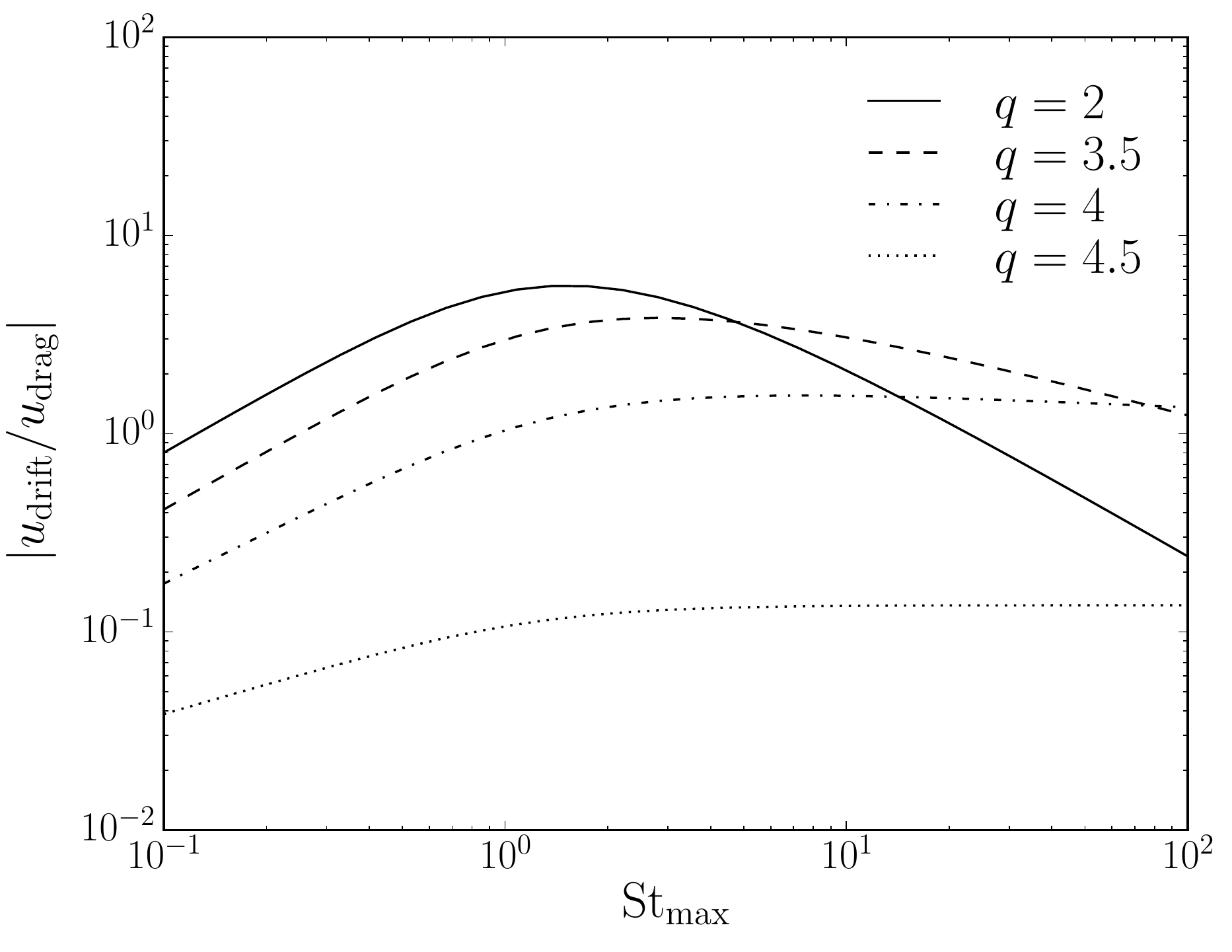}
\caption{Ratio of the drift and viscous drag component of the gas radial velocity for different vales of $q$, assuming $\epsilon=0.01$ and $\alpha=10^{-3}$ (top-left panel of Fig.~\ref{fig:vrgsa}). For $\mathrm{St_{max}}\gg1$, the drift-to-drag ratio decreases with increasing $q$ and spreads over larger values of $\mathrm{St_{max}}$, resembling the dependences with $\mathrm{St_{max}}$ shown in Eq.~\ref{eq:caseratio}. For $q>4$, the ratio depends on $\mathrm{St_{min}}$ and is $\ll 1$ for typical values of $\mathrm{St_{min}}$ and $\epsilon/\alpha$.}
\label{fig:ratiolambda}
\end{center}
\end{figure}

Fig.~\ref{fig:vrgsa} shows the ratio of the gas radial velocity with respect to the viscous (i.e. dust-free) gas velocity (Eq.~\ref{eq:vviscz0}) computed at $r=1$ au for different values of the dust-to-gas ratio $\epsilon=[0.01,0.05,0.1]$ and viscosity $\alpha=[10^{-3},10^{-2}]$. All the grains in our model aerodynamically interact with the gas in the Epstein regime (first expression in Eq.~\ref{eq:stoppingtimecase}). It can be noticed that, even assuming the canonical ISM dust-to-gas ratio (i.e. $\epsilon=0.01$, see left panels), the cumulative effect of the back-reaction leads to an increase of the gas velocity over the gas viscous velocity. This effect is maximized for smaller index of the grain-size distribution (i.e. $q\to 2$), since in these cases most of the mass is embodied in larger grains which mostly affect the gas motion. 
Fig.~\ref{fig:vrgsa} shows that grains with $\mathrm{St_{\mathrm{max}}}\sim [0.5,10]$ are mostly able to affect the gas motion, with a maximum efficiency of grains in the range $\mathrm{St_{\mathrm{max}}}\sim [1,3]$, depending on the value of $q$. 
The range of Stokes numbers where the effect of the back-reaction is non-negligible increases with increasing index of the power law for $q\leq 4$. Interestingly, for $q\gtrsim 4$ the amount of dust mass embodied in large dust grains that are responsible for the drift term decreases, leading to a reduction of the drift component with respect to the viscous drag component, regardless from the value of $\mathrm{St_{\mathrm{max}}}$. This results in a reduction of the gas motion under the viscous dust-free motion (see Eq.~\ref{eq:vgrvisc}) due to the resistance to the gas viscous flow of non-migrating dust particles. However, since the viscous drag component is typically a factor $\alpha$ lower than the pressure component (see Eqs.~\ref{eq:magnvp} and \ref{eq:magnvvisc}), the reduction is negligible (less than 10\%).
Increasing the value of $\alpha$ (see bottom panels of Fig.~\ref{fig:vrgsa}) results in an increase of the viscous drag component with respect to the drift component, leading to a reduced outflow of the gas compared to the low viscosity case. Moreover, the transition from drift to viscous drag dominated regime occurs at lower values of $q$.

The radial dynamics of gas and tightly-coupled dust grains can be understood by investigating the ratio $\left| u_{\mathrm{drift}} / u_{\mathrm{drag}} \right|$ with different grain-size distributions.  Adopting the estimates shown in Eq.~\ref{eq:casedeltah} to evaluate $\lambda_k$ (Eq.~\ref{eq:lambda_int2}), we can infer the dependence of the drift-to-drag ratio for different values of the maximum and minimum sizes of the distribution, assuming $\mathrm{St}_{\rm min} \ll 1$ and $\mathrm{St}_{\rm max} \gg 1$ in the Epstein regime. We obtain
\begin{equation}
\left | \frac{u_{\mathrm{drift}}}{u_{\mathrm{drag}}} \right | \simeq   \frac{\lambda_1}{\alpha \left(1+\lambda_0 \right)} \propto \frac{\epsilon}{\alpha}
\begin{cases}
\mathrm{St_{min}}^{q-4}  , & \text{$q >4$,} \\  
\mathrm{St_{max}}^{q-4}   , & \text{$3<q<4$,}\\
 \mathrm{St_{max}}^{-1}, & \text{$q<3$.}
\end{cases}
\label{eq:caseratio}
\end{equation}
For the specific cases $q=[2,3,4]$, there is not a simple dependence of this ratio with  $\mathrm{St_{max}}$ and $\mathrm{St_{min}}$. Fig.~\ref{fig:ratiolambda} shows the ratio between the drift and viscous drag component for different values of $q$, assuming $\epsilon=0.01$ and $\alpha=10^{-3}$. For shallow distributions ($q<4$), the drift-to-drag ratio peaks at intermediate $\mathrm{St_{max}}$ and falls off as $\mathrm{St_{max}}$ increases. For distributions with $q$ in the range $[3,4]$, the decrease of the drift-to-gas ratio with increasing $\mathrm{St_{max}}$ becomes less steep compared to the case where $q<3$, as shown in Fig.~\ref{fig:vrgsa}.
On the other hand, for steeper distributions ($q>4$), since most of the mass is embodied in small grains, the drift-to-drag ratio depends on $\mathrm{St_{min}}$ and, for typical protoplanetary discs with $\epsilon \lesssim \alpha$, it is smaller than unity and decreases with increasing $q$. 
As shown in Figs.~\ref{fig:vrgsa} and \ref{fig:ratiolambda}, by increasing the value of $q$, the drift component decreases with respect to the viscous drag component and spreads over larger values of the Stokes number, resembling the dependences with $\mathrm{St_{max}}$ shown in Eq.~\ref{eq:caseratio} for $\mathrm{St_{max}}\gg1$. 
Moreover, the drift-to-drag ratio increases with $\epsilon/\alpha$, implying that, since $\alpha\sim \left[10^{-2}, 10^{-4} \right]$ in typical protoplanetary discs, the drift can dominate the viscous drag component even for the canonical ISM dust-to-gas ratio.

%
\begin{figure*}
\begin{center}
\includegraphics[height=0.1954\textwidth,trim={0 0 0 0},clip]{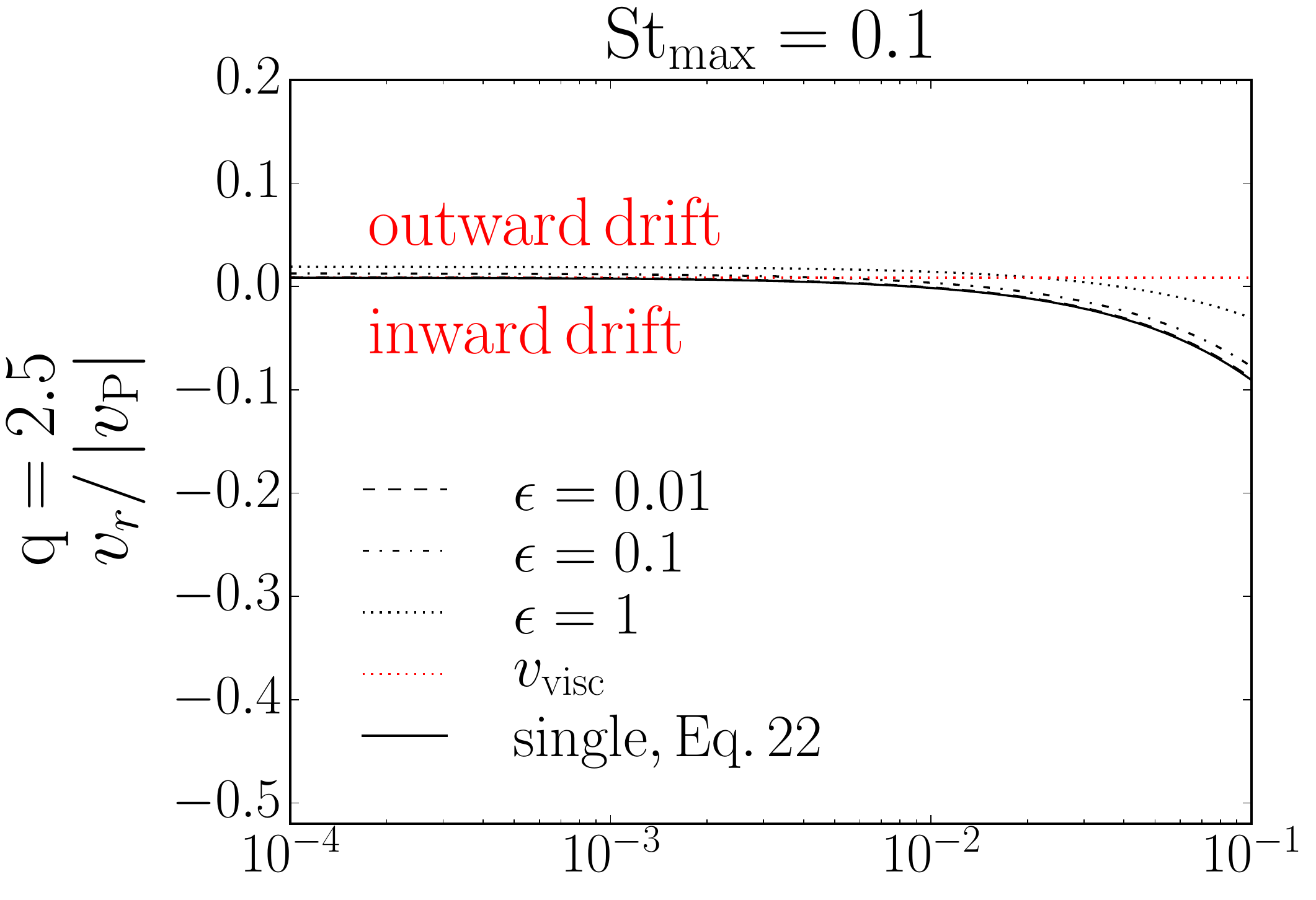}
\includegraphics[height=0.1954\textwidth,trim={1.3cm 0 0 0},clip]{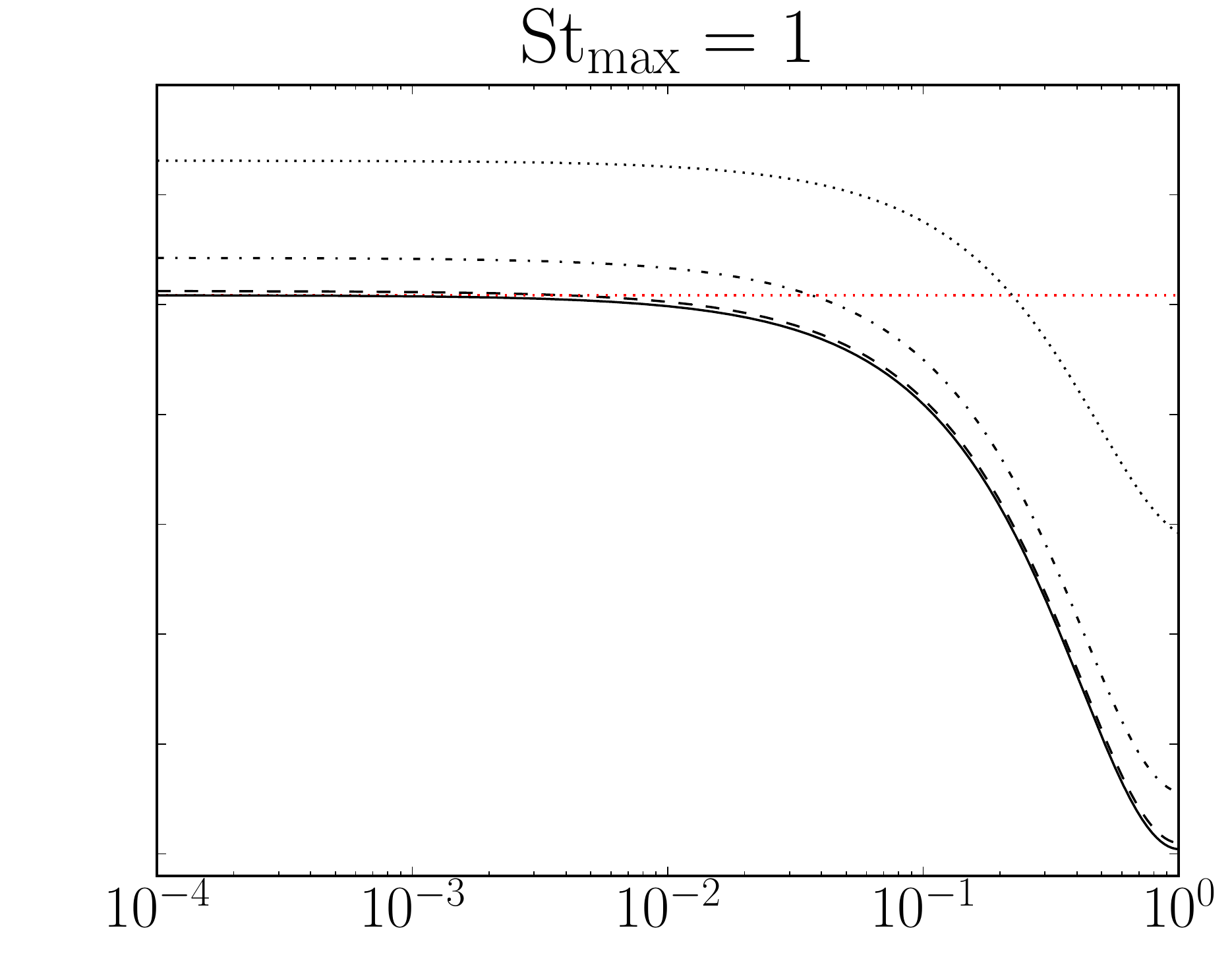}
\includegraphics[height=0.1954\textwidth,trim={1.3cm 0 0 0},clip]{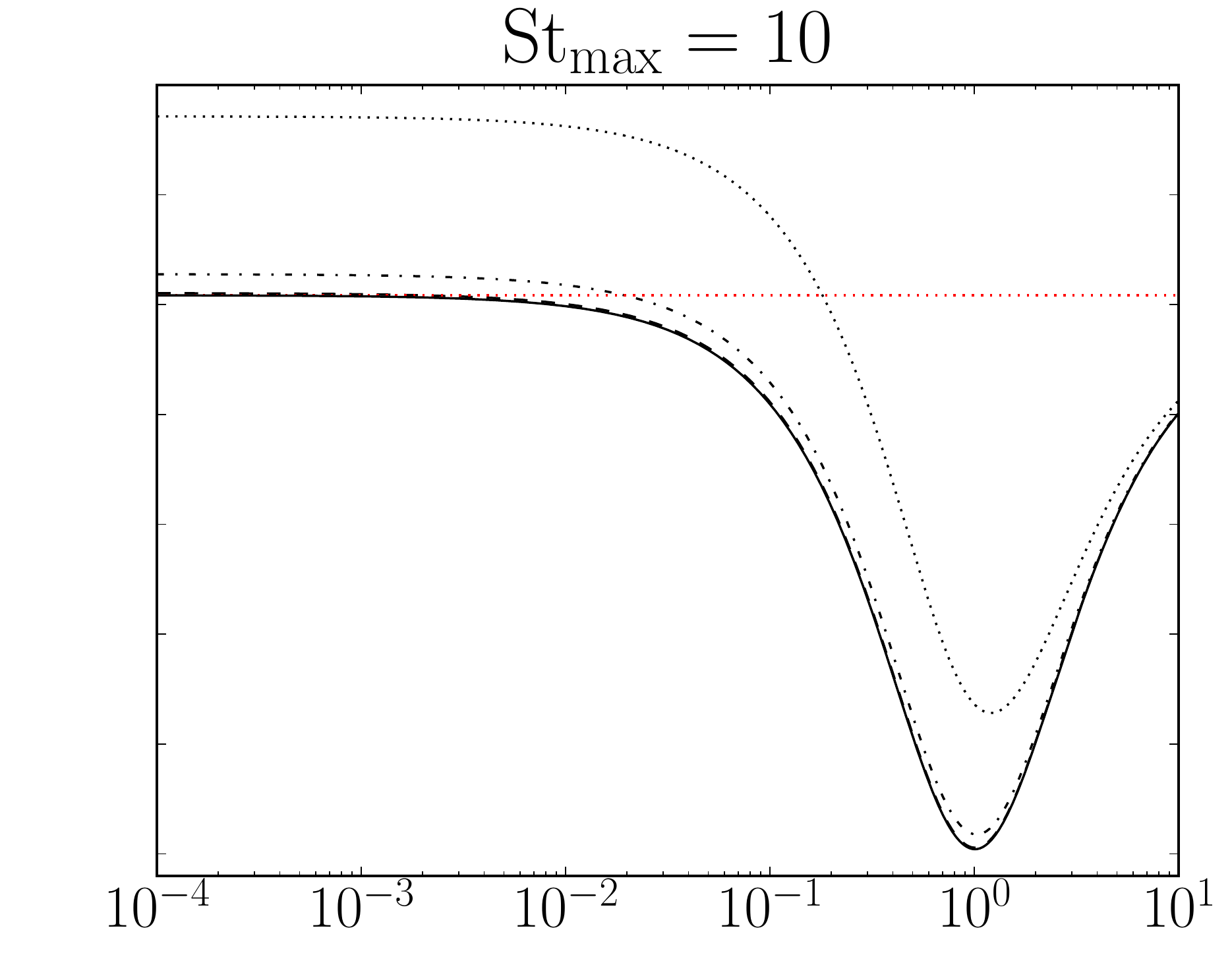}
\includegraphics[height=0.1954\textwidth,trim={1.3cm 0 0 0},clip]{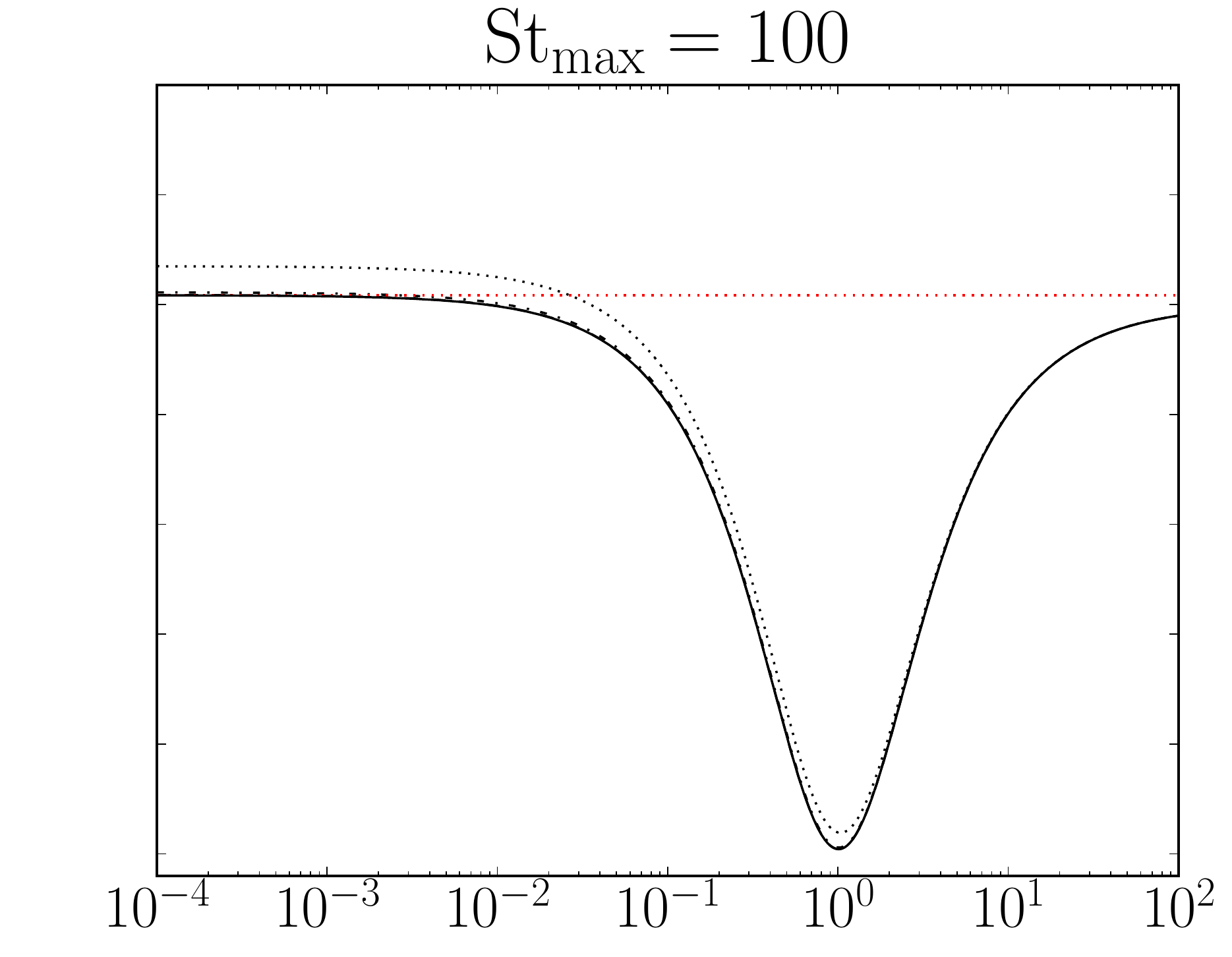}
\includegraphics[height=0.1869\textwidth,trim={0 0 0 0},clip]{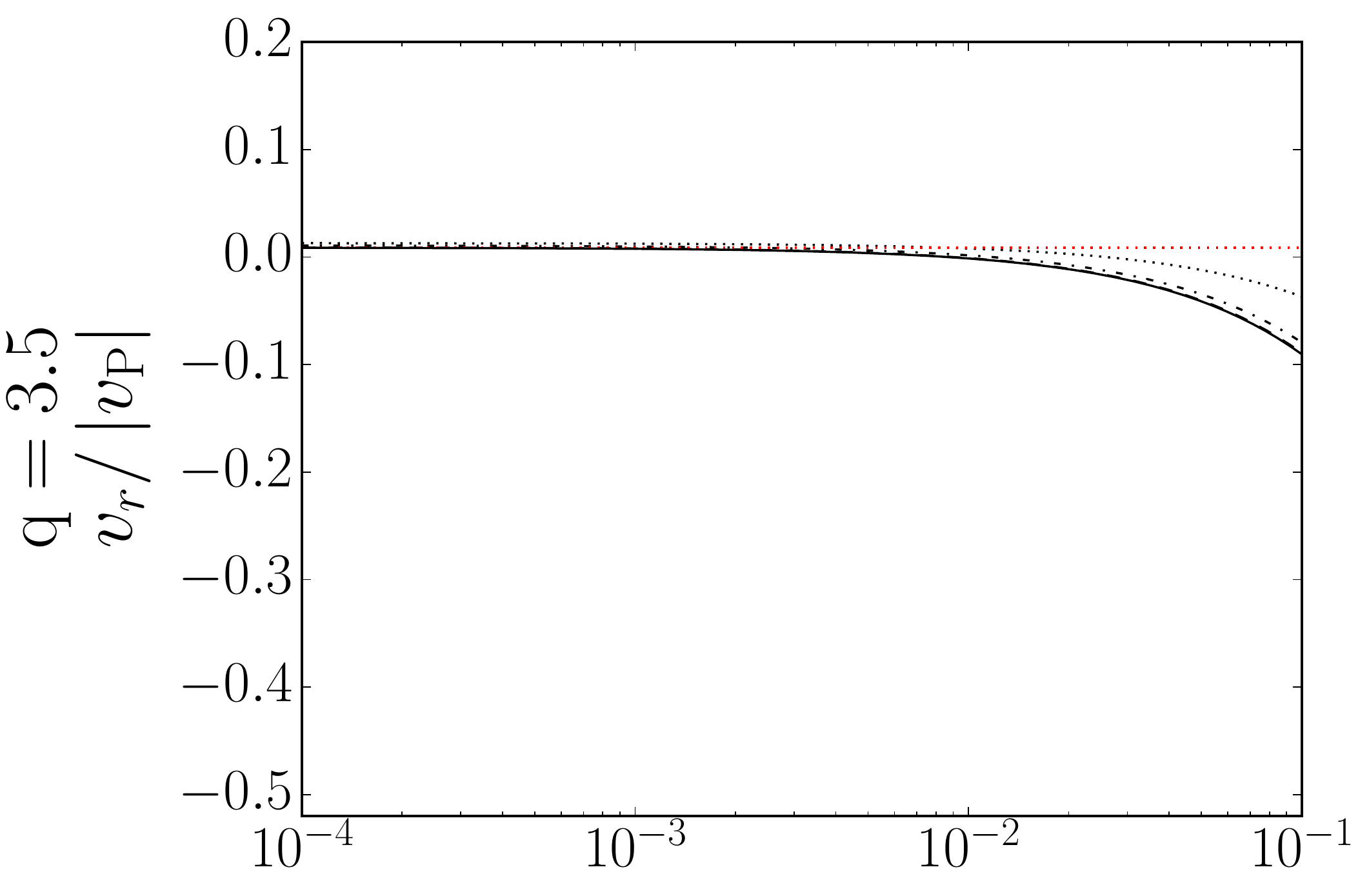}
\includegraphics[height=0.1869\textwidth,trim={1.3cm 0 0 0},clip]{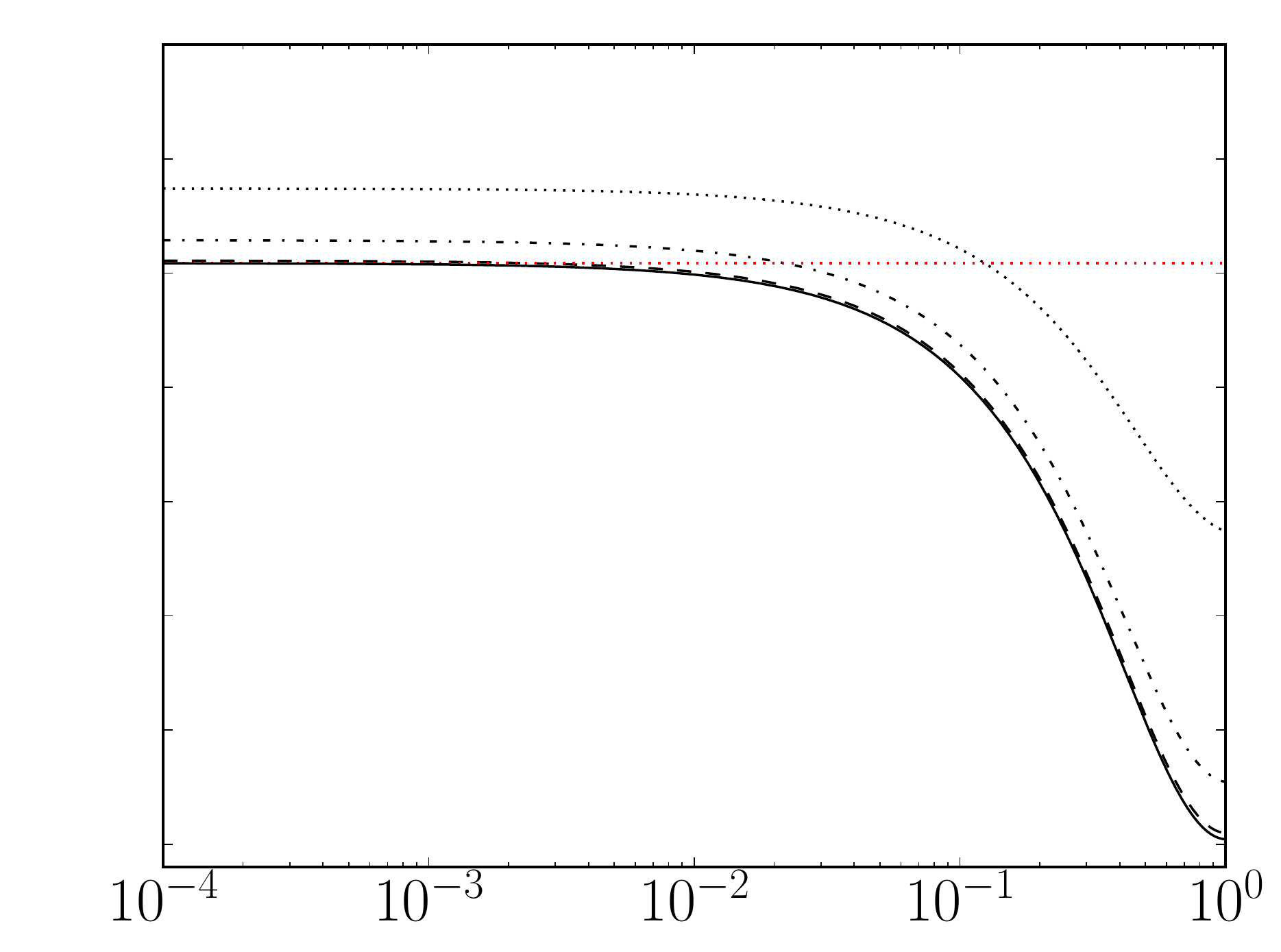}
\includegraphics[height=0.1869\textwidth,trim={1.3cm 0 0 0},clip]{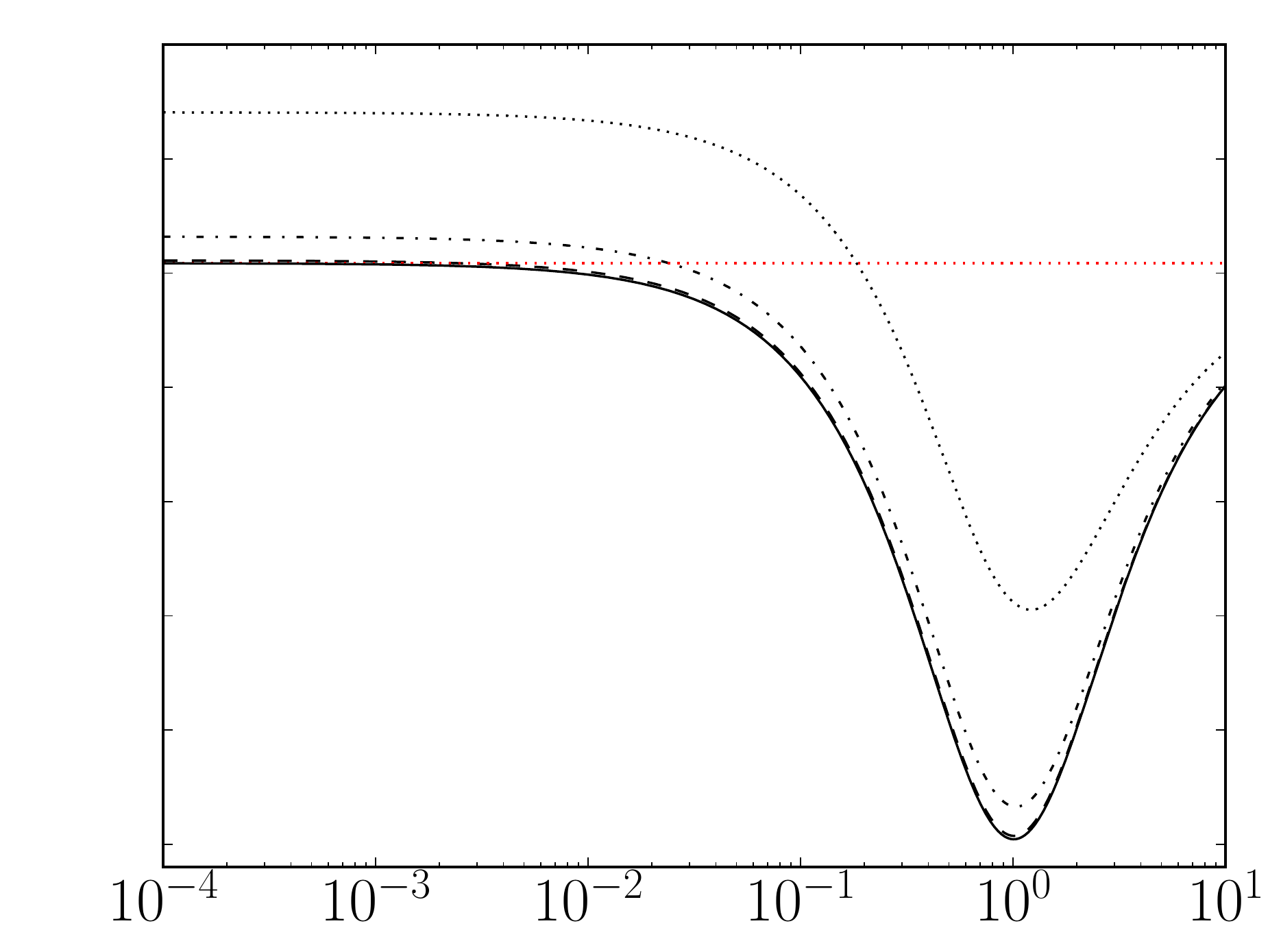}
\includegraphics[height=0.1869\textwidth,trim={1.3cm 0 0 0},clip]{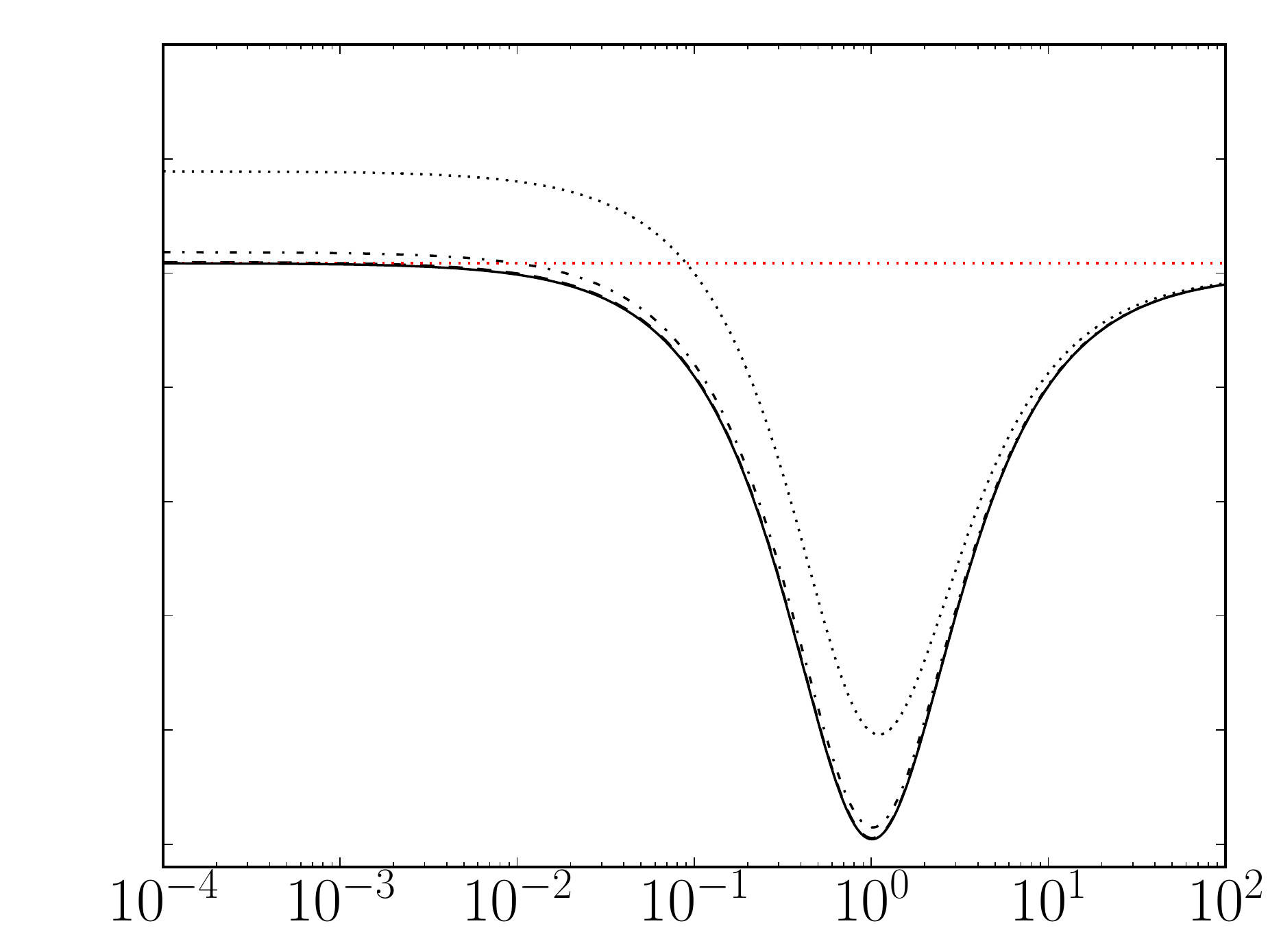}
\includegraphics[height=0.2033\textwidth,trim={0 0 0 0},clip]{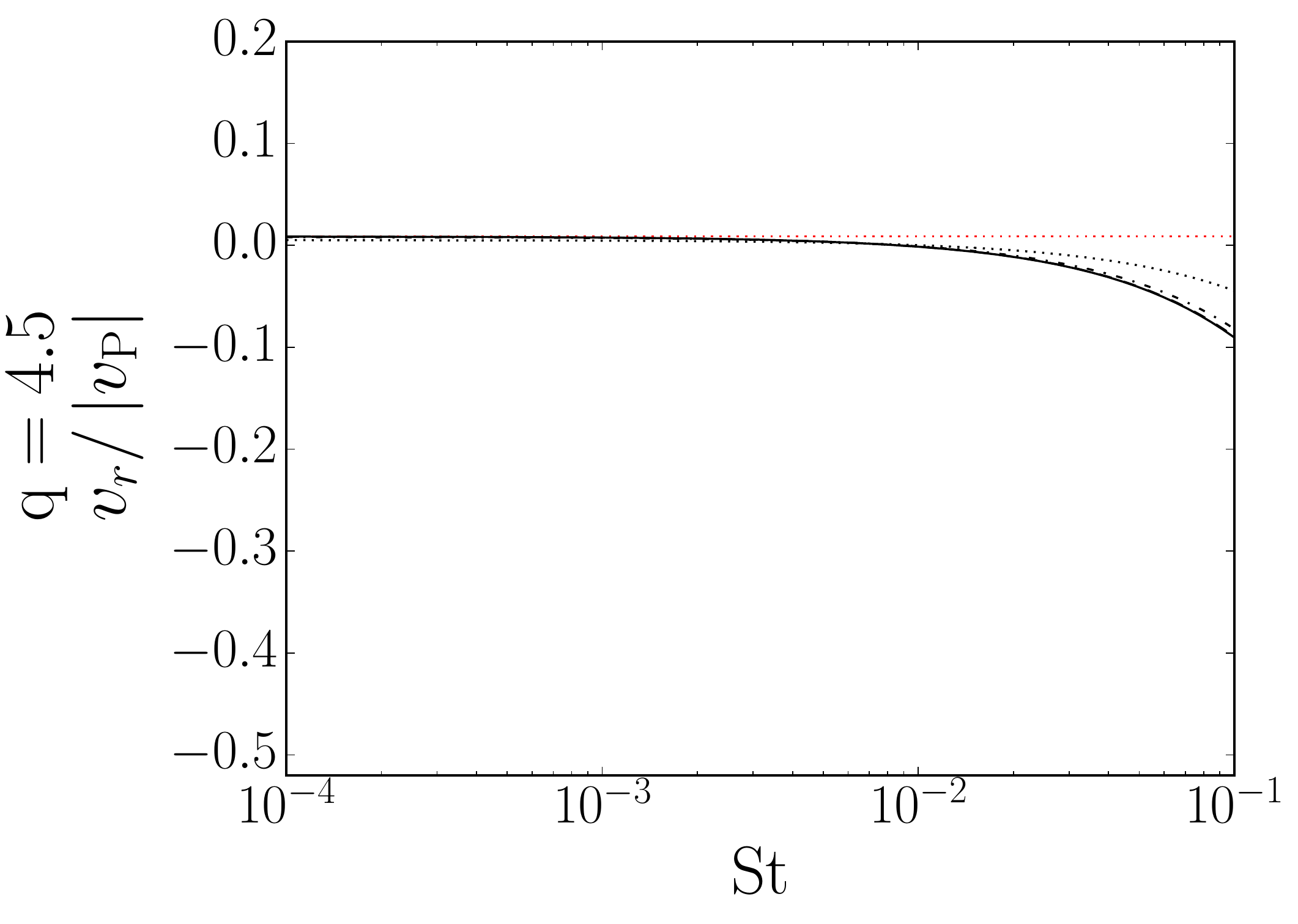}
\includegraphics[height=0.2033\textwidth,trim={1.3cm 0 0 0},clip]{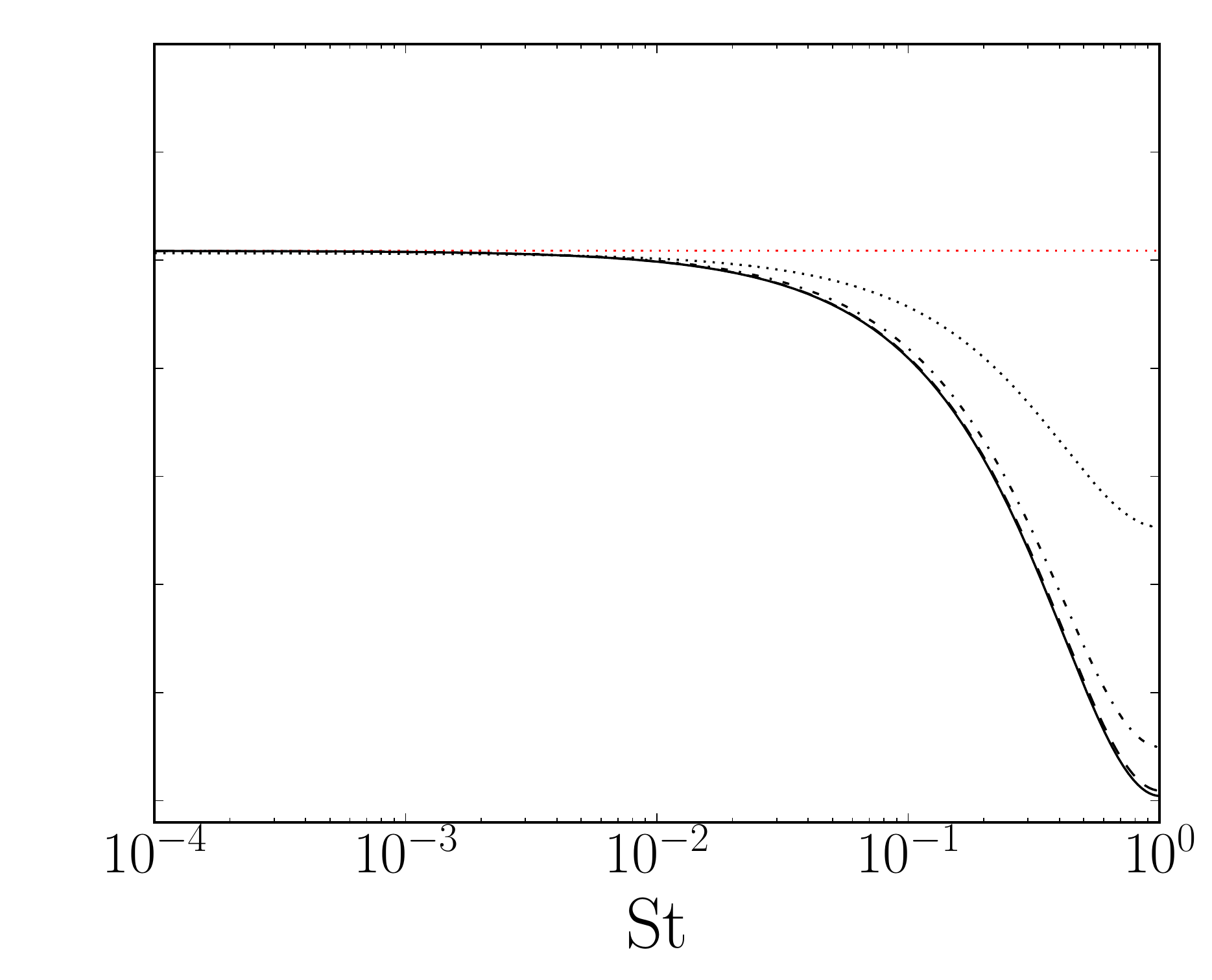}
\includegraphics[height=0.2033\textwidth,trim={1.3cm 0 0 0},clip]{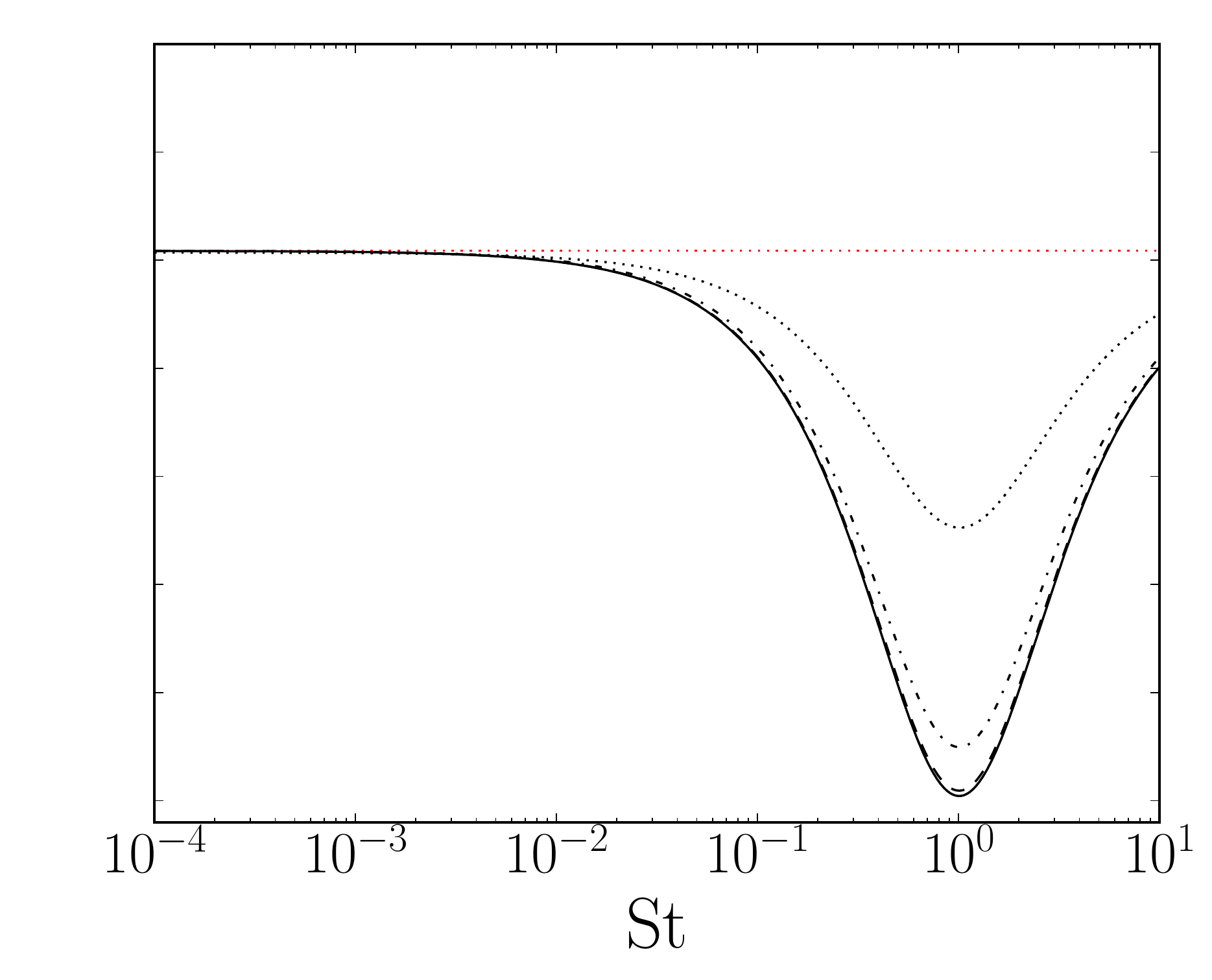}
\includegraphics[height=0.2033\textwidth,trim={1.3cm 0 0 0},clip]{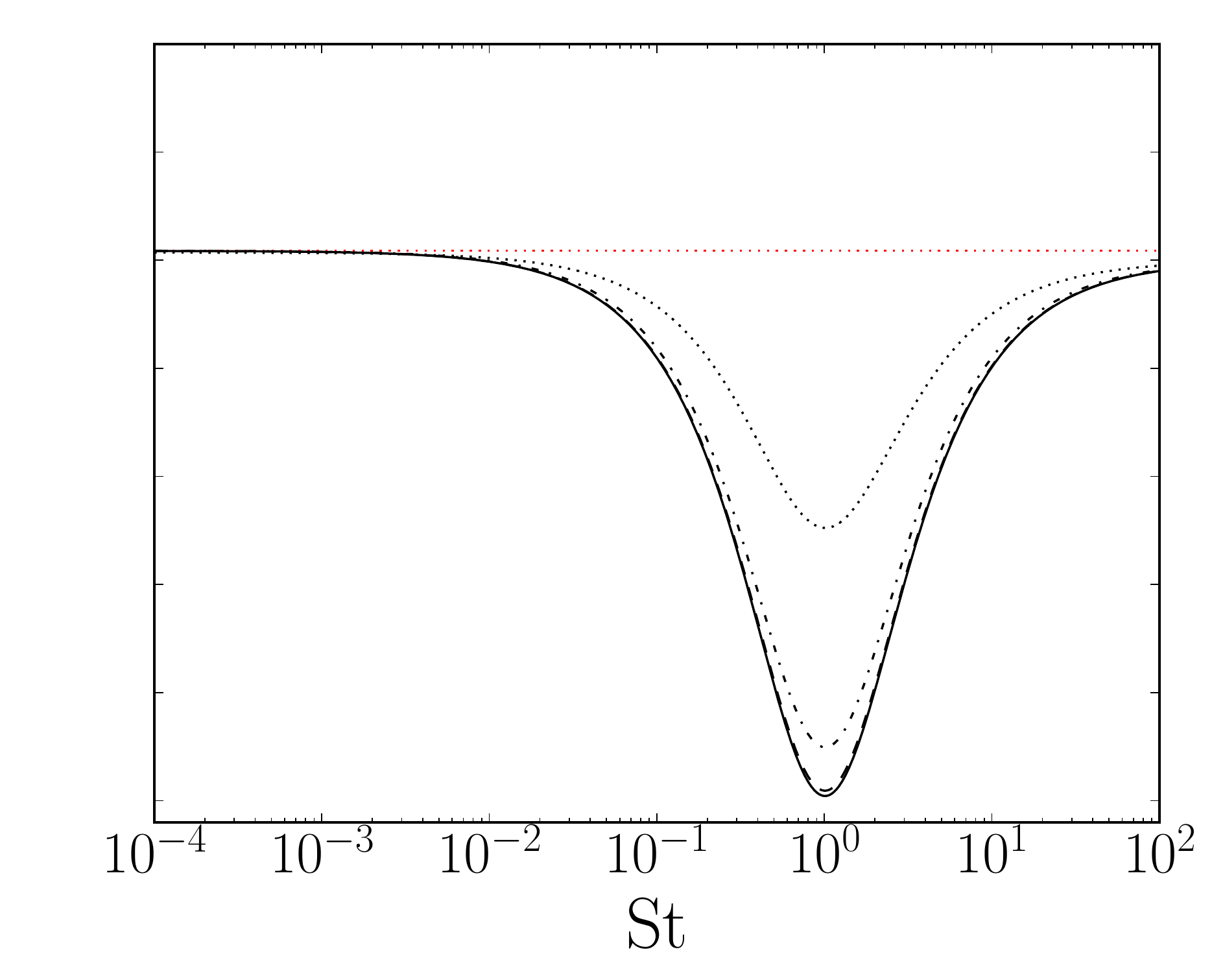}
\caption{Radial drift velocities for dust species in populations with different properties. Columns (\textit{left} to \textit{right}) vary the maximum Stokes numbers, $\mathrm{St_{max}}=[0.1,1,10,100]$, while rows (\textit{top} to \textit{bottom}) vary the index of the power-law size distribution $q=[2.5,3.5,4.5]$. Solid lines denote the radial drift velocity expected for an equilibrium with single-species (Eq.~\ref{eq:vradustsimp}). The dotted red line denotes the viscous dust-free gas velocity expressed in Eq.~\ref{eq:vviscz0}. While the radial drift velocities of large grains are reduced compared to the single-phase population, tightly-coupled particles follow the outward gas motion induced by the back-reaction, especially in those cases characterized by large values of $\epsilon$, size distributions with $q < 4$ and $\mathrm{St_{max}}\sim 1$--$10$.}
\label{fig:vdrchangest}
\end{center}
\end{figure*}

\subsection{Dust dynamics}
Fig.~\ref{fig:vdrchangest} compares the dust velocities from single- and multi-phase dust populations, with more or less steep grain-size distributions and values of the total dust-to-gas ratios that are relevant for protoplanetary discs. It can be noticed that the radial drift velocities of large grains are reduced compared to the single-phase dust population. The effect becomes more important with increasing $\epsilon$, as previously noted by \citet{bai10a}. Fig.~\ref{fig:vdrchangest} shows that this decrease is most pronounced for distributions satisfying $\mathrm{St_{max}}\sim 1-10$ and small values of the exponent $q$. In particular, steep size distributions with $q\geq 4$ do not exhibit any noticeable effect on the dust radial drift for any value of $\mathrm{St_{max}}$, since most of the mass is embodied in small grains which do not affect the orbital motion of the gas (see Fig.~\ref{fig:vrgsa}). The role played by dust back-reaction is apparent when comparing the radial motion of small grains of Fig.~\ref{fig:vdrchangest} with the dust-free gas velocity induced by viscous torques (see the dotted red line). When the dust-to-gas ratio is sufficiently large and the size distribution is moderately steep ($q < 4$) and extends to values of $\mathrm{St_{max}}\gtrsim 1$, the motion of small grains is dominated by the outward motion of the gas induced by the dust back-reaction (Sect.~\ref{sect:gasdyn2D}).
As a result, while tightly-coupled particles follow the outward gas motion, marginally-coupled grains feel a smaller headwind due to the increased orbital velocity of the gas, leading to a decrease of their exchange of angular momentum with the gas and a consequent reduction of their inward motion (see, e.g., the second and third plot from the left on the top panels in Fig.~\ref{fig:vdrchangest}). The efficiency of these two effects depends on that part of dust mass embodied in grains able to affect the gas motion, which increases with decreasing $q$. 
With increasing dust-to-gas ratio the gas becomes more entrained by the solids, leading to stronger reduction of the drift velocity for large particles and an increase of the outward migration of gas and tightly-coupled particles.

\section{Vertically stratified discs}
\label{sect:dynamicsvert}
In this Section we extend the analysis described in Sect.~\ref{sect:cumeffback} to compute the vertically averaged steady-state gas and dust velocities, adopting typical models for vertically stratified protoplanetary discs. The aim is to infer the disc conditions that favour the role of the back-reaction in shaping the gas and dust evolution in typical protoplanetary discs with realistic grain-size distributions.

\subsection{Vertical structure of the gas}
\label{sect:gasvertstruct}

\subsubsection{Density}
\label{sect:gasdensity}
We assume a vertically isothermal disc in which the dust distribution has relaxed towards a steady-state (see Sect.~\ref{sect:dustvert}). Vertical hydrostatic equilibrium arises from the balance between gravity from the central star, gas pressure and dust back-reaction. Hence \citep{nakagawa86a},
\begin{equation}
\frac{\partial \rho_{\mathrm{g}}}{\partial z}=-\frac{\rho_{\mathrm{g}}}{c_s^2} \left( 1 + \epsilon \right)\frac{\mathcal{G}M_{\star} z}{\left(r^{2} +  z^{2}\right)^{3/2}} ,
\label{eq:gasdensvert}
\end{equation}
where $M_{\star}$ is the mass of the central star and $\mathcal{G}$ is the gravitational constant.
Eq.~\ref{eq:gasdensvert} shows that gas pressure accounts for the weight of the dust, potentially affecting the dust dynamics via back-reaction. However, since $\epsilon\ll1$, we can safely make the approximation $1 + \epsilon \sim 1$ and neglect back-reaction in the vertical direction (see Appendix B of \citealt{kanagawa17a}). The typical scale height $H_{\mathrm{g}}$ resulting from this balance is 
\begin{equation}
H_{\mathrm{g}} \left(r \right) \equiv \frac{c_{\mathrm{s}}}{\Omega_{\mathrm{k,mid}}} ,
\label{eq:height_gas}
\end{equation}
where $\Omega_{\mathrm{k,mid}} \equiv \sqrt{\mathcal{G}M_{\star}/r^3}$ is the Keplerian angular velocity in the midplane. Since $H_{\rm g} / r \ll 1$, the solution to Eq.~\ref{eq:gasdensvert} can be expanded to first order in $z^{2} / r^{2}$, which gives
\begin{equation}
\rho_{\mathrm{g}}\left (r,z\right)=\rho_{\mathrm{g}0} \left (r \right)\exp{\left (-\frac{z^2}{2H_{\mathrm{g}}^2} \right )} .
\label{eq:rhoz}
\end{equation}
For a thin protoplanetary disc, the departure of $\rho_{\mathrm{g}}$ (computed by solving Eq.~\ref{eq:gasdensvert}) from the Gaussian profile given by Eq.~\ref{eq:rhoz} is negligible (see Fig.~3 in \citealt{armitage15a}).

\subsubsection{Velocities}
\label{sect:gasvel3D}
In the absence of dust, the centrifugal balance of forces in the radial direction yields
\begin{equation}
\frac{u_{ \phi}^2}{r}= \frac{\mathcal{G} M_{\star} r}{\left(r^2+z^2\right)^{3/2}} +\frac{1}{\rho_{\mathrm{g}}}\frac{\partial P}{\partial r} .
\label{eq:vphigas2}
\end{equation}
%
Combining Eqs.~\ref{eq:rhoz} and \ref{eq:vphigas2}, the angular velocity of the gas $\Omega_{\mathrm{g}}\equiv r u_{ \phi}$ can be written as (\citealt{takeuchi02a}, see Appendix~\ref{app:steady} for further details)
\begin{equation}
\Omega_{\mathrm{g}}\left(r,z\right) = \Omega_{\mathrm{k,mid}} \sqrt{1- \left( \frac{H_{\mathrm{g}}}{r} \right)^2 \left [p+\frac{m+3}{2} + \frac{m}{2} \left(\frac{z^2}{H_{\mathrm{g}}^2}\right)\right ]} ,
\label{eq:omegakz}
\end{equation}
where we assume $\Sigma_{\mathrm{g}}\propto r^{-p}$ and $T\propto r^{-m}$. Eq.~\ref{eq:omegakz} shows that vertical stratification modifies the orbital velocity in a way which is \textit{not} perturbative, even for a moderate fraction of the scale height above the midplane. Similarly, the viscous velocity depends on $z$ according to (\citealt{takeuchi02a}, see Appendix~\ref{app:steady} for further details)
\begin{equation}
v_{\mathrm{visc}}(r,z)=\frac{\nu}{2r} \left [6p+m-3+\left(5m-9\right) \left(\frac{z^2}{H_{\mathrm{g}}^2}\right) \right].
\label{eq:vviscrz}
\end{equation}
Eq.~\ref{eq:vviscrz} shows that gas flows outwards close to the midplane and inwards above the height 
\begin{equation}
\left | z \right | \sim \left | \frac{6p+m-3}{9-5m} \right |^{1/2} H_{\mathrm{g}}.
\end{equation}
For typical disc parameters ($p=1$ and $m=1/2$), this critical height occurs at $\left | z \right | \sim 0.7 H_{\mathrm{g}}$ (see Fig.~1 in \citealt{takeuchi02a}). \newtext{In laminar viscous discs (contrary to discs with turbulence driven by magneto-rotational and vertical shear instabilities, see \citealt{flock11a} and \citealt{stoll17b}), the gas is therefore expected to flow outward at the disc midplane and inward in the disc upper layers, producing the so-called \textit{meridional circulation} \citep{urpin84a,kley92a,philippov17a}. This motion can represent a natural way of driving global outward radial transport of dust particles and could explain the refractory inclusions in meteorites \citep{takeuchi02a,ciesla09a,hughes10a}.} Hence, Eqs.~\ref{eq:omegakz} and \ref{eq:vviscrz} show that both the radial and the orbital velocities of the gas vary in the vertical direction, which affects the efficiency of gas-dust viscous drag and drift.  

\subsection{Vertical structure of the dust}
\label{sect:dustvert}
\subsubsection{Density}
\label{sect:destvertdens}
Since in our analysis we assume $\epsilon \ll1$, we can neglect the effect of dust back-reaction on to the dust vertical distribution and safely treat the particles as trace contaminants \citep[e.g.][]{morfill84a,clarke88a}. An important consequence of gas stratification is that the Stokes number of the particles depends on the vertical height in the disc. With the gas density given by Eq.~\ref{eq:rhoz}, the Stokes number of the $i^{\mathrm{th}}$ dust species in the Epstein regime (see the first expression in Eq.~\ref{eq:stoppingtimecase}) is given by
\begin{equation}
\mathrm{St}_i = \frac{\pi}{2}  \frac{ \rho_{\mathrm{grain}} s_i }{\Sigma_{\mathrm{g}} }  \exp{\left [\frac{z^2}{2H_{\mathrm{g}}^2} \right ]} \equiv \mathrm{St}_{\mathrm{mid},i}\exp{\left [\frac{z^2}{2H_{\mathrm{g}}^2} \right ]},
\label{eq:stokevarz}
\end{equation}
where $\mathrm{St}_{\mathrm{mid},i}$ denotes the Stokes number in the midplane. Hence, the same grain can be tightly-coupled to the gas in the midplane, but decoupled as it reaches heights of, e.g., $\sim 2$--$3\,H_{\rm g}$. From a dynamical point of view, drag damps the vertical motion of the particles and makes them settle towards the midplane of the disc. On the other hand, drag also couples the grains to any stochastic fluctuations of gas velocity  \citep{weidenschilling84a}, which are presumably due to gas turbulence. 
We assume the vertical turbulent diffusivity of the gas is of order of the turbulent kinematic viscosity $\nu$ (see Fig.~16 of \citealt{zhu15c}), although the gas turbulence is not expected to be homogeneous in vertically stratified protoplanetary discs (\citealt{stoll17b}, see Sect.~3.2 of \citealt{fromang09a}). Moreover, we simply assume that the dust diffusivity in the vertical direction is well approximated by the turbulent kinematic viscosity $\nu$.
The vertical evolution of the dust distribution can therefore be described by a Fokker-Planck equation \citep{carballido06a,fromang09a}. The steady-state density distribution, reached after a typical settling time, is given by 
\begin{equation}
\rho_{\mathrm{d}i}=\rho_{\mathrm{d}i,0} \exp{\left \lbrace-\frac{z^2}{2H_{\mathrm{g}}^2}-\frac{\mathrm{St}_{\mathrm{mid},i}}{\alpha} \left [ \exp{\left (\frac{z^2}{2H_{\mathrm{g}}^2} \right)}-1 \right ] \right\rbrace} ,
\label{eq:dustvertprofdiff}
\end{equation}
where $\rho_{\mathrm{d}i,0}$ denotes the density of the $i^{\mathrm{th}}$ dust species in the midplane. Eq.~\ref{eq:dustvertprofdiff} shows that the dust distribution is almost Gaussian and concentrated close the midplane when $\mathrm{St}_{\mathrm{mid},i} \gtrsim \alpha$ and becomes more step-like with an effective width of $\sim 3 H_{\rm g}$ for $\mathrm{St}_{\mathrm{mid},i} < \alpha$ (see Fig.~4 of \citealt{takeuchi02a}). The typical scale height of the $i^{\mathrm{th}}$ dust phase is defined according to 
\begin{equation}
H_{\mathrm{d}i} \equiv \sqrt{ \frac{1}{\rho_{\mathrm{d}i,0}}\int_{-\infty}^{+\infty} z^{2} \rho_{\mathrm{d}i}\left( z \right) \mathrm{d}z } . 
\label{eq:Hd_def}
\end{equation}
Although no closed-form analytical expression exists for Eq.~\ref{eq:Hd_def} with the dust distribution given by Eq.~\ref{eq:dustvertprofdiff}, the following convenient estimate is usually adopted \citep{dubrulle95a,youdin07a}
\begin{equation}
H_{\mathrm{d}i}=H_{\mathrm{g}}\sqrt{\frac{\alpha}{\alpha+\mathrm{St}_{\mathrm{mid},i}}}.
\end{equation}
In this model, we remark that grains do not interact with each other. When sticking and destructive collisions are considered, small grains may be trapped within thin layers when the time-scale for which small particles are swept up by large grains is shorter than the mixing time-scale, which results in an enhanced amount of small particles close to the midplane \citep{krijt16a}. 

\subsubsection{Velocities}
\label{sect:dustvelmodel}
In the absence of gas, dust orbits the star at the local Keplerian velocity. The actual inclination of the orbit with respect to the midplane of the disc can be approximated to order $ z^{2}/r^{2}$ according to
\begin{equation}
\Omega_{\mathrm{d}}\left(r,z\right) = \sqrt{\frac{\mathcal{G} M_{\star} r}{\left(r^2+z^2\right)^{3/2}}} \simeq \Omega_{\mathrm {k,mid}} \left(1 - \frac{3}{4}\frac{z^{2}}{r^{2}} \right).
\label{eq:omegakvert}
\end{equation}
To the same order, the differential orbital velocity between the gas (Eq.~\ref{eq:omegakz}) and the dust (Eq.~\ref{eq:omegakvert}) is therefore given by
\begin{equation}
\frac{\Omega_{\mathrm{g}}-\Omega_{\mathrm{d}}}{\Omega_{\mathrm {k,mid}}}  \simeq \frac{1}{2} \frac{v_{\mathrm{P}}}{v_{\mathrm{k,mid}}} ,
\label{eq:diffvgasvdustphi}
\end{equation}
where
\begin{equation}
\frac{v_{\mathrm{P}}}{v_{\mathrm{k,mid}}} \equiv -\left( \frac{H_{\mathrm{g}}}{r} \right)^2 \left [p+\frac{m+3}{2} + \frac{m-3}{2} \left(\frac{z^2}{H_{\mathrm{g}}^2}\right)\right ]
\label{eq:vpvkz}
\end{equation}
is the drift velocity (Eq.~\ref{eq:vp}, \citealt{takeuchi02a}), assuming $\Sigma_{\mathrm{g}}\propto r^{-p}$ and $T\propto r^{-m}$, while $v_{\mathrm{k,mid}}\equiv r \Omega_{\mathrm{k,mid}}$ is the Keplerian velocity at the disc midplane. 
Eq.~\ref{eq:diffvgasvdustphi} shows that, in typical discs where $p+\left(m+3 \right)/2>0$, the gas is sub-Keplerian close to the midplane and super-Keplerian above the height
\begin{equation}
\left | z \right | \sim \left | \frac{2p+m+3}{3-m} \right |^{1/2} H_{\mathrm{g}} ,
\label{eq:condvppo}
\end{equation}
which, for typical disc parameters ($p=1$ and $m=1/2$), occurs at $\left | z \right | \sim 1.5 H_{\mathrm{g}}$. 
Therefore, at high altitudes, dust grains rotate slower than the gas, resulting in dust drifting outwards and gas drifting inwards. Moreover, Eq.~\ref{eq:vviscrz} shows that the crosswind experienced by dust grains under the action of the viscous accretion flow decreases with altitude. This leads to a fast decrease of the direct viscous drag on dust grains as $z$ increases. Hence, dust back-reaction is therefore expected to reinforce the outward dust-free gas accretion flow close to the midplane, while for  $\left | z \right | \gtrsim 0.7 H_{\mathrm{g}}$ the dust back-reaction tends to act against it, with an efficiency that decreases with $z^2$ and changes sign above $\left | z \right | \sim 1.5 H_{\mathrm{g}}$. We remark that our model accounts for deviations to the dust- and gas-free dynamics to order $ z^2/r^2 $, which play a crucial role for the exchange of the angular momentum among all the phases and can nevertheless produce features that might be detectable with high resolution and high sensitivity observations \citep{testi14a}.
\subsection{Disc models}
\label{sect:discmodels}
As previously adopted in Sect.~\ref{sect:discmodel1d}, we consider a locally isothermal gas disc with mass $M_{\mathrm{disc}}=0.01 \, M_{\odot}$ orbiting around a central star with mass $M_{\star}=1\,M_{\odot}$. We adopt a surface density profile $\Sigma_{\mathrm{g}} \propto r^{-1}$, a temperature $T\propto r^{-0.5}$ and an aspect ratio $H_{\mathrm{g}}/r\propto r^{1/4}$, where $H(r_0)/r_0=0.05$ at $r_0=1\,$au, such that the disc is in vertical hydrostatic equilibrium (Sect.~\ref{sect:gasdensity}). We consider a disc with a radial extent of $r \in [r_{\mathrm{in}},r_{\mathrm{out}}]$, where $r_{\mathrm{out}}=100\,\mathrm{ au}$ and the inner radius $r_{\mathrm{in}}=0.1\,\mathrm{ au}$ is set to be consistent with the assumption that the local temperature (assuming that the gas and dust are thermally coupled) should not be higher than the silicate grains sublimation temperature ($T\sim 1500$--$2000$ K, \citealt{helling01a}). 
Turbulence is parametrised by the parameter $\alpha$ in the range $[10^{-3},10^{-2}]$. 

A dusty disc with the similar radial extent and spatial distribution is superimposed on top of the gas. The total dust mass is set by the value of the dust-to-gas ratio $\epsilon$, which takes the values $[0.01,0.05,0.1]$. This fixes typical values for the ratio $\epsilon/\alpha$ to be of order $\sim 1$--$100$. We model a dust density per unit grain size of the form
\begin{align}
\frac{\mathrm{d} \rho_{\mathrm{d}}}{\mathrm{d}s} \left (r,z\right )&= C \,m\left(s\right) \left( \frac{s}{s_{\mathrm{max}}} \right)^{-q}  \times \nonumber \\ 
&\exp{\left \lbrace-\frac{z^2}{2H_{\mathrm{g}}^2} -\frac{\mathrm{St_{\mathrm{mid}}\left (s\right)}}{\alpha} \left [ \exp{\left (\frac{z^2}{2H_{\mathrm{g}}^2} \right)}-1 \right ] \right\rbrace},
\label{eq:modeldustsz}
\end{align}
where $C$ is a normalisation factor obtained by integrating  Eq.~\ref{eq:modeldustsz} over $s$ and $z$ and setting the result equal to $\epsilon \Sigma_{\rm g}$. Hence, the grain-size distribution resembles a power-law form in the midplane, but is multiplied by factors that account for settling in the vertical direction, as shown by Eq.~\ref{eq:dustvertprofdiff}. The exponent $q$ parametrising the size distribution takes the values $q = [2.5, 3.5, 4.5]$. We assume a material grain density $\rho_{\mathrm{grain}}=1\, \mathrm{g\, cm^{-3}}$. We fix the minimum size of the distribution to $s_{\rm{min}}=0.1 \,\mathrm{\mu} \text{m}$, although the exact value does almost not affect our results for $s_{\rm{min}} < 1 \,\mathrm{\mu} \text{m}$. 
\newtext{The maximum size $s_{\rm max}$ of a steady-state grain-size distribution is typically set by fragmentation and/or radial drift processes \citep[e.g.][]{birnstiel12b,laibe14b}. 
In the inner disc regions of typical protoplanetary discs, fragmentation tends to dominate by limiting the maximum grain size below a threshold at which the impact velocities between colliding grains become larger than the fragmentation velocity threshold. This value corresponds to a value of the Stokes number given by \citep[e.g.][]{ormel07a,birnstiel12b}
\begin{equation}
\mathrm{St}_{\rm{fragm}}= \frac{2}{3}\frac{v^2_{\rm{frag}}}{ \alpha  c_{\rm{s}}^2} ,
\label{eq:amax}
\end{equation}
where $v_{\rm{frag}}$ is the fragmentation threshold velocity. The exact value of $v_{\rm{frag}}$ is not well-known since it depends on the internal properties of the grains, but a reasonable order-of-magnitude estimate is $10\, \mathrm{m \,s^{-1}}$.
In the outer disc regions, drift may prevent grains reaching very large sizes by transporting the marginally-coupled solids inward, especially in low viscosity discs \citep{birnstiel12b,lambrechts14b,okuzumi16a}. A typical estimate for the drift-induced maximum size is given by \citep[e.g.][]{birnstiel12b}
\begin{equation}
\mathrm{St}_{\rm{drift}}=\epsilon \left(\frac{v_{\mathrm{k,mid}}}{c_\mathrm{s}}\right)^2 \left | \frac{\partial \log P_0}{\partial \log r} \right|^{-1} .
\label{eq:amaxdrift}
\end{equation}
The derivation of the drift limited size is based on equating the time within which particles are removed by drift with the growth time-scale using a local approach \citep[e.g.][]{birnstiel12b}. 
However, global effects can emerge from this local dynamics such as strong pile-ups  \citep{cuzzi93a,stepinski96a,laibe14b}. Moreover, care is also required since Eqs.~\ref{eq:amax} and \ref{eq:amaxdrift} have been derived by assuming a mono-disperse size distribution of particles with $\mathrm{St}<1$ without taking into account the back-reaction on to the growth and the drift of dust grains (see Sects.~5.4 and 6.2 of \citealt{laibe14b}). \citet{gonzalez17a} suggest that, by taking into account the reduction of the dust drift induced by the back-reaction, dust grains are able to pile-up and break the radial drift barrier, with the fragmentation barrier still limiting the maximum grain size. 
Therefore, it is not straightforward to define consistently the maximum size of a steady-state dust distribution induced by the interplay between grain growth and drift motions. A detailed treatment of this problem requires numerical simulations.
Since the main purpose of the paper is to investigate the radial drift of gas and dust grains in protoplanetary discs, we assume, as a reference case, that the maximum value $s_{\rm max}$ of the grain-size distribution is set by fragmentation processes (i.e. $\mathrm{St}_{\rm{max}} = \mathrm{St}_{\rm{fragm}}$). 
We also assume that the limiting effects on the grain-size distribution essentially occur close to the midplane, with the resulting grains being redistributed in the vertical direction by turbulence. We therefore compute the value for $s_{\mathrm{max}}$ corresponding to the Stokes number $\mathrm{St}_{\mathrm{max}} =\mathrm{St}_{\rm{fragm}}$ by solving  Eqs.~\ref{eq:amax} and \ref{eq:stoppingtimecase} \textit{in the midplane}.
Moreover, we investigate alternative cases  for the maximum grain size of the steady-state dust distribution by considering different values of $s_{\rm max}$ corresponding to Stokes numbers $\mathrm{St}_{\mathrm{max}} = [\mathrm{St}_{\rm{fragm}}, \min(\mathrm{St}_{\rm{fragm}},\mathrm{St}_{\rm{drift}}),0.1,1]$. 
In our disc models, the drift limits the grain-size distribution (i.e. $\mathrm{St}_{\rm{drift}} < \mathrm{St}_{\rm{fragm}}$) for
\begin{equation}
\epsilon < \frac{2}{3 \alpha} \left( \frac{v_{\rm{frag}}}{ v_{\mathrm{k,mid}}}\right)^2 \left | \frac{\partial \log P_0}{\partial \log r} \right|,
\label{eq:sizelimstok}
\end{equation}
which correspond to $\epsilon$ smaller than values in the range $[2\times 10^{-5}, 0.02]$. 
Therefore, since in our model we assume $\epsilon = [0.01,0.05,0.1]$, the maximum size $s_{\rm max}$ of the grain-size distribution is limited by the drift only in the outer regions of the disc model with $\epsilon=0.01$. 
}

\begin{figure}
\begin{center}
\includegraphics[height=0.39\textwidth]{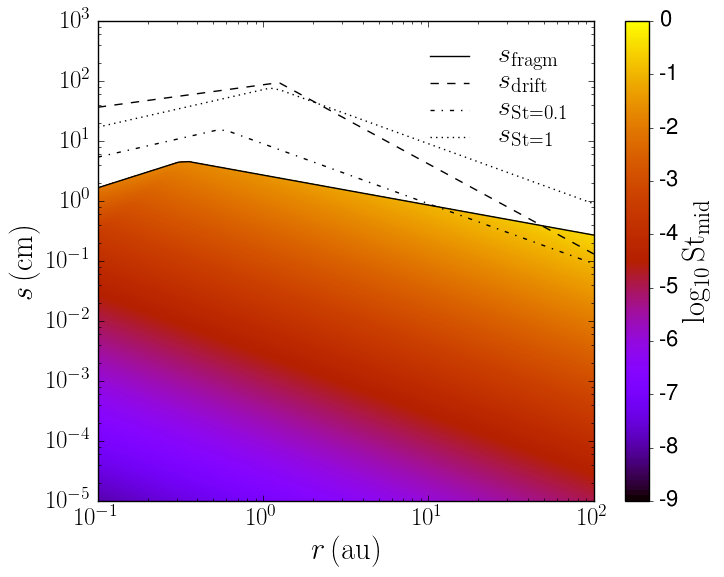}
\caption{Stokes numbers of each dust species at the disc midplane in the reference disc model ($q=3.5$, $\epsilon=0.01$, $\alpha=10^{-3}$ and $\mathrm{St}_{\mathrm{max}} = \mathrm{St}_{\rm{fragm}}$). The maximum grain size in the dust distribution $s_{\mathrm{fragm}}$ corresponds to the Stokes number shown in Eq.~\ref{eq:amax}, while the minimum value is $0.1\, \mathrm{\mu} \text{m}$. \newtext{The dashed line $s_{\mathrm{drift}}$ indicates the sizes corresponding to the Stokes number expressed by Eq.~\ref{eq:amaxdrift}, while the dotted-dashed and the dotted lines indicate the sizes where the Stokes number at the midplane is 0.1 and 1 respectively.}
Most of the dust grains in the inner disc are tightly-coupled with the gas ($\mathrm{St}\ll 1$), while millimetre-sized grains in the outer disc regions are marginally-coupled ($\mathrm{St}\sim 1$).}
\label{fig:stokez}
\end{center}
\end{figure}

Fig.~\ref{fig:stokez} shows the Stokes number of the dust grains in the reference model ($q=3.5$, $\epsilon=0.01$, $\alpha=10^{-3}$ and $\mathrm{St}_{\mathrm{max}} = \mathrm{St}_{\rm{fragm}}$) at the midplane. Most of the dust grains in the inner disc are tightly-coupled with the gas at the disc midplane, while millimetre-sized grains in the outer disc regions are nearly marginally-coupled ($\mathrm{St}\sim 1$). At a radius $r\lesssim 0.3$ au, the drag law changes from Epstein to Stokes regime and the maximum grain size of the distribution decreases with decreasing distance from the star, due to the different scalings of the Stokes number with the local disc properties (see Eq.~\ref{eq:stoppingtimecase}). Precisely, $s_{\mathrm{max}}\propto \Sigma_{\mathrm{g}} c_{\mathrm{s}}^{-2}\propto r^{-1/2}$ in the Epstein regime, while in the Stokes regime $s_{\mathrm{max}}\propto \left( c_{\mathrm{s}} \Omega_{\mathrm{k}}\right)^{-1/2}\propto r^{7/8}$. \newtext{At a distance from the central star $r \gtrsim 50$ au, the drift-limited maximum size (dashed line in Fig.~\ref{fig:stokez}) is smaller the one given by the fragmentation.}
In the vertical direction, Eq.~\ref{eq:stokevarz} shows that particles that are well coupled with the gas in the midplane can be loosely-coupled in the upper disc layers, especially in the outer disc regions. The Stokes number increases $\propto\mathrm{e}^{z^2/2H_{\rm g}^{2}}$, e.g. by a factor $\sim 8$ at $z=2H_{\mathrm{g}}$. However, only a small amount of marginally-coupled grains are found far from the midplane, since vertical settling efficiently removes these grains to lower regions with higher aerodynamic coupling (see Sect.~\ref{sect:destvertdens}).
\begin{figure*}
\begin{center}
\includegraphics[height=0.2578\textwidth]{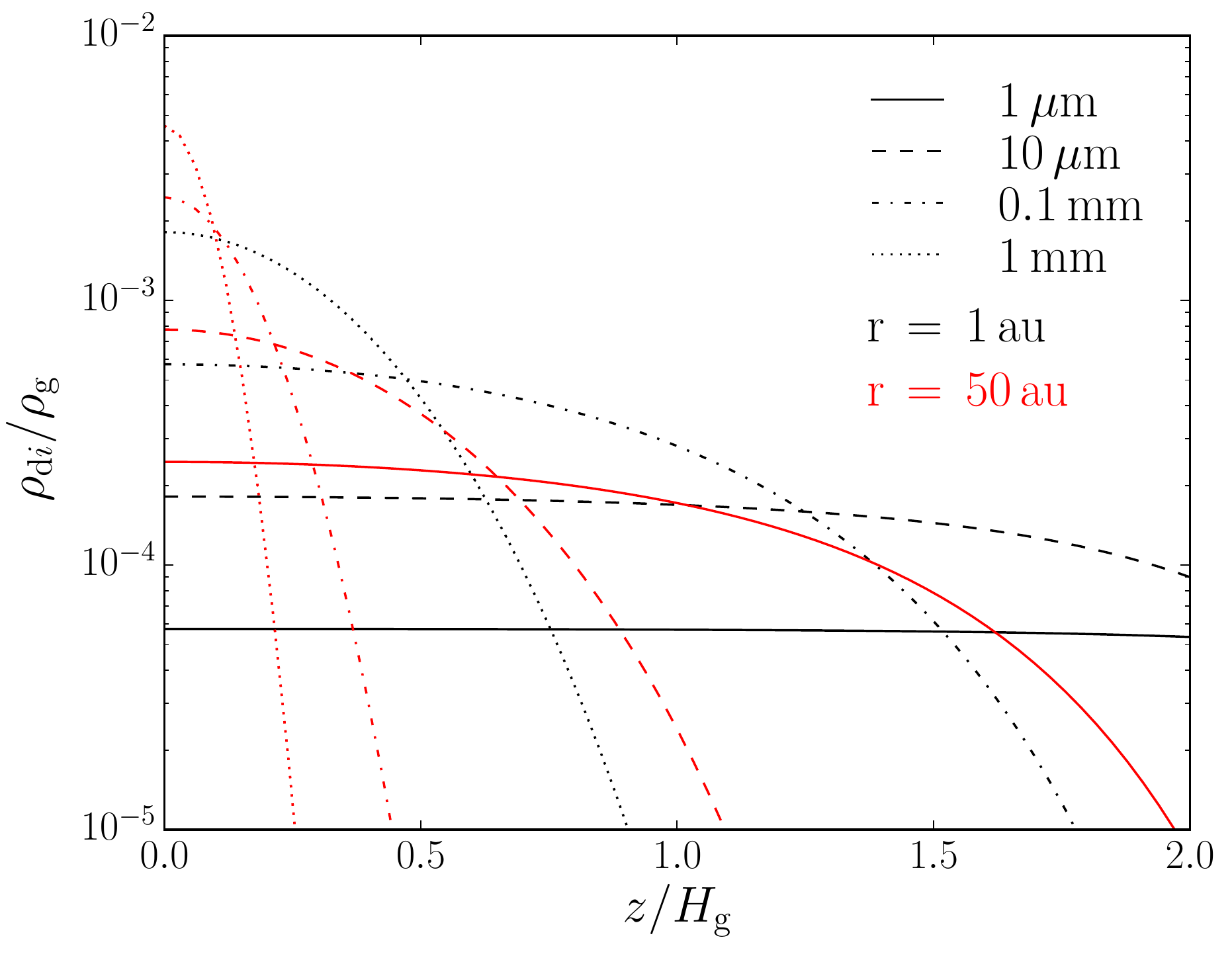}
\includegraphics[height=0.2578\textwidth]{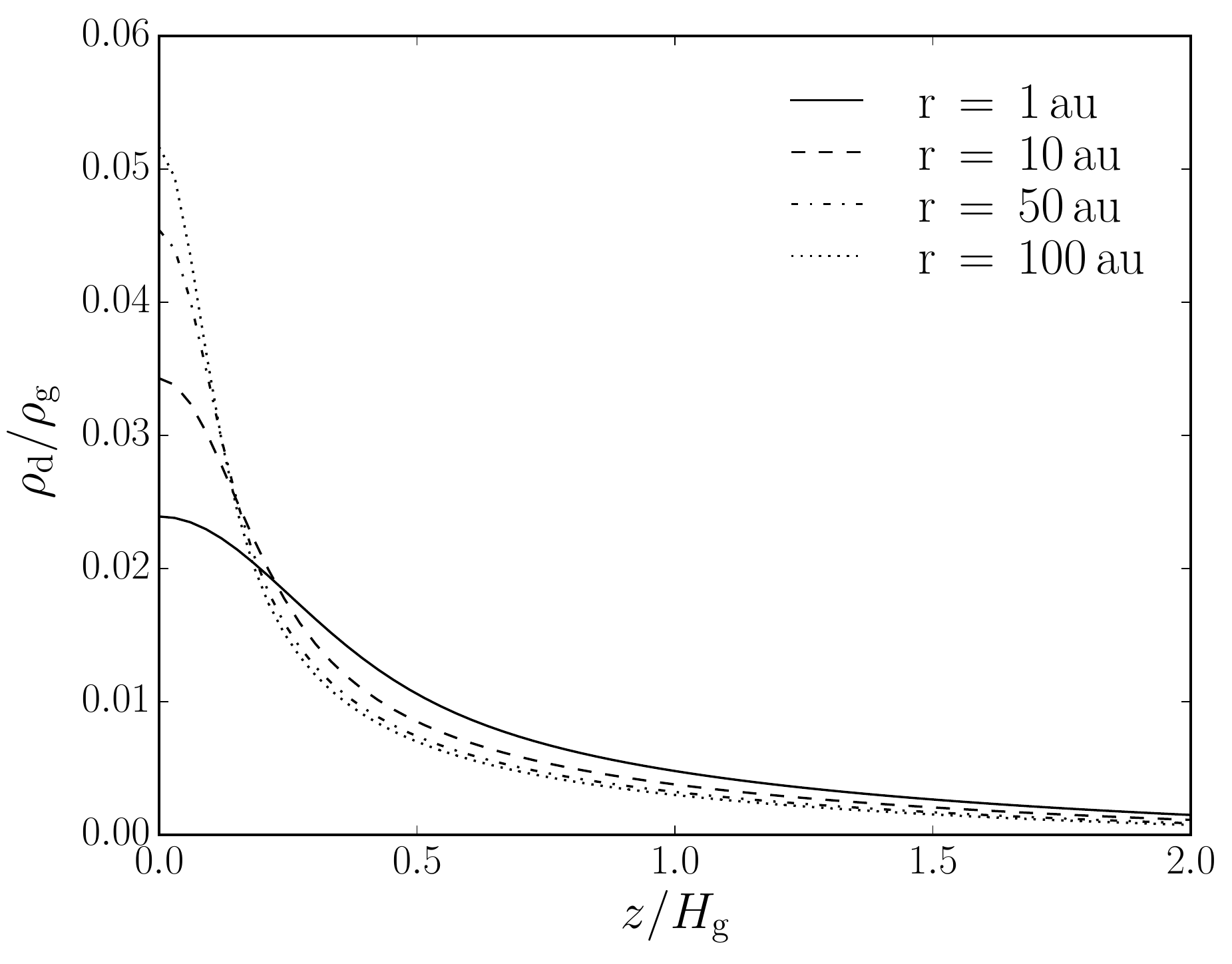}
\includegraphics[height=0.2578\textwidth]{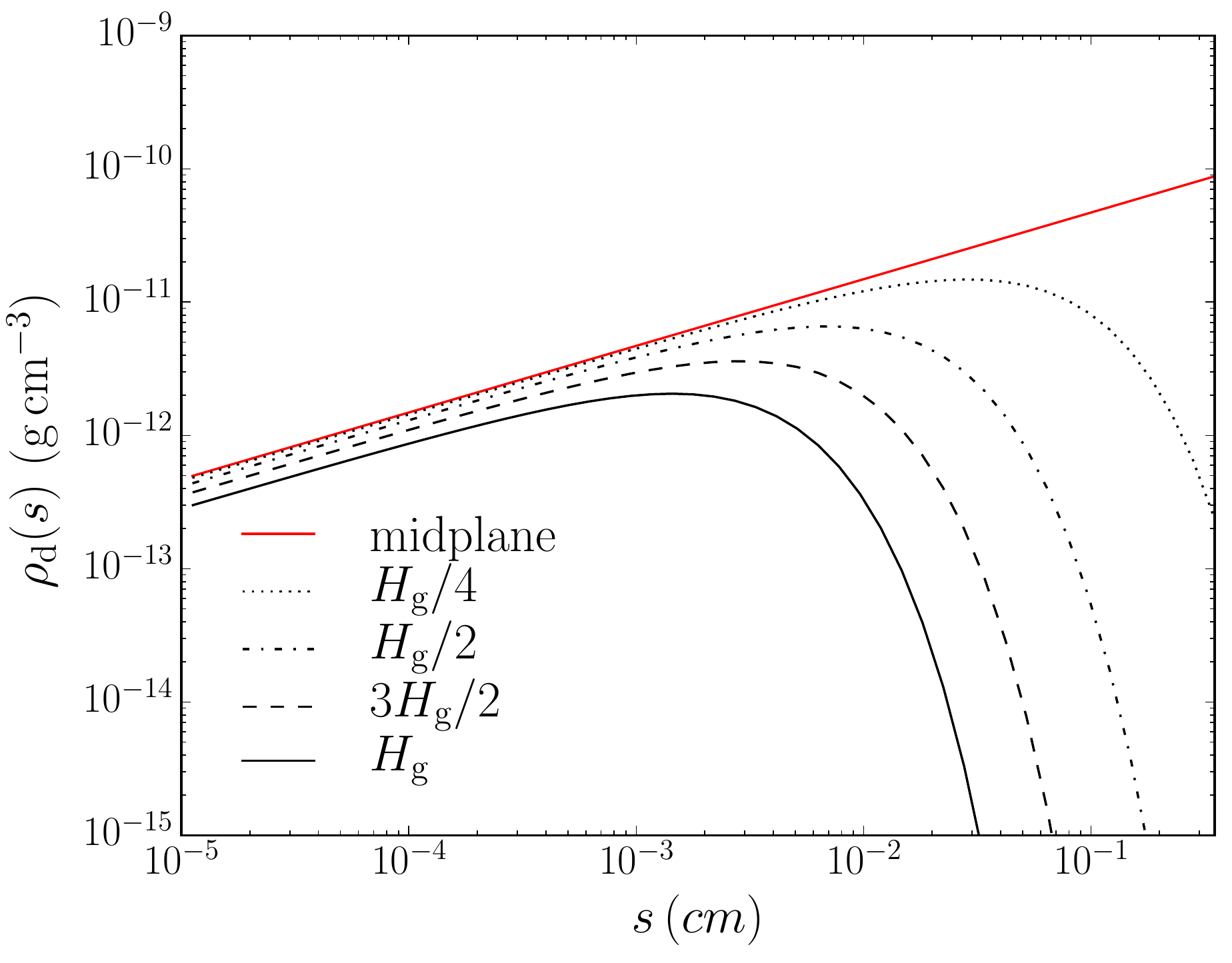}
\caption{\textit{Left}: vertical profile of the dust-to-gas ratio for grains with different sizes in the range $\mathrm{[1 \,\mu m, 10\,\mu m, 0.1 \, mm,1\,mm]}$ at 1 au (black) and 50 au (red). \textit{Centre}: vertical profile of the total dust-to-gas ratio at different distances from the central star. \textit{Right}: density of dust grains at different altitudes from the midplane at a distance of 50 au from the central star. For each plot we assume a disc model with $q=3.5$, $\epsilon=0.01$, $\alpha=10^{-3}$ and $\mathrm{St}_{\mathrm{max}} = \mathrm{St}_{\rm{fragm}}$.
The height above the midplane $z$ in the left and central panels is normalised by the disc scale height $H_{\mathrm{g}}$ at each radius. As distance from the central star increases, particles become less coupled to the gas and settle toward the midplane (Eq.~\ref{eq:modeldustsz}). The dust density retains a power-law dependence in the midplane (Eq.~\ref{eq:modeldustsz}), while dust settling reduces the density of large grains in the upper layers.}
\label{fig:rhodustz}
\end{center}
\end{figure*}

Fig.~\ref{fig:rhodustz} shows that dust grains close to the star are well mixed with the gas, since the large gas density generates strong drag (see Fig.~\ref{fig:stokez}). As the distance from the central star increases, dust grains become less coupled and populate regions closer to the midplane, leading to an increase of the local dust-to-gas ratio (see the central panel of Fig.~\ref{fig:rhodustz}). The right panel of Fig.~\ref{fig:rhodustz} shows that the dust density retains a power-law dependence in the midplane (Eq.~\ref{eq:modeldustsz}), while dust settling reduces the density of large grains in the upper layers (note that the plot shows the dust density in each size bin: $\propto m\left(s\right) s^{1-q} \propto s^{0.5}$ for $q=3.5$). The exact value of the dust-to-gas ratio of different particles shown in Fig.~\ref{fig:rhodustz} depends on the size of the bin over which we integrate the dust-to-gas density ratio per unit grain size (Eq.~\ref{depsds}) or, equivalently, on the number of size bins we use to discretise the size distribution. We enforce $N_{\mathrm{bins}} \gtrsim 50$ to ensure numerical convergence.

\subsection{Steady-state velocities}
\label{sect:netveloc}
The vertical stratification of the disc implies that the dust-free viscous gas velocity, the relative orbital velocities between the phases, the drag and thus, the steady-state gas and dust velocities depend on $z$. Using Eqs.~\ref{eq:vradg} and \ref{eq:vraddi} we obtain the  vertical profile of the radial velocity of the gas and each dust phases by taking into account the vertical stratification of all the quantities involved \newtext{(see Appendix~\ref{app:approx} for further details about the extension of our formalism in vertically stratified discs)}.

\begin{figure*}
\begin{center}
\includegraphics[width=0.428\textwidth,trim={0 1.15cm 0 0},clip]{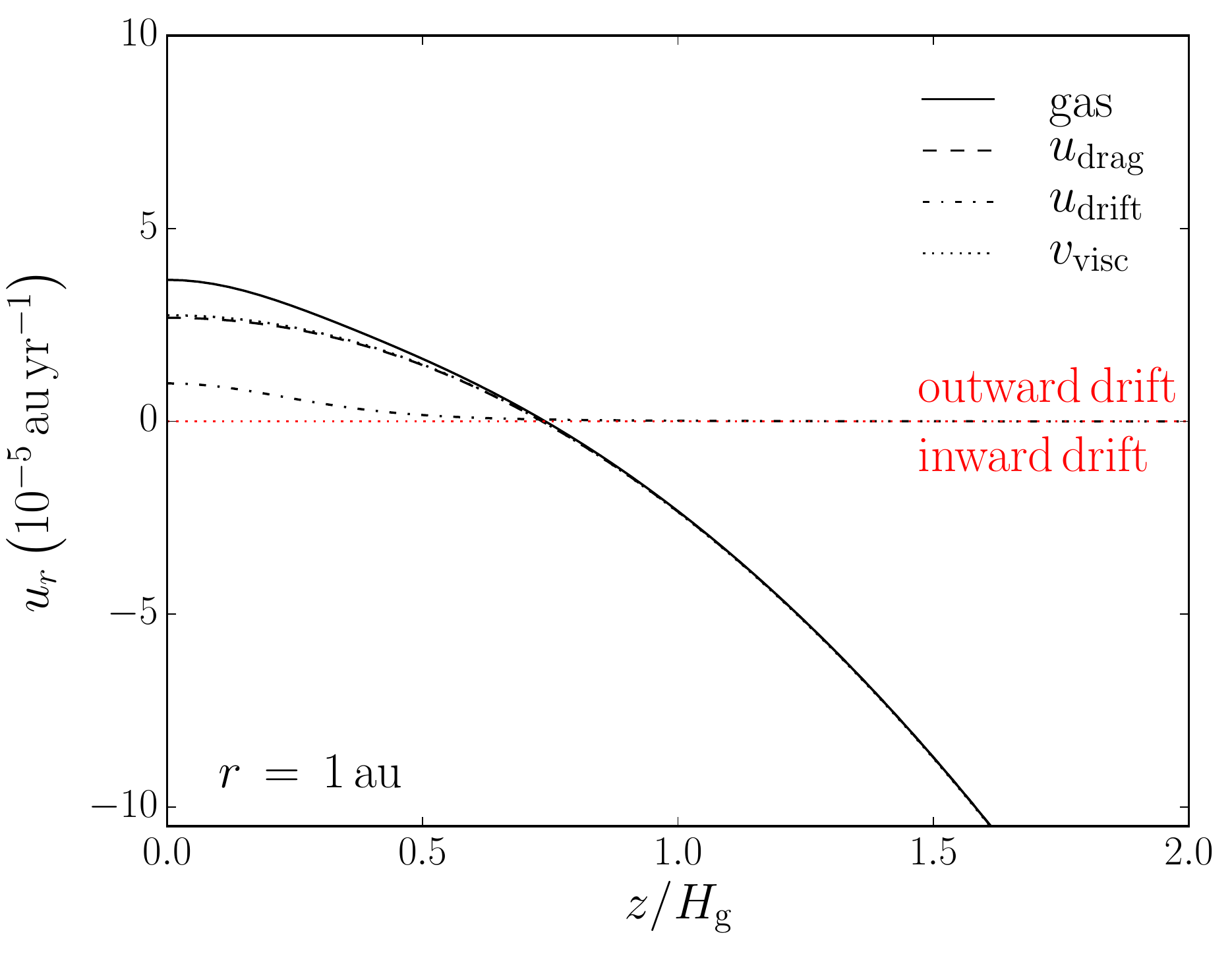}
\includegraphics[width=0.418\textwidth,trim={0.4cm 1.15cm 0 0},clip]{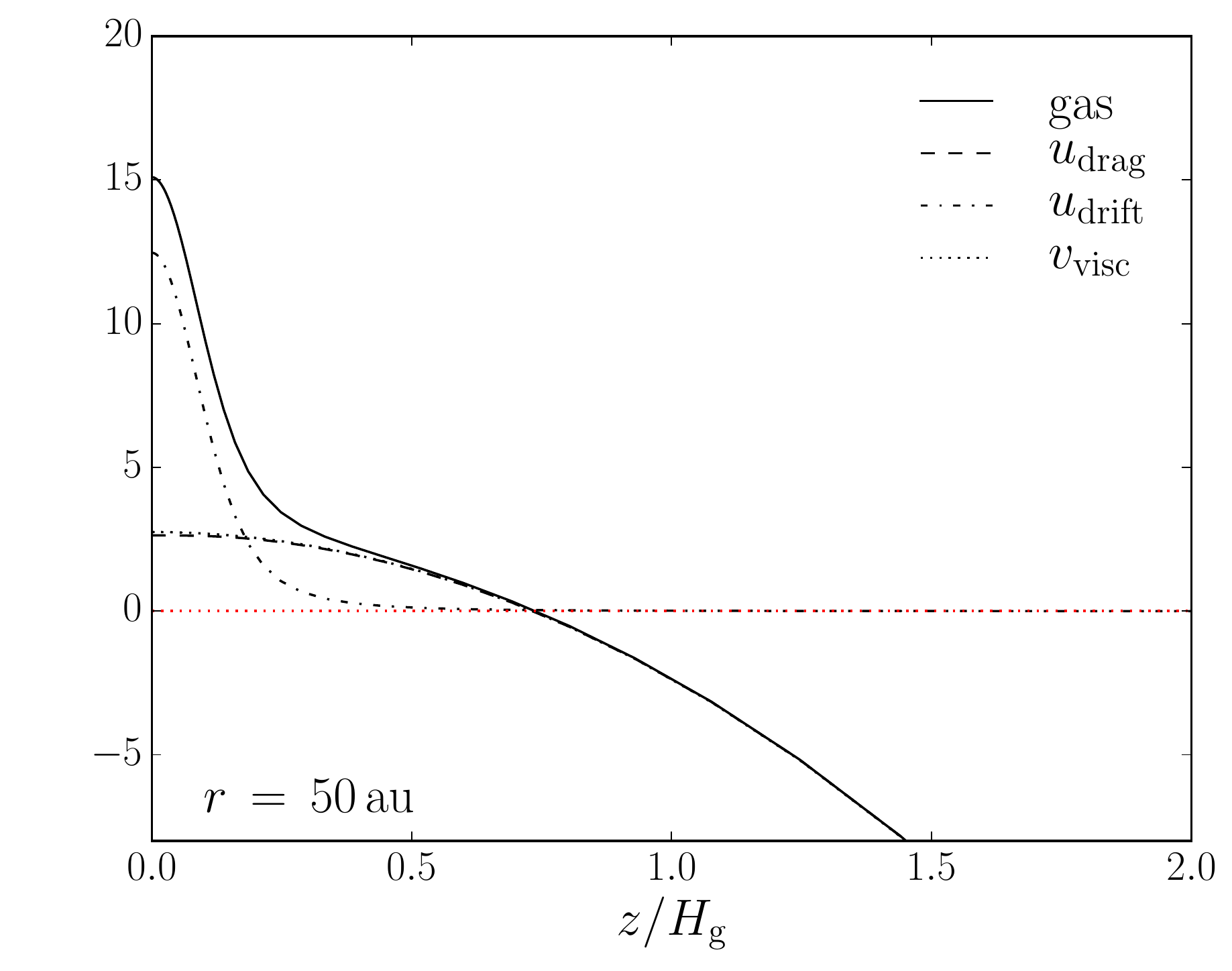}
\includegraphics[width=0.428\textwidth]{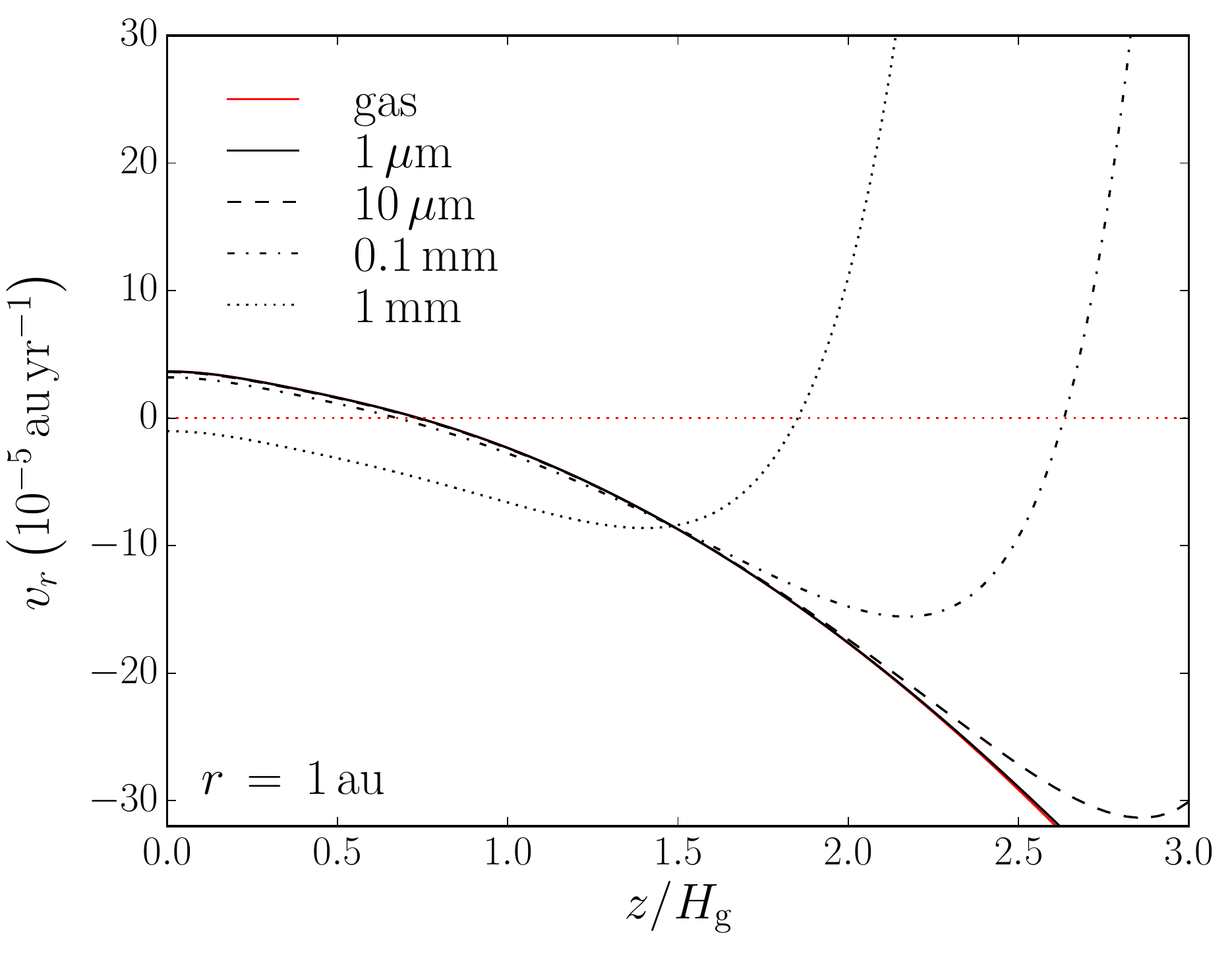}
\includegraphics[width=0.421\textwidth,trim={0.7cm 0 0 0},clip]{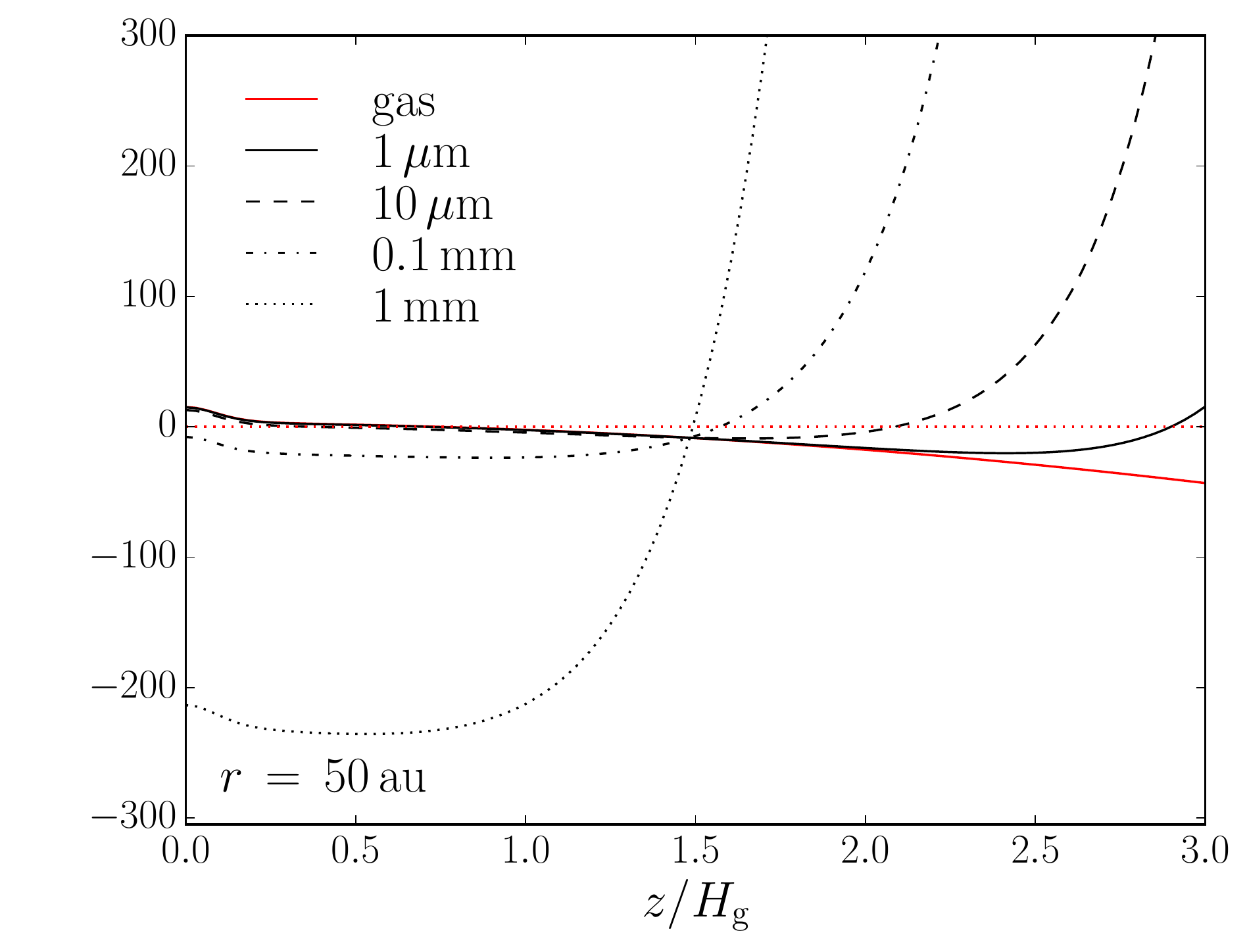}
\caption{Vertical profile of the radial (\textit{top}) gas velocity and (\textit{bottom}) dust velocity of particles with sizes in the range $[1\,\mathrm{\mu} \text{m},10\,\mathrm{\mu} \text{m}, 0.1\,\text{mm},1\,\text{mm} ]$ (black lines) compared to the gas velocity (solid red line). The left panels refer to velocities computed at 1 au, while the right panels at 50 au from the central star. In these plots we consider the disc model with $q=3.5$, $\epsilon=0.01$ and $\alpha=10^{-3}$.
In the top panels, the dashed and dot-dashed line indicates the two contributions of the radial gas velocity, $u_{\mathrm{drag}}$ and $u_{\mathrm{drift}}$ (expressed in Eqs.~\ref{eq:vgrvisc} and \ref{eq:udrift} respectively), while the dotted black line is the expected dust-free gas velocity $v_{\mathrm{visc}}$ (Eq.~\ref{eq:vviscrz}). In the bottom panels the solid red line indicates the radial gas velocity, indicated with the solid black lines in the top panels. \newtext{Note that the red solid line in the \textit{bottom-left} panel is hidden behind the solid black line.}
The distance from the midplane $z$ is normalized by the disc scale-height $H_{\mathrm{g}}$ at each radius. The horizontal dotted red line indicates the null value. While close to the central star (left panels), gas and dust grains share the same drag-induced motion due to the high gas density, in the outer disc regions the dust dynamics is dominated by the drift contribution, especially close to the midplane.}
\label{fig:vrgasprofz}
\end{center}
\end{figure*}

The top panels of Fig.~\ref{fig:vrgasprofz} show the vertical profile of the radial velocity of the gas $u_r$ at steady-state (Eq.~\ref{eq:vradg}) for our reference disc model ($q=3.5$, $\epsilon=0.01$, $\alpha=10^{-3}$ and $\mathrm{St}_{\mathrm{max}} = \mathrm{St}_{\rm{fragm}}$). We compare the relative contributions of the direct viscous drag (dashed line, Eq.~\ref{eq:vgrvisc}), the drift (dotted-dashed line, Eq.~\ref{eq:udrift}) and the expected dust-free gas motion (dotted line, Eq.~\ref{eq:vviscrz}) at two different distances from the star. Close to the central star (top-left panel), the Stokes number of a grain with a given size is smaller than in the outer regions of the disc since the gas is denser. Hence, the drift induced by the tailwind experienced by gas is negligible with respect to the contribution from the viscous drag. The small contribution of the drift originates from the low dust-to-gas ratio in the midplane, since even large grains are scattered out of the midplane by turbulence at these large gas densities (see the black lines in Fig.~\ref{fig:rhodustz}). As the distance from the star increases (see the top-right panel of Fig.~\ref{fig:vrgasprofz}), since the gas density drops, the grains become less coupled and settle close to the midplane (see Fig.~\ref{fig:stokez} and \ref{fig:rhodustz}). Therefore, the dust-to-gas ratio of large grains increases close the midplane (Eq.~\ref{eq:modeldustsz}), leading the drift induced by back-reaction to overcome the viscous drag term (Eq.~\ref{eq:caseratio}). 
In any case, drift efficiency decreases at large heights due to both a reduced dust-to-gas ratio of grains with $\rm{St} \sim 1$ (since they are settled) and the decrease of the relative velocity of the gas with respect to the Keplerian motion (Eq.~\ref{eq:vpvkz}). This decrease is only slightly mitigated by the increase of the Stokes number (Eq.~\ref{eq:stokevarz}) for particles of a given size. Far from the midplane, the gas velocity can equal the one of the dust-free viscous motion, since dust back-reaction does not have a strong contribution in these regions of low dust-to-gas ratio (see Sect.~\ref{sect:gasdyn2D}).

The vertical profiles of the radial velocities of the dust grains $v_{i,r}$ (Eq.~\ref{eq:vraddi}) are shown in the bottom panels of Fig.~\ref{fig:vrgasprofz}. In the inner disc regions (bottom-left panel), small particles (i.e. $\sim 1 - 10 \,\mathrm{\mu} \text{m}$) are tightly-coupled with the gas and are vertically well-mixed. Gas and small dust grains move almost together over the entire disc scale height with the gas velocity (denoted by the solid red line in the bottom panels of Fig.~\ref{fig:vrgasprofz}). This corresponds to an outward motion at low altitude and an inward motion at high altitude. In other words, in these cases the viscous drag  dominates over drift (Eq.~\ref{eq:vraddi}). Larger particles with size $\sim 1 \,\text{mm}$ decouple with the gas closer to the midplane compared to smaller particles (Fig.~\ref{fig:stokez}) and start drifting towards the star. The inward drift velocity of large grains increases in absolute value with altitude (when $z \lesssim 1.5 H_{\mathrm{g}}$) due to the increase of the Stokes number toward values of order unity (see Eq.~\ref{eq:stokevarz}). In the upper layers ($z \gtrsim 1.5 H_{\mathrm{g}}$), large dust particles ($0.1$--$1 \,\text{mm}$) are loosely-coupled with the gas, but in this case drift outwards due to the tailwind they experience from orbiting slower than the gas (see Eq.~\ref{eq:vpvkz}). These results are roughly consistent with the single-species calculations shown in \citet{takeuchi02a} (see their Fig.~3), especially far from the midplane where back-reaction (not considered in \citealt{takeuchi02a}) is negligible. In the outer disc regions (bottom-right panel of Fig.~\ref{fig:vrgasprofz}), dust particles are generally less coupled with the gas compared to the inner regions, strengthening the role of the drift in their dynamics (first term in the right hand side of Eq.~\ref{eq:vraddi}). In particular, drift dominates their motion close to the midplane, where the radial pressure gradient is maximized, while in the upper layers the dust radial motion is reversed due to the changes in the relative orbital motion between the gas and dust phases (see Eq.~\ref{eq:vpvkz}). In the outer disc regions and very close to the midplane ($z\lesssim 0.3 H_{\mathrm{g}}$), the bottom-right panel of Fig.~\ref{fig:vrgasprofz} shows that both gas and dust velocities increases with decreasing altitude. This is related to the effect of the dust back-reaction (see top-right panel of Fig.~\ref{fig:vrgasprofz}), due to the increasing values of the dust-to-gas ratio with decreasing altitudes induced by the settling of large grains (see the red lines in the left panel of Fig.~\ref{fig:rhodustz}).

\subsubsection{Averaged velocities}
To gain insights about the net motion of gas and multiple species of dust grains, we compute the vertical average of the gas and dust velocities, taking into account the vertical concentration of all the phases (Eqs.~\ref{eq:rhoz}--\ref{eq:modeldustsz}).
\begin{figure*}
\begin{center}
\includegraphics[height=0.315\textwidth,trim={0 1cm 0 0},clip]{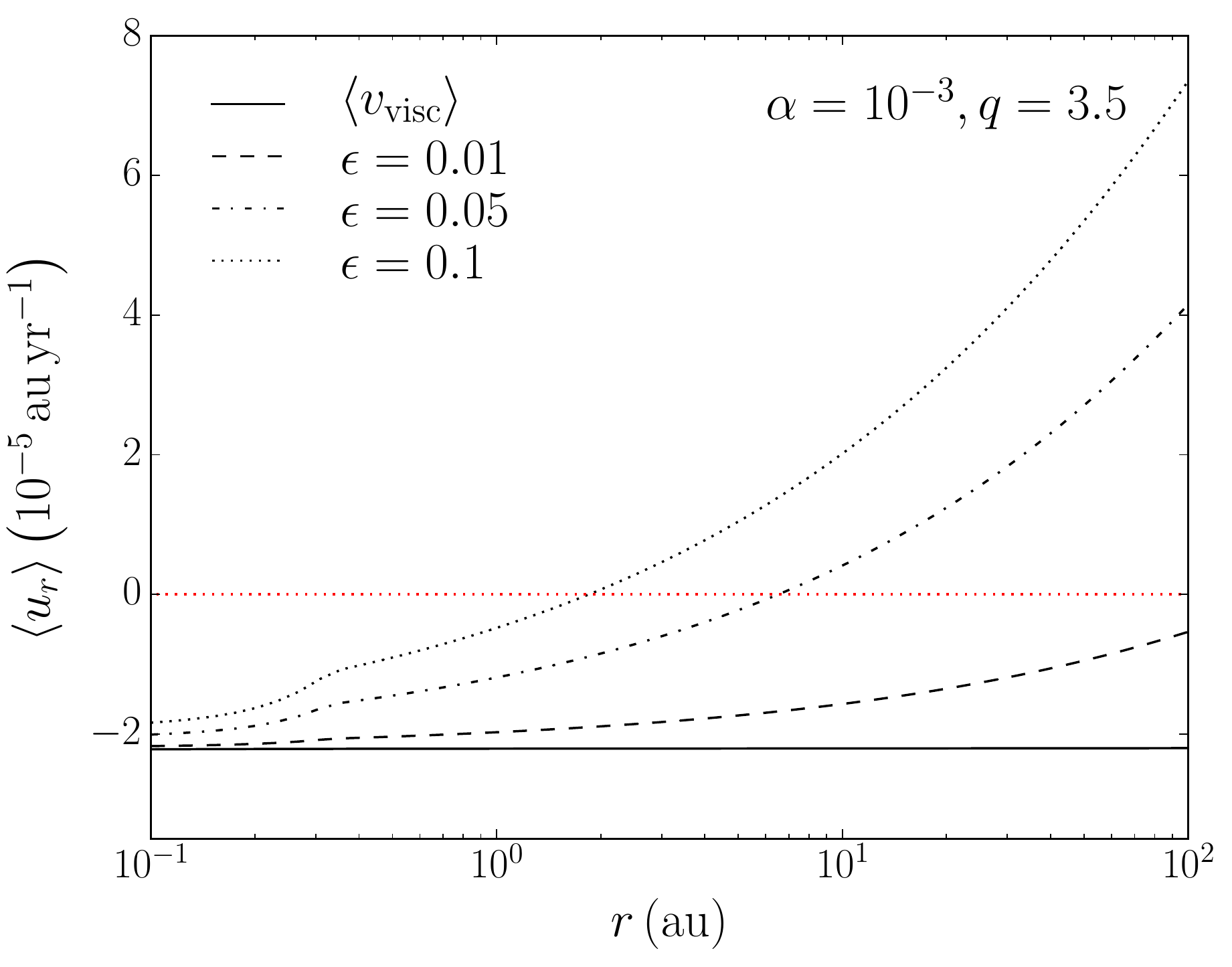}
\includegraphics[height=0.315\textwidth,trim={1.4cm 1cm 0 0},clip]{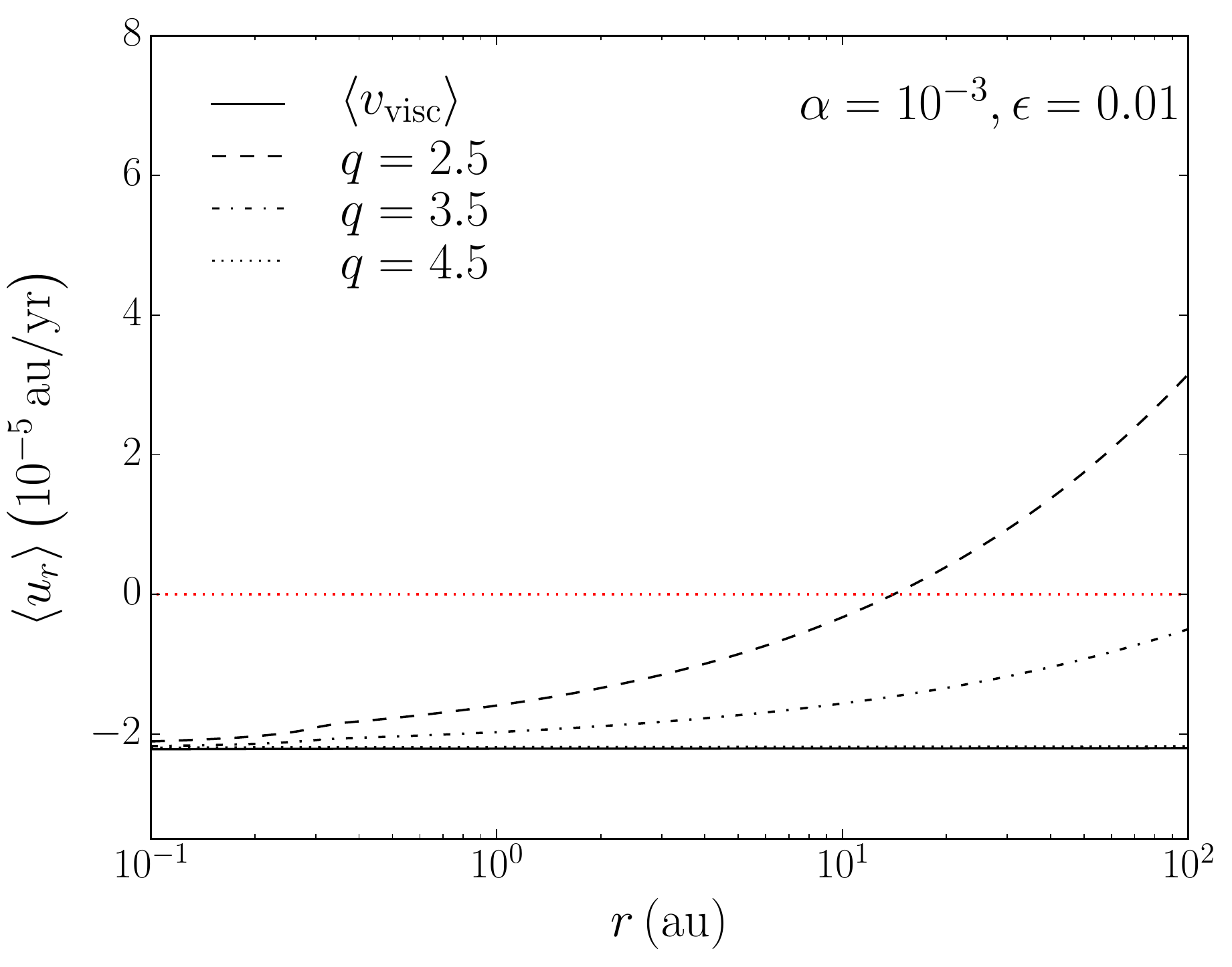}
\includegraphics[height=0.336\textwidth,trim={0 0 0 0},clip]{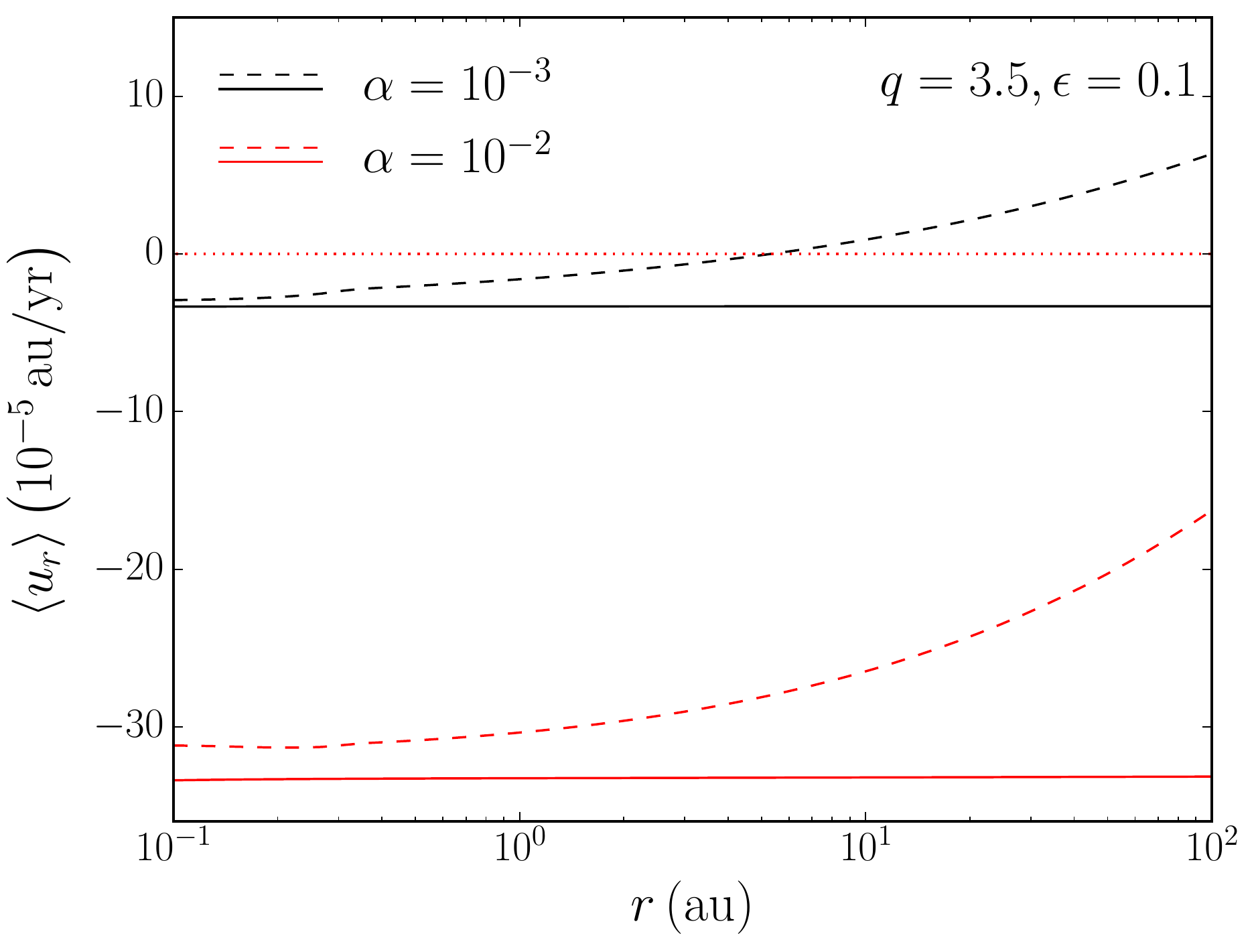}
\includegraphics[height=0.341\textwidth,trim={1.4cm 0 0 0},clip]{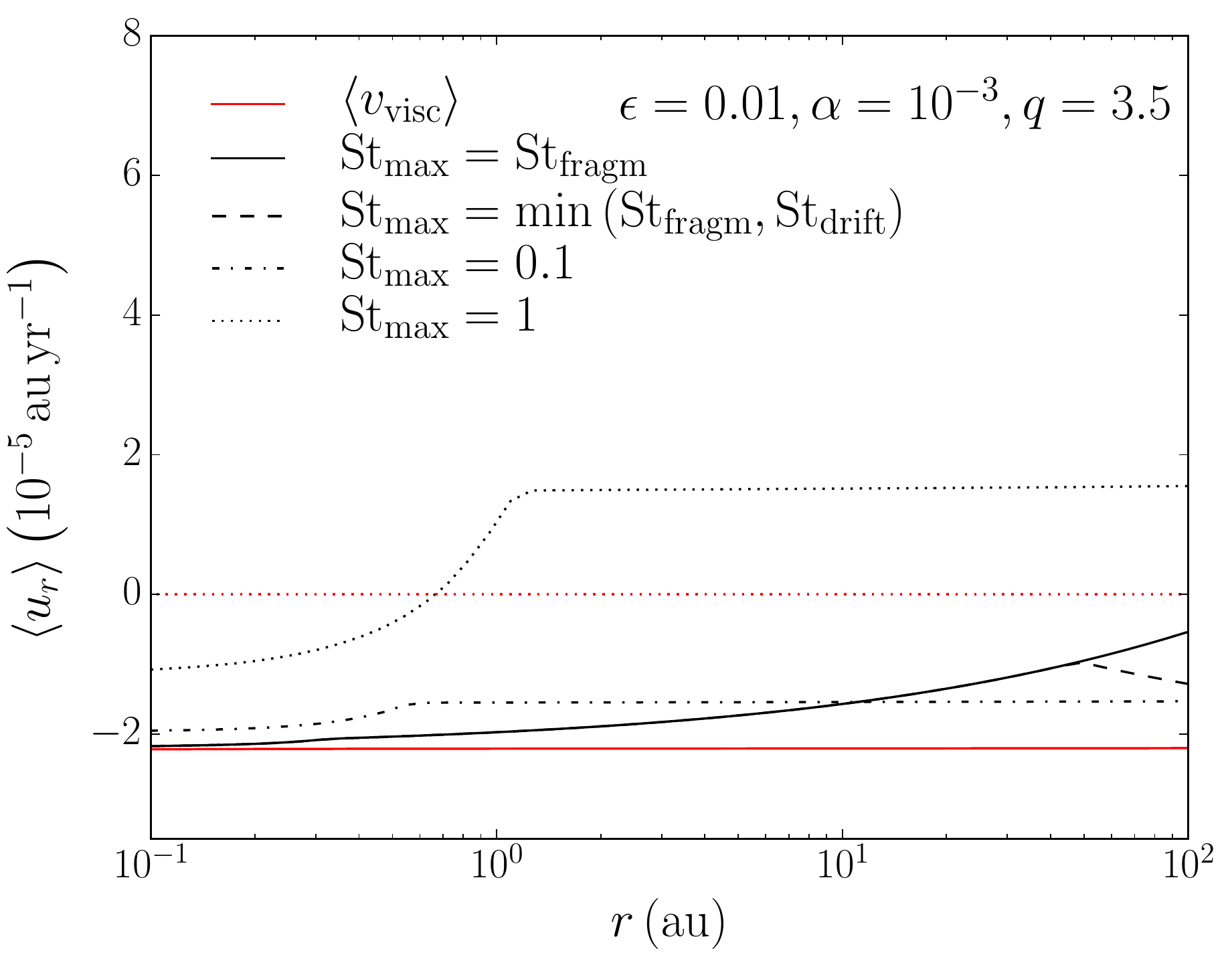}
\caption{\newtext{Averaged radial gas velocities as a function of distance from the star while individually varying the fiducial parameters in our disc model ($\epsilon=0.01$, $q=3.5$, $\alpha=10^{-3}$ and $\mathrm{St}_{\mathrm{max}} =\mathrm{St}_{\rm{fragm}}$) as follows: (\textit{top-left}) total dust-to-gas ratio, $\epsilon=[0.01,0.05,0.1]$; (\textit{top-right}) power-law exponent of the size distribution, $q=[2.5,3.5,4.5]$;  (\textit{bottom-left}) viscosity, $\alpha=[10^{-3},10^{-2}]$ and (\textit{bottom-right}) maximum Stokes number $\mathrm{St}_{\mathrm{max}} = [\mathrm{St}_{\rm{fragm}},\min(\mathrm{St}_{\rm{fragm}},\mathrm{St}_{\rm{drift}}),0.1,1]$.
The solid lines (apart from the solid black line in the \textit{bottom-right} panel) denote the dust-free averaged gas velocity (Eq.~\ref{eq:vviscaverage}).} The horizontal dotted red line indicates the null value. The gas radial velocity deviates from the expected dust-free viscous velocity due to the effect of the back-reaction, especially for higher $\epsilon$ and lower values of $q$ and $\alpha$ in the outer disc regions.}
\label{fig:vragsavg}
\end{center}
\end{figure*}
The averaged radial gas velocity is defined as 
\begin{equation}
\langle u_r\rangle=\frac{1}{\Sigma_{\mathrm{g}}} \int_{-\infty}^{\infty}\rho_{\mathrm{g}}(z)\, u_r \, \mathrm{d}z.
\label{eq:urnet}
\end{equation}
In order to infer the averaged effect of the back-reaction on to the gas motion, we define the averaged viscous dust-free gas velocity as 
\begin{equation}
\langle v_{\mathrm{visc}}\rangle=\frac{1}{\Sigma_{\mathrm{g}}} \int_{-\infty}^{\infty}\rho_{\mathrm{g}}(z)\,v_{\mathrm{visc}} (z) \, \mathrm{d}z,
\label{eq:vviscaverage}
\end{equation}
where $v_{\mathrm{visc}} (z)$ is expressed by Eq.~\ref{eq:vviscrz}. 
The top-left panel of Fig.~\ref{fig:vragsavg} shows the averaged gas radial velocity for three different values of the total dust-to-gas ratio, compared to the dust-free viscous velocity (solid line). It can be noticed that the dust-free viscous velocity (Eq.~\ref{eq:vviscaverage}) is negative, implying that the radial gas viscous inflow at high $z$ carries more mass inward than is transported out by the midplane outflow (Eq.~\ref{eq:vviscrz}).
In all cases, the gas radial velocity deviates from the expected dust-free viscous velocity due to the effect of the back-reaction, even for small values of the total dust-to-gas ratio. 
Close to the central star, the gas velocity approaches the dust-free values due to the low level of dust settling and the tight coupling of all the dust grains with the gas (see Fig.~\ref{fig:stokez}). 
In the outer disc regions, the increase of the Stokes number and the settling-induced higher dust-to-gas ratio in the midplane (Fig.~\ref{fig:rhodustz}) enhance the effect of the back-reaction that tends to both decrease the viscous motion (Eq.~\ref{eq:vgrvisc}) and push the gas towards the outer regions if $v_{\mathrm{P}}<0$ (Eq.~\ref{eq:vpvkz}), i.e. close to the midplane where most of the mass resides (Eq.~\ref{eq:udrift}). Importantly, even for low dust-to-gas ratio and small Stokes number close to the midplane, the averaged gas motion is significantly perturbed by the dust back-reaction compared to the viscous dust-free gas motion.
As the distance from the star increases, the averaged radial gas velocity increases and, for $\epsilon > 0.01$, changes signs at a distance from the star $\sim 1-10$ au. Consistently with the single-species dust calculations presented in \citet{kanagawa17a}, the gas outward motion is enhanced for larger dust-to-gas ratios 
(see their Fig.~3). 
Although most of the dust grains in our model are tightly-coupled with the gas in the midplane ($\mathrm{St_{mid}}\ll 1$, Fig.~\ref{fig:stokez}) and the integrated dust-to-gas ratio is distributed over a wide range of grains with different sizes, the cumulative effect of the back-reaction can reduce and even reverse the averaged radial motion of the gas in the outer disc regions -- even for $\epsilon \ll 1$. 
Moreover, the averaged radial motion of the gas changes with different values of $q$ (see the top-right panel of Fig.~\ref{fig:vragsavg}), especially in the outer disc regions. As shown in Sect.~\ref{sect:gasdyn2D}, dust back-reaction does not affect the gas dynamics if the grain-size distribution is sufficiently steep ($q \gtrsim 4$), due to low mass embodied in marginally-coupled grains. Decreasing the value of $q$ results in an enhanced effect of the back-reaction, leading to an outward gas migration in the outer disc regions.
We also explore the values of the averaged gas velocity in discs with different values of $\alpha$ (see the bottom-left panel of Fig.~\ref{fig:vragsavg}) by considering the same grain-size distribution (we use the same value of $s_{\mathrm{max}}$  computed from Eq.~\ref{eq:amax} assuming $\alpha=10^{-3}$ to highlight the role of dust back-reaction on the same dust population). 
In highly viscous discs ($\alpha=10^{-2}$) the effect of the back-reaction is not sufficient to trigger outward gas migration and the viscous-driven motion dominates the gas dynamics. Moreover, we find that the back-reaction effect is reduced in highly viscous disc due to also the lower level of dust settling.
These results are consistent with those shown in \citet{kanagawa17a}, although the dust grains in our model are more tightly-coupled with the gas at the midplane and the total dust-to-gas ratio is distributed over a wide range of grains with different size.
\newtext{The bottom-right panel of Fig.~\ref{fig:vragsavg} shows the averaged gas radial velocity for four different maximum values of the grain-size distribution. As observed in Fig.~\ref{fig:vrgsa}, the effect of the back reaction increases with increasing $\mathrm{St_{max}}$. For $\mathrm{St_{max}}=\min(\mathrm{St}_{\rm{fragm}},\mathrm{St}_{\rm{drift}})$, Fig.~\ref{fig:vragsavg} shows that the averaged gas radial velocity deviates from the reference case (solid line, $\mathrm{St_{max}}=\mathrm{St}_{\rm{fragm}}$, equal to the dashed line in the top-left panel) for $r\gtrsim 50$ au, where the drift limits the grain-size distribution (see Fig.~\ref{fig:stokez}). In this region, although the radial drift velocity decreases with increasing distance from the star, it still deviates from the expected dust-free viscous velocity due to the effect of the back-reaction. In the cases where $\mathrm{St_{max}}=[0.1,1]$, the radial gas velocity increases with distance from the star and reaches a constant value at the transition between the Stokes and Epstein regime (see the dotted-dashed and dotted lines in Fig.~\ref{fig:stokez}). For $\mathrm{St_{max}}=1$ the effect of the back-reaction is maximized and the gas migrates outwards over a large region of the disc.
From this analysis we infer that the minimum value of $\mathrm{St_{max}}$ to produce outward gas migration by back-reaction is $\sim 0.5$ in our reference disc model ($q=3.5$, $\epsilon=0.01$ and $\alpha=10^{-3}$). }
\begin{figure*}
\begin{center}
\includegraphics[height=0.254\textwidth,trim={0 1cm 0 0},clip]{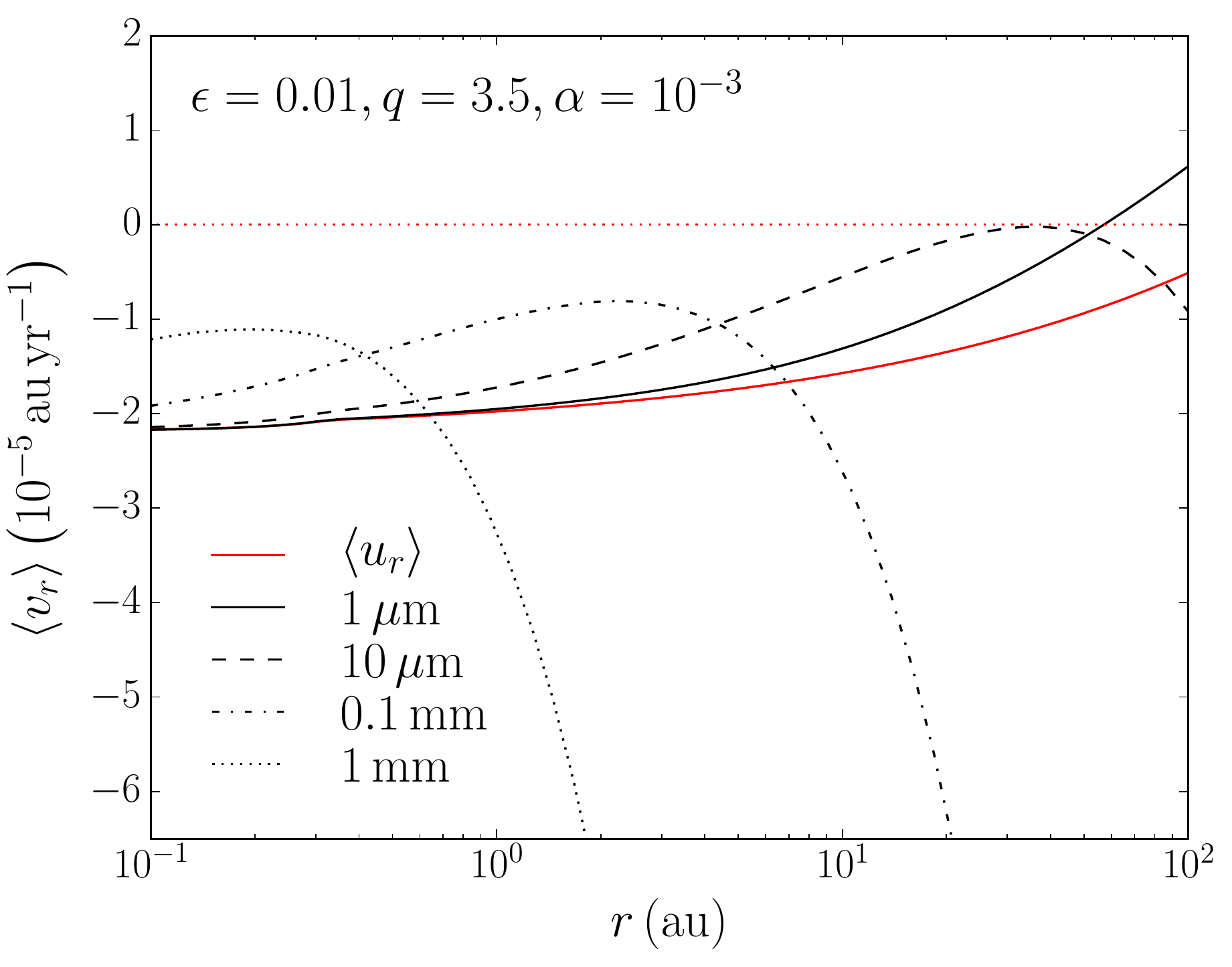}
\includegraphics[height=0.254\textwidth,trim={1.4cm 1cm 0 0},clip]{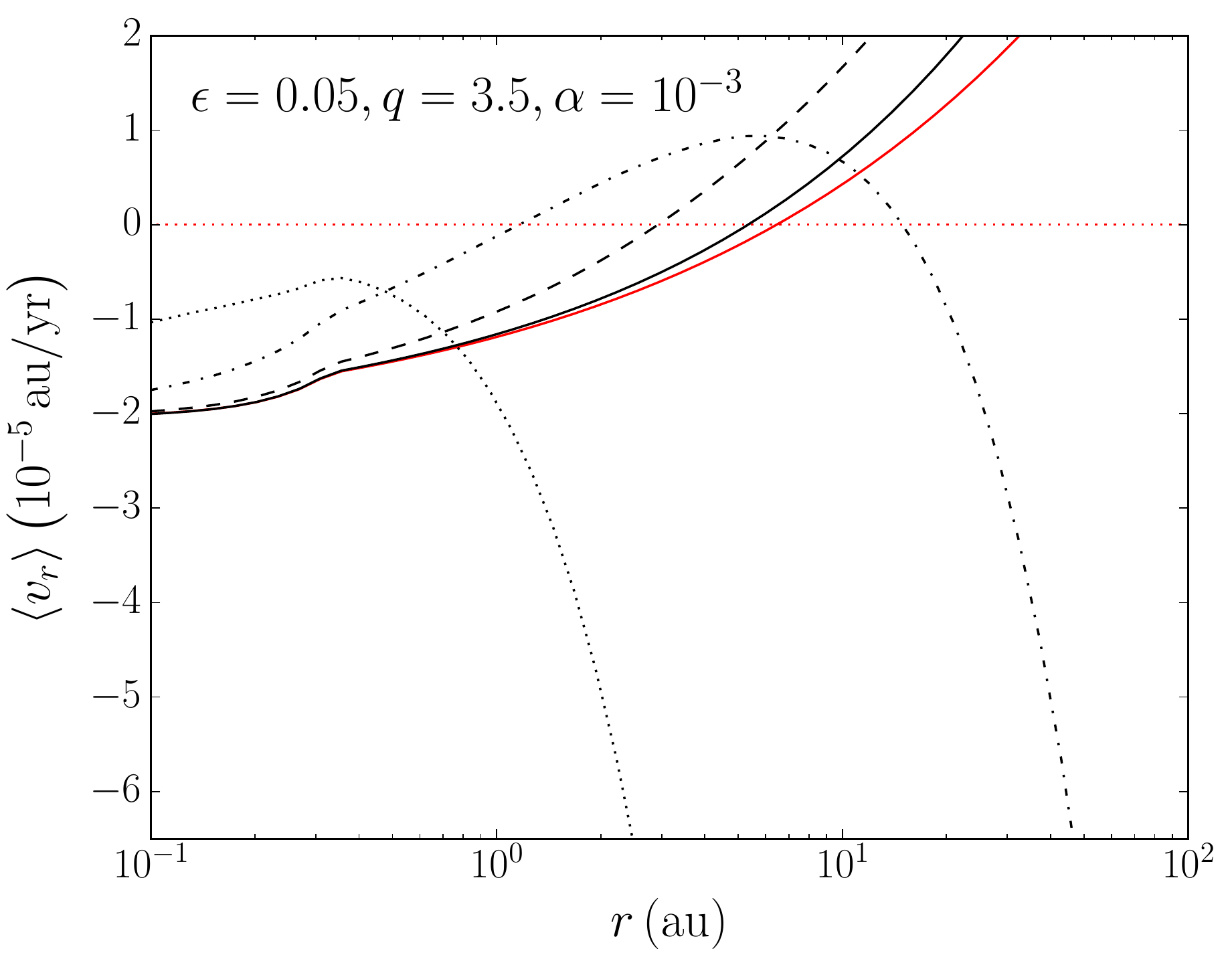}
\includegraphics[height=0.249\textwidth,trim={1.4cm 1cm 0 0},clip]{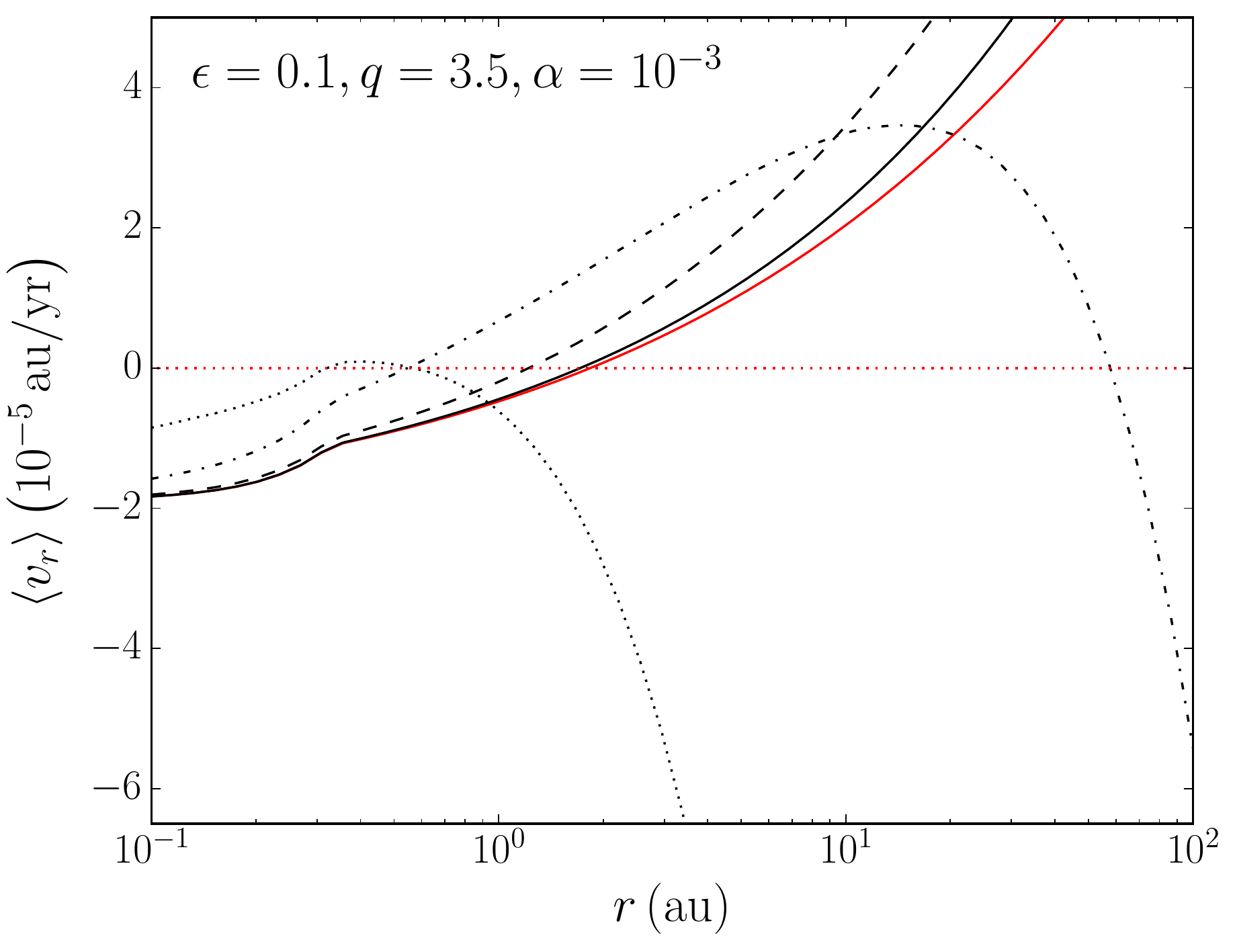}
\includegraphics[height=0.2715\textwidth,trim={0 0 0 0},clip]{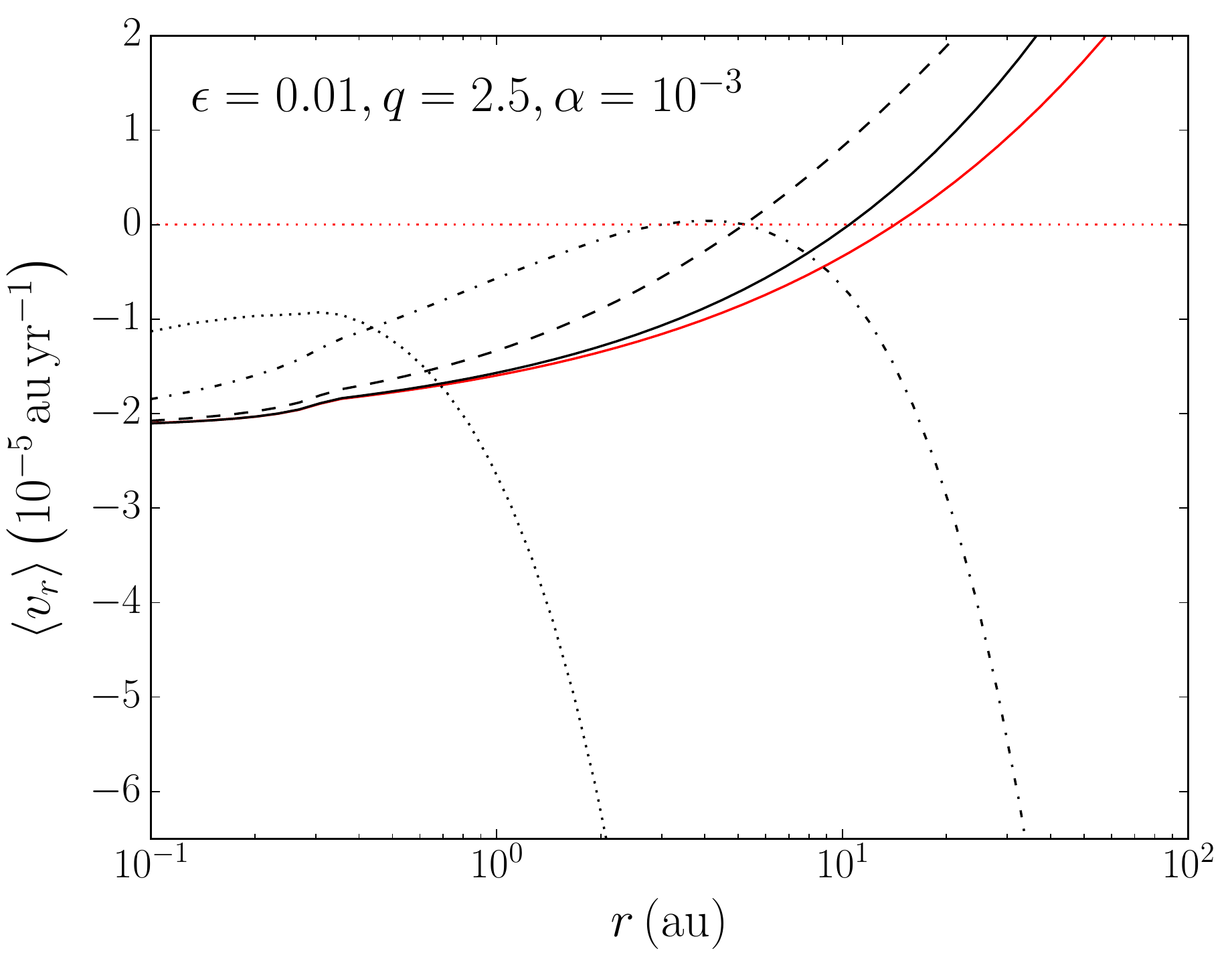}
\includegraphics[height=0.2715\textwidth,trim={1.4cm 0 0 0},clip]{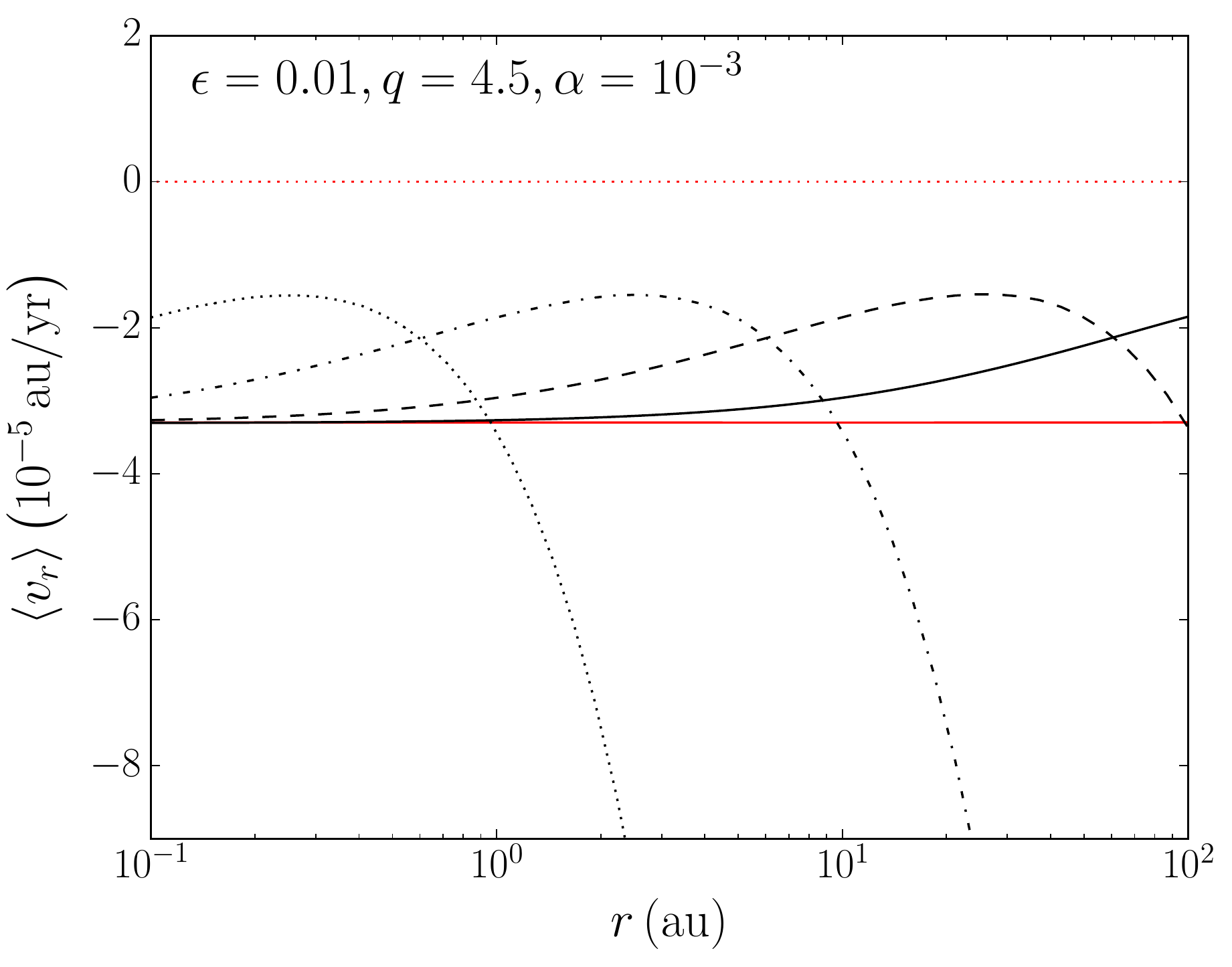}
\includegraphics[height=0.2715\textwidth,trim={1.4cm 0 0 0},clip]{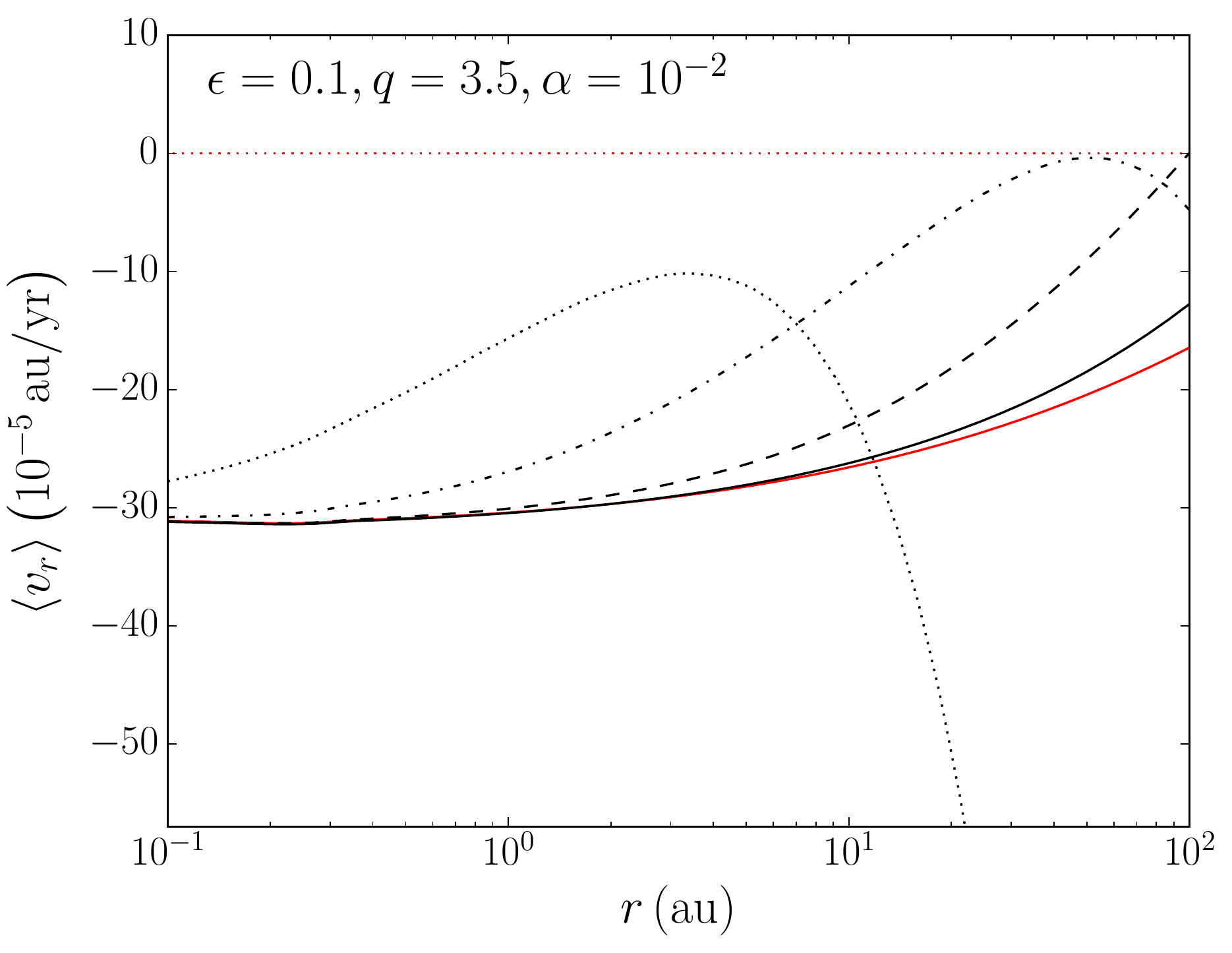}
\caption{Averaged radial velocity of gas (solid red line) and dust (black lines) with grain sizes $[1\,\mathrm{\mu} \text{m},10\,\mathrm{\mu} \text{m}, 0.1\,\text{mm},1\,\text{mm} ]$ assuming the model disc parameters listed in the top-left corner of each panel and $\mathrm{St_{max}}=\mathrm{St}_{\rm{fragm}}$. Our fiducial model is located in the \textit{top-left} panel. The two adjacent panels on top increase the dust-to-gas ratio ($\epsilon = [0.05,0.1]$), the \textit{bottom-left}/\textit{centre} panels sample a smaller/larger power-law index for the grain-size distribution ($q=[2.5,4.5]$) and the \textit{bottom-right} panel increases the dust-to-gas ratio ($\epsilon = 0.1$) and the viscosity ($\alpha=10^{-2}$). In each panel, the horizontal dotted red line indicates the null value. Dust grains have a slower inward velocity compared to the gas in the inner disc, while they decouple from the gas and drift inwards due to the drift-induced motion in the outer disc. In some cases, the effect of the back-reaction leads to an outward motion of dust grains in the intermediate--outer disc.}
\label{fig:vrdustprofr}
\end{center}
\end{figure*}


Similarly to Eq.~\ref{eq:urnet}, the averaged radial velocity of dust grains with size $s$ is defined as
\begin{equation}
\langle v_{r} (s) \rangle=\frac{1}{\Sigma_{\mathrm{d}}(s) } \int_{-\infty}^{\infty}\rho_{\mathrm{d}} (s, z )\,v_{r}(s,z) \,\mathrm{d}z,
\end{equation}
where 
\begin{equation}
\Sigma_{\mathrm{d}}(s)=\int_{-\infty}^{\infty}\rho_{\mathrm{d}}(s,z) \,\mathrm{d}z
\end{equation}
is the surface density, with $\rho_{\mathrm{d}}(s,z)$ defined per unit grain size in Eq.~\ref{eq:modeldustsz}.
%
Fig.~\ref{fig:vrdustprofr} shows the radial profile of the averaged radial velocities of gas (solid red line) and dust grains (black lines) with sizes in the range $[1\,\mathrm{\mu} \text{m},10\,\mathrm{\mu} \text{m}, 0.1\,\text{mm},1\,\text{mm}]$ in different disc models.
These results, although qualitatively similar to those shown in \citet{takeuchi02a} (see their Fig.~7), reveal additional features that need to be considered.
For all cases shown in Fig.~\ref{fig:vrdustprofr}, the averaged inward velocity for the gas (red solid line) and small dust grains (black lines) converge to the same value in the inner disc. This is due to the enhanced effect of the direct gas viscous drag compared to the drift contribution (Eq.~\ref{eq:vraddi}), especially for smaller particles which are more tightly-coupled with the gas in the inner regions and tends therefore to move with the inward gas flow (solid red lines). In all the cases, dust grains have a slower inward velocity compared to the gas in the inner disc. This is related to the fact that in the inner disc, since dust grains are slightly more concentrated around the midplane (see the black lines in the left panel of Fig.~\ref{fig:rhodustz}), they feel more the viscous drag effect of the outflowing gas around the midplane (for $z\lesssim 0.7 H_{\mathrm{g}}$, Eq.~\ref{eq:vviscrz}), compared to the case in which the particle vertical distribution is the same as the gas. Moreover, in all the cases shown in Fig.~\ref{fig:vrdustprofr}, drift dominates the dust motion over the direct viscous drag contribution in the outer disc regions, leading to a fast radial migration of dust grains toward the inner disc. Larger grains drift faster than smaller grains due to both the loss of coupling with the gas (i.e. the Stokes number approaches unity, Fig.~\ref{fig:stokez}) and the high level of dust settling (see the red lines in the left panel of Fig.~\ref{fig:rhodustz}). Due to the enhanced dust settling, the number of particles experiencing the maximum headwind from the gas around the midplane (see Eq.~\ref{eq:vpvkz}) is therefore larger than in the case in which the particle vertical distribution is the same as the gas.

As the dust-to-gas ratio increases (top panels of Fig.~\ref{fig:vrdustprofr}), the direct drag from the gas, strongly influenced by the back-reaction (outward gas motion at $r\gtrsim 10$ au, see the top-left panel of Fig.~\ref{fig:vragsavg}), forces the dust particles to drift outward in the intermediate disc regions, in contrast with the expectations of the single-species dust dynamics \citep{takeuchi02a}.
For example,  $0.1$ mm-sized particles in the disc model with $\epsilon=0.05$ (dotted-dashed line in the top-central panel of Fig.~\ref{fig:vrdustprofr}) drift inwards in regions with $r\lesssim 1$ au and $r\gtrsim 20$ au and outwards in intermediate regions. 
This inward motion at the inner and outer disc regions of these grains can be ascribed to two different contributions. 
While in the outer discs ($r\gtrsim 20$ au) these grains decouple from the gas and drift inwards due to the drift-induced motion, in the inner regions ($r\lesssim 1$ au) they start experiencing the direct viscous drag from the averaged inward gas motion (solid red line). In the intermediate disc regions ($1 \,\text{au} \lesssim r \lesssim 20$ au), the $0.1$ mm sized particles settle effectively to the midplane and are coupled enough to feel the direct viscous drag from the outward gas motion close to the midplane ($z\lesssim 0.7 H_{\mathrm{g}}$) enhanced by the back-reaction (see the dotted-dashed line in the top-left panel of Fig.~\ref{fig:vragsavg}).
For the disc model with the highest dust-to-gas ratio ($\epsilon=0.1$, see the top-right panel of Fig.~\ref{fig:vrdustprofr}), even the dust radial motion of millimetre-sized particles is strongly damped in a narrow disc region. 
Decreasing the steepness of the grain-size distribution (see bottom-left panel of Fig.~\ref{fig:vrdustprofr}) results in an increased effect of the back-reaction on to the gas radial motion and, indirectly, to tightly-coupled grains. Comparing the results shown in the left panels of Fig.~\ref{fig:vrdustprofr}, the dust radial velocity is smaller (in absolute value) for shallower grain-size distribution, due to the enhanced effect of the back-reaction (see the top-right panel of Fig.~\ref{fig:vragsavg} or compare the red lines in the left panels of Fig.~\ref{fig:vrdustprofr}). For steeper grain-size distributions (bottom-central panel of Fig.~\ref{fig:vrdustprofr}), the back-reaction is strongly damped (see the dotted line in the top-right panel of Fig.~\ref{fig:vragsavg}) and our results recover qualitatively those shown in Fig.~7 of \citet{takeuchi02a}. Moreover, increasing the viscosity (bottom-right panel of Fig.~\ref{fig:vrdustprofr}) results in a reduction of the effect of the back-reaction compared to the less viscous disc cases (compare the right panels in Fig.~\ref{fig:vrdustprofr}). In this case, the back-reaction is not sufficient to trigger outward migration of gas and small dust grains. 


\section{Discussion}
\label{sect:discussion}

The dynamics in a mixture of gas and dust is richer when more than one dust phase is present. As first remarked by \citet{bai10a}, multiple dust phases should not be studied by treating the dust phases as independently coupled to the gas and the conventional solution of single species dust dynamics \citep{nakagawa86a,dipierro17a,kanagawa17a} does not simply generalise to the case with multiple species of particles.
Our analysis clearly shows that the cumulative interaction between gas and dust grains of different sizes has a non-negligible effect on the evolution of both phases in typical protoplanetary discs. 
Even for low dust-to-gas ratios, the dynamical feedback from the dust can dominate over the viscously driven motion of the gas in a large portion of the disc, leading to a decrease of the inward gas velocities compared to the dust-free gas motion. 

\subsection{Gas and dust depletion in the inner disc regions: consequences for disc evolution}
The net effect of the cumulative back-reaction is to decrease the inward gas velocity (compared to the dust-free viscously-driven motion), especially in regions where the dust is not tightly-coupled to the gas. This effect becomes more pronounced with increasing distance from the star and for increased dust-to-gas ratios. For large dust-to-gas ratios ($\epsilon\gtrsim 0.05$), the back-reaction can even produce an outward gas motion. Although the solutions presented here assume steady-state, our results show that the back-reaction cannot be neglected in disc evolution calculations, and is likely to modify the disc structure significantly on secular time-scales ($\gtrsim10^5$--$10^6$\,yr). The back-reaction is therefore likely to produce gaps in discs, especially where the viscosity is low (i.e., $\alpha \sim 10^{-3} - 10^{-4}$) and in regions where dust settling increases the dust-to-gas ratio at the midplane. 

These results also confirm that the enhanced effect of the drag due to the increased coupling in the inner disc regions halts the inward drift of the particles that make the grains follow the gas motion, consistent with single-species calculations \citep{laibe12c}.
The inward drift of large grains is always slower than the inward gas flow, leading to a long-time retention of large grains in the inner disc.
On secular time-scales this naturally leads to an increased dust-to-gas ratio, which tends to provide a negative feedback on the radial inward drift of large dust grains \citep{drazkowska16a,kanagawa17a}. Therefore, increasing the dust-to-gas ratio tends to (i) decrease the inward dust motion and (ii) enhance the outward gas flow.
Together, these effects result in a net outward motion of both gas and dust in the intermediate disc regions where dust grains are tightly-coupled enough to follow the gas motion, but still able to sediment around the midplane (and therefore experience the direct drag of the viscously-driven outward gas flow enhanced by  back-reaction).
This process could drive a global redistribution of gas and solids in protoplanetary discs and may lead to a pile-up of pebbles in the inner disc regions \citep{pinte14a,drazkowska16a,gonzalez17a}. 
The resulting increase of the dust-to-gas ratios and the reduction of dust inward drift velocities in the inner disc regions create conditions favourable for overcoming the difficulties in growing centimetre-sized objects to kilometre sizes, enhancing the efficiency of grain growth.
These results also show that the effect of the back-reaction is closely linked to the local disc conditions. It then follows that the intrinsic gradients in the gas and dust quantities could lead to radial gradients in the effect of the back-reaction on the net gas motion, which in turn might produce peculiar features in the gas and dust density distribution \citep{gonzalez15a,gonzalez17a}.

While we must of course be cautious in drawing conclusions about disc evolution from steady-state solutions, our results point naturally towards a scenario where the back-reaction creates a gap or cavity in the inner disc, in both gas and dust. This has important implications for disc evolution, and has two likely consequences. First, the reduction of the inward gas velocity means that gas accretion is suppressed at intermediate radii in the disc. These radii ($\sim$ few to tens of au) are the same locations where we expect significant mass-loss from photoevaporative and/or magnetically-launched winds \citep[e.g.,][]{alexander14a,ercolano17a}. If gas accretion is suppressed at these radii then mass-loss overcomes the accretion flow much more readily, at earlier times in the disc's evolution, and the dust back-reaction may even ``trigger'' early disc dispersal \citep[similar to the way in which giant planets can trigger photoevaporative clearing,][]{alexander09a,rosotti13a}. Moreover, disc winds efficiently entrain small, micron-sized dust grains while larger dust particles are left behind in the disc \citep{takeuchi05b,gorti16a,hutchison16a,hutchison16b}. On secular time-scales this invariably increases the dust-to-gas ratio in larger grains, increasing the efficiency of the dust feedback and further suppressing the inward motion of the gas.

Second, the creation of inner cavities due to differential gas-dust dynamics has important consequences for our understanding of so-called ``transitional'' discs. In recent years high-resolution mm-wavelength observations have discovered large numbers of discs with large ($\gtrsim$ tens of au) cavities in their dust discs \citep[e.g.,][]{brown08a,andrews11a,van-der-marel15b}, and perhaps the most popular interpretation is that such structures are evidence of dynamical clearing by giant planets. However, the relatively high incidence of these cavities ($\gtrsim$ 10\%, and perhaps 2--3 times higher among the sub-set of mm-bright of transitional discs, \citealt{van-der-marel-18a}) is increasingly difficult to reconcile with dynamical clearing by planets, as exoplanet surveys have now firmly established that the incidence of massive exoplanets at wide separations is low ($\lesssim$5\%; e.g., \citealt{vigan17a}). Our results suggest that differential dust-gas motions provide a more generally applicable mechanism for opening cavities, the conditions for which are likely to be met at some point during the evolution of most protoplanetary discs. Dust feedback on the gas dynamics can be therefore considered as a potential mechanism to produce cavities, and might explain the recent link in the evolution of dust and molecular gas in the inner disc regions \citep{banzatti17a}.
Time-dependent calculations are still required, but these results have potentially profound implications for both the dynamical and secular evolution of planet-forming discs.

\subsection{Limitations of the model}
\label{sec:limitation}
The fact that we consider only steady-state solutions is the main limitation of this study. We can extrapolate from our steady-state velocities to estimate how the gas and dust evolution in protoplanetary discs, but time-dependent solutions obtained by integrating the equations of motion are necessary to understand the full implications of our results. In addition, our solutions make a number of symmetry assumptions (most notably azimuthal symmetry) and full 3D calculations will be required to understand the dynamics induced by the back-reaction. This however requires sophisticated gas and dust simulations, the tools for which have only recently begun to be developed \citep{gonzalez17a,ricci18a,humphries18a,hutchison18a,weber18a}.

The other notable limitation of our work is that the gas and dust density evolution on long time-scales are expected to be affected by grain growth. Evolution of the dust size distribution naturally leads to a wider range of drag interaction regimes between the gas and dust. Even though we assume a fragmentation-limited grain-size distribution, the dynamics of dust grains investigated in this paper affects the relative velocities between colliding grains and their relative density in different part of the discs \citep{drazkowska16a}, which may have consequences for grain growth.

Finally, we assume in our model that the dust-free gas evolution in the disc is governed by viscous angular momentum transport. This approach is commonly used to model angular momentum transport driven by turbulence generated by, e.g., magneto-rotational or gravitational instabilities \citep{pringle81a,balbus91a,lodato04a}. It remains unclear to what degree the properties of disc turbulence can be characterized by a kinematic viscosity \citep[e.g.,][]{balbus99a,turner14a} and recent work suggests that accretion may be driven by non-diffusive mechanisms, such as magnetohydrodynamic winds or spiral density waves \citep{bai16a,fung17a}. Thus, in the absence of a full description of the disc microphysics, we restrict our calculations to the viscous disc formalism.


\section{Conclusions}
\label{sec:conclusion}
In this paper we investigate the steady-state dynamics of a disc consisting of a mixture of a single gas phase and a collection of dust species covering a wide range of grain sizes and drag regimes. The coexistence of multiple dust species with the gas in protoplanetary discs is the natural outcome of grain growth across the entire disc extent.
We adopt a model where the gas exchanges angular momentum with the entire dust mixture (as opposed to each grain size individually), thereby indirectly coupling the dust species together via their combined feedback on the gas. Despite showing similar drag-induced trends, we find that the radial migration of gas and dust when multiple dust species are included is remarkably different from the single-species case \citep{kanagawa17a}. Our results are summarized as follows:

\begin{itemize}
\item We derive the steady-state velocities of a mixture of gas and a population of dust particles with different sizes, taking into account the gas viscosity. We validate the equations using SPH gas/dust simulations, finding a remarkable agreement between the analytical results and the numerical calculations. 
\item The cumulative effect of the dust back-reaction leads to a decrease of the gas motion compared to the pure viscously-driven flow, with an efficiency that depends on the mass embodied in marginally-coupled grains, dust-to-gas  ratio and disc viscosity. 
\item In typical protoplanetary discs, the dust feedback strongly affects the gas dynamics -- even for small values of the total dust-to-gas ratio -- leading to gas outflow in the outer disc regions.
Dust feedback is especially effective in the outer disc regions of low viscosity discs, where particles are less coupled with the gas and efficiently settle to the midplane. 
\item In the inner and intermediate regions of typical protoplanetary discs, the averaged dust inward motion is reduced below the value of the gas inward motion due to the combined effects of strong drag and settling. In cases with high dust-to-gas ratio, the motion of large particles can be directed outward, due to the drag-induced motion of the gas, which in turn is influenced by the dust back-reaction.
\end{itemize}
We highlight that the cumulative effect of the back-reaction can play a crucial role in shaping the gas and dust density evolution in protoplanetary discs. 
Our results show that the dust back-reaction is a viable mechanism for driving gas outflow and halting or slowing down the radial migration of dust grains. This may allow protoplanetary disc to retain their large dust grains, as recently seen in high resolution observations \citep{testi14a}. 

On long time-scales the general effect of the back-reaction is to enhance inside-out disc evolution, even early in the disc's lifetime. It is therefore crucial to include the dynamical effect of multiple dust species in {\it gas} disc evolution models, and to calculate the impact of the dust back-reaction on disc evolution and dispersal. The dust back-reaction cannot be neglected in disc evolution models, and needs to be taken into account in numerical simulations of dusty discs \citep{gonzalez17a,hutchison18a}.

\section*{Acknowledgements}
We wish to thank the referee for an insightful report of the manuscript.
This project has received funding from the European Research Council (ERC) under the European Union's Horizon 2020 research and innovation programme (grant agreement No 681601). GL acknowledges financial support from PNP, PNPS, PCMI of CNRS/INSU, CEA and CNES, France. This project was supported by the IDEXLyon project (contract n°ANR-16-IDEX-0005) under University of Lyon auspices. The numerical SPH simulations have been run on the Piz Daint supercomputer hosted at the Swiss National Computational Centre and were carried out within the framework of the National Centre for Competence in Research PlanetS, supported by the Swiss National Science Foundation.

\bibliography{biblio}

\appendix

\newtext{
\section{Notations}
\label{app:notations}

The notations used throughout this paper paper are summarized in Table~\ref{tab:not}.

\begin{table}
	\centering
	\caption{Notations used in the paper.}
	\label{tab:not}
	\begin{tabular}{ll} 
		\hline
		\hline
		Symbol & Meaning\\
		\hline
		$\mathbf{u}$ & gas velocity\\
		$\mathbf{v_i}$ & velocity of the $i^{th}$ dust phase\\
		$K_i$ & drag coefficient of the $i^{th}$ dust phase\\
		$M_{\star}$ & mass of the central star \\
		$\mathcal{G}$ & gravitational constant\\
		$\Phi$ & gravitational potential\\
		$P$ & gas pressure\\
		$P_0$ & gas pressure at $z=0$\\
		$c_{\mathrm{s}}$ & sound speed \\
		$\sigma$ & viscous stress tensor\\
		$\nu$ & kinematic viscosity\\
		$\alpha$ & Shakura-Sunyaev turbulence parameter\\
 		$\rho_{\mathrm{g}}$ & gas volume density\\
		$\rho_{\mathrm{g}0}$ & gas volume density at $z=0$\\
		$\rho_{\mathrm{d}}$ & dust volume density\\
		$\rho_{\mathrm{d}i}$ & $i^{th}$ dust phase volume density\\
		$\rho_{\mathrm{d}i,0}$ & $i^{th}$ dust phase volume density at $z=0$\\
		$\Sigma_{\mathrm{g}}$ & gas surface density\\
		$p$ & radial surface density exponent, $\Sigma_{\mathrm{g}}\propto r^{-p}$\\
		$T$ & gas temperature\\
		$m$ & radial surface temperature exponent, $T\propto r^{-m}$\\
		$t_i^{\mathrm{s}}$ & drag stopping time of the $i^{th}$ dust phase\\ 
		$\epsilon$ & dust-to-gas ratio\\
		$\epsilon_i$ & dust-to-gas ratio of the $i^{th}$ dust phase\\
		$H_{\mathrm{g}}$ & gas scale height\\
		$H_{\mathrm{d}i}$ & scale height of the $i^{th}$ dust phase\\
		$\Omega_{\mathrm{g}}$ & gas angular velocity \\
		$\Omega_{\mathrm{k}}$ & Keplerian angular velocity \\
		$\Omega_{\mathrm{k,mid}}$ & Keplerian angular velocity at $z=0$\\
		$v_{\mathrm{k}}$ & Keplerian velocity \\
		$v_{\mathrm{k,mid}}$ & Keplerian velocity at $z=0$ \\
		$\mathrm{St}_i$& Stokes number of the $i^{th}$ dust phase\\
		$\mathrm{St}_{\mathrm{mid},i}$& Stokes number of the $i^{th}$ dust phase at $z=0$\\
		$s_i$&  grain size of the $i^{th}$ dust phase\\
		$\rho_{\mathrm{grain}}$ & material grain density\\
		$\lambda_{\mathrm{mfp}}$ & mean free path\\
                 $n(s)$ & grain-size distribution\\
                 $q$ & grain-size distribution exponent, $n(s)\propto s^{-q}$\\
                 $s_{\mathrm{max}}$ & maximum grain size\\
                 $s_{\mathrm{min}}$ & minimum grain size\\
                 $m(s)$ & mass of a dust grain of size $s$\\
                 $\mathrm{St_{fragm}}$& fragmentation-limited Stokes number\\
                  $\mathrm{St_{drift}}$& drift-limited Stokes number\\
		\hline
		\hline
	\end{tabular}
\end{table}
}

\section{Numerical test}
\label{app:numericaltests}
To validate our expressions for the steady-state gas and dust velocities in viscous discs, we performed 3D gas and dust simulations using the Smoothed Particle Hydrodynamics (SPH) code \textsc{phantom} and its newly implemented \textsc{multigrain} algorithm \citep{hutchison18a}. The \textsc{multigrain} algorithm extends the one-fluid, terminal velocity approach for high drag regimes \citep[e.g.][]{price15a,ballabio18a} to account for the cumulative back-reaction of multiple small dust species on to the gas, thereby allowing us to recover the radial migration velocities predicted by Eqs.~\ref{eq:vradg} and \ref{eq:vraddi}. During the original testing of the \textsc{multigrain} algorithm, \citet{hutchison18a} attempted a similar test and benchmarked their results using the analytical solutions for an inviscid disc presented by \citet{bai10a}. However, apart from good fits obtained with carefully constructed initial conditions, they found that the diffusive nature of SPH caused the numerical solution to relax into a substantially different steady-state than that predicted by the inviscid equations. Now, by properly accounting for the effects of viscosity, we repeat the test and use it as a mutual verification of our equations and the \textsc{multigrain} algorithm.

\subsection{Setup}
\label{sec:numericaltests_setup}

We place a gas/dust disc orbiting around a central star with mass $M_{\star}=1\,M_{\odot}$ with an inner and outer disc radius of $r_{\mathrm{in}} = 1\,\text{au}$ and $r_{\mathrm{out}} = 150\,\text{au}$, respectively. We assume an initial power-law gas and dust surface density profile $\Sigma \propto r^{-1}$ and model the disc with $2 \times 10^6$ SPH particles. We adopt a locally isothermal equation of state with a temperature power-law index $m = 1/2$. The reference values for the gas model parameters at $1\,\text{au}$ are as follows:  $c_{\mathrm{s},1\text{au}} \approx 1.5 \, \mathrm{km s}^{-1}$, $H_{\mathrm{g},1\text{au}} = 0.05 \,\text{au}$, and $\Sigma_{\mathrm{g},1\text{au}} \approx 475 \,\mathrm{g\,cm}^{-2}$ such that the total gas mass in the disc is $0.05\,M_\odot$. We initially set the radial velocity to zero and the azimuthal velocities to the midplane Keplerian velocity, $v_\mathrm{k,mid}$, and correct for the radial pressure gradient in the disc (see Sect.~\ref{sect:gasvel3D}). To model the viscosity in the disc, we indirectly set a Shakura-Sunyaev disc viscosity with $\alpha \sim 0.01$ \citep{shakura73a} via the artificial dissipation parameter $\alpha^\mathrm{AV}$ as follows \citep{lodato10a}:
\begin{equation}
	\alpha \approx \frac{\alpha^\mathrm{AV}}{10} \frac{\langle h \rangle}{H_{\mathrm{g}}},
	\label{eq:artificial_viscosity}
\end{equation}
where $\langle h \rangle$ is the mean smoothing length on particles in a cylindrical ring at a given radius. For two million particles and  $\alpha^\mathrm{AV} \approx 0.74$, we obtain a representative $\alpha \approx 0.01$ at the radial midpoint in the disc.

The dust disc is composed of spherical grains with a uniform intrinsic density of $3\,\mathrm{g\,cm}^{-3}$, a grain-size distribution that ranges from $s_{\rm min} = 0.1\,\mathrm{\mu}\text{m}$ to $s_{\rm max} = 1\,\text{mm}$, and a number density power-law index of $q = 3.5$ (Eq.~\ref{grainsizedistrib}). The total dust mass is distributed into $10$ representative dust species, logarithmically spaced between $s_{\rm min}$ and $s_{\rm max}$, as described by \citet{hutchison18a}. All dust phases are distributed equally with the gas in space, but scaled in mass by the dust fractions listed in Table~\ref{tab:dust_vals_migration}, such that the total dust-to-gas ratio is $\epsilon = 0.5$. To avoid having to characterize time-dependent dust enhancements/depletions caused by vertical settling and/or radial migration of dust, we force the dust-to-gas ratio of the individual particles to remain constant for the duration of the simulation. Finally, although we initially set the dust velocity equal to the gas, the velocities of the different dust phases are automatically updated when the forces are initialised for the first time at $t = 0$.
\begin{table}
	\centering
	\caption{Grain sizes $s_i$, dust-to-gas ratios $\epsilon_i$ and representative Stokes numbers $\mathrm{St}^*_i$ (calculated at the disc's radial midpoint) for the dust in the radial migration test shown in Fig.~\ref{fig:radial_drift}. Summing the individual dust-to-gas ratios of each of the dust phases gives a total dust-to-gas ratio of 0.5.}
	\label{tab:dust_vals_migration}
	\sisetup{table-format = 1.2,table-auto-round = true}
	\begin{tabular*}{\columnwidth}
		{@{\extracolsep{\stretch{1}}}
			S[table-format=2.0]
			S[table-format=1.2e-1,scientific-notation=true]
			S[table-format=1.2e-2,scientific-notation=true]
			S[table-format=1.2e-2,scientific-notation=true]
		@{}} \toprule
		{  $i$} 	&	{$s_i\,$(cm)}			&	{$\epsilon_i$}	&	{$\mathrm{St}^*_i$}				\\\midrule
		1	&	1.5848931924611141e-05		&	2.9540060225e-03	&	0.792E-05		\\
		2	&	3.9810717055349735e-05	&	4.6817840356e-03	&	0.199E-04		\\
		3	&	1.0000000000000000e-04	&	7.4201276466e-03	&	0.500E-04		\\
		4	&	2.5118864315095806e-04		&	1.1760109794e-02	&	0.126E-03		\\
		5	&	6.3095734448019331e-04	&	1.8638517956e-02	&	0.315E-03		\\
		6	&	1.5848931924611132e-03		&	2.9540060225e-02	&	0.792E-03		\\
		7	&	3.9810717055349708e-03	&	4.6817840356e-02	&	0.199E-02		\\
		8	&	9.9999999999999985e-03	&	7.4201276466e-02	&	0.500E-02		\\
		9	&	2.5118864315095805e-02		&	1.1760109794e-01	&	0.126E-01		\\
		10	&	6.3095734448019331e-02	&	1.8638517956e-01	&	0.315E-01		\\\midrule
	\end{tabular*}
\end{table}

\subsection{Results}
\label{sec:numericaltests_results}
\begin{figure}
	\centering{\includegraphics[width=\columnwidth]{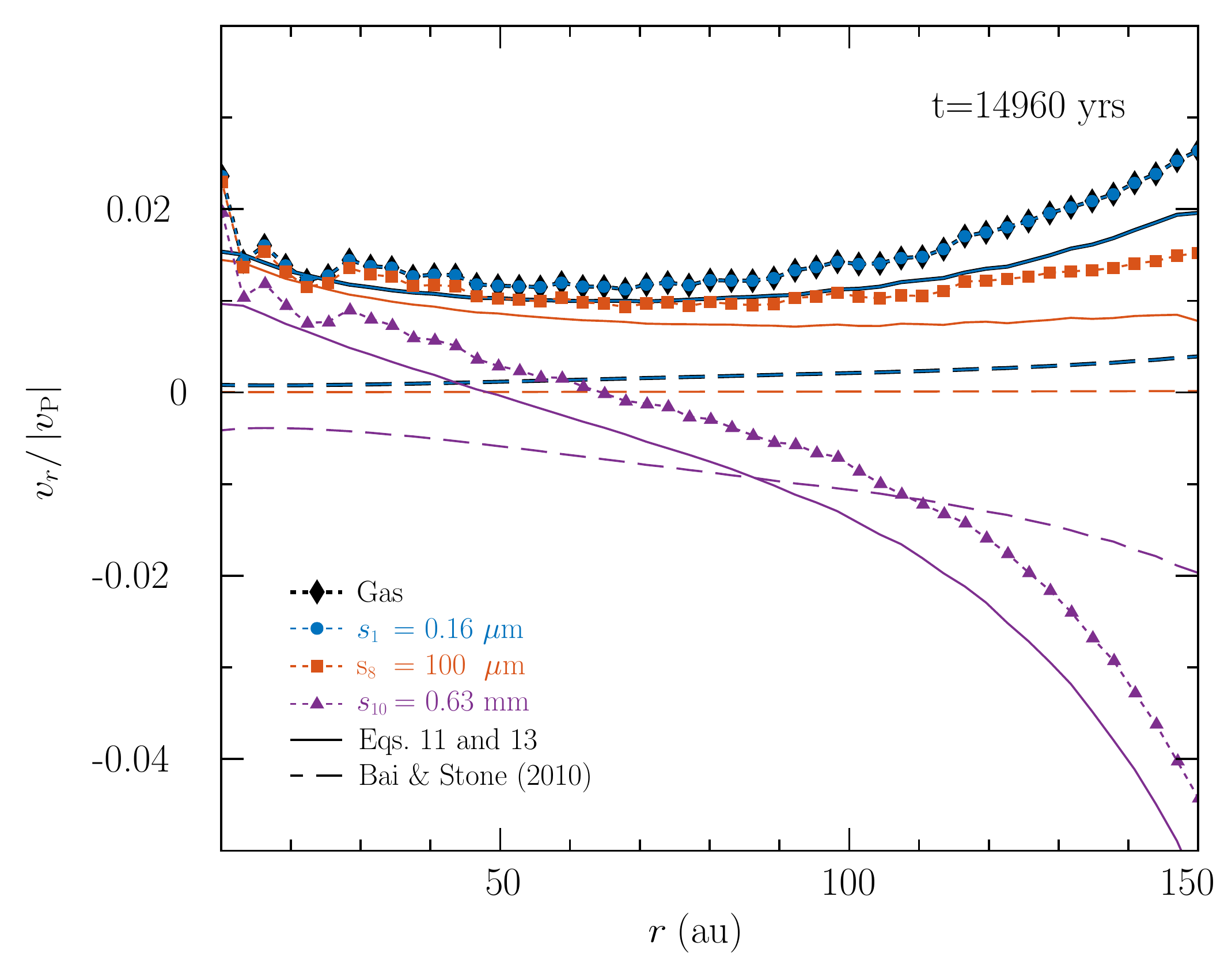}}%
	\caption{Mean radial drift velocities for gas (black) and dust (blue, orange, purple) near the midplane of a viscously evolving protoplanetary disc. The dust is composed of 10 different phases that all interact with the gas but are constrained to keep their initial dust-to-gas ratio. For clarity, only phases $[s_{1},s_{8},s_{10}] = [0.16\,\mu \text{m}, 100\,\mu \text{m}, 0.63\, \text{mm}]$ are shown. Solid lines represent the analytic solutions from Eqs.~\ref{eq:vradg} and \ref{eq:vraddi} while the dashed lines give the inviscid solutions from \citet{bai10a}. The systematic offset between the numerical and analytic solutions is most likely due to inaccuracies in the viscosity approximation in Eq.~\ref{eq:artificial_viscosity} (see text). Importantly, Eqs.~\ref{eq:vradg} and \ref{eq:vraddi} provide a marked improvement over the inviscid equations from \citet{bai10a} when computing the steady-state radial velocities for the gas and dust in a viscous disc.}
	\label{fig:radial_drift}
\end{figure}
We relax the disc for \newtext{$14\,960\,\textrm{yrs}$} (\newtext{$\sim 8$} orbits at $r_\mathrm{out}$) to allow the fluctuations in the radial velocity to damp. We then bin the particles radially into 50 uniformly spaced bins and average the radial velocities both azimuthally and vertically within each bin. Because the radial velocities, densities and Stokes numbers change with $z$, we only include particles within $|z| < H_\mathrm{g}$ in the averaging process. In practice, the thickness over which we average the velocities has only a minor impact on the final mean velocity profile for each grain size (as long as we remain within $|z| \lesssim H_\mathrm{g}$). In contrast, the averaging thickness has a noticeable effect on the average $\mathrm{St}_i$ computed for each bin, due to its exponential dependence on $z$ through $\rho_{\mathrm{g}}$. By averaging over successively thinner slices, we find that $\mathrm{St}_i$ asymptotically converges to a set of values that are $\sim 0.7$ times smaller than those in our original cut. Ideally, we would use these thinner cuts to calculate the radial velocities as well, but the dispersion in the velocities (the standard deviation ranges from $\sim 0.05$--$1\,v_{\mathrm{P}}$) and the low number statistics caused by a radially decreasing number density combine to make noisy data. Therefore, we continue to average the velocities over the local scale height of the disc and manually scale the Stoke numbers by this factor of $0.7$.

The steady-state radial velocities for gas (black) and dust (colours) are shown in Fig.~\ref{fig:radial_drift}. Symbols connected by a dotted line are the SPH velocities while solid lines are the corresponding analytic solutions from Eqs.~\ref{eq:vradg} and \ref{eq:vraddi}. For comparison, the inviscid solutions from \citet{bai10a} are overlaid using dashed lines. By construction, the initial disc conditions mirror that of an inviscid disc such that the numerical solution matches the solution from \citet{bai10a} at $t = 0$. Once evolution begins and the effects of viscosity are felt, the numerical solution quickly departs from the inviscid solution and settles near our new  equations (Eqs.~\ref{eq:vradg}--\ref{eq:vraddi}) from the inside out on a time-scale of approximately five to six orbital periods of the local disc.

A systematic offset persists between our analytic and numerical solutions, even in regions that have reached a steady-state. As already mentioned, the way in which we construct the 2D midplane properties for the analytic solution out of our 3D SPH simulation data can affect some fluid quantities more than others. These variations are then passed on to the analytic solution, causing errors in the fit. For example, the asymptotically corrected $\mathrm{St}_i$ values improve our fit by $\sim 20\%$. Although it is possible that averaging effects from other parameters contribute to the offset, we suspect that the remaining offset is primarily related to the conversion from artificial to physical viscosity using the approximation in Eq.~\ref{eq:artificial_viscosity}. This suspicion is based on the good agreement we see with the \citet{bai10a} solutions at $t = 0$, since a problem with the averaging would have also showed up in the inviscid solutions. Thus, with the viscosity being the only difference between the solutions, it stands to reason that the viscosity is also culprit. Indeed, uniformly scaling the $\alpha$ parameter by a factor of $\sim 1.3$ corrects the offset entirely. However, without an independent method of measuring the true $\alpha$ in the simulation, we cannot be sure that our viscosity model is responsible for the remaining offset or not.

As a final test, we ran a more extreme case with $\alpha = 0.1$ where the viscous and inviscid solutions are even more separated. The high viscosity leads to significant processing of the inner/outer disc through viscous accretion/spreading, but we find similar agreement between our analytic and numerical solutions in the disc interior. Thus we are confident that our analytic solution and the \textsc{multigrain} algorithm are both working as expected and that these offsets are a result of an inconsistency in how we post-process and compare our solutions. At the very least, our new analytic solution provides a significant improvement over the inviscid solution from \citet{bai10a} when computing the steady-state radial velocities of the gas and dust in viscous discs.

\section{Steady-state dust-free gas velocities}
\label{app:steady}

The usual way to derive the steady-state velocities of the gas in a stratified viscous disc consists in performing a linear expansion with respect to the Keplerian velocity by keeping the leading terms in $\alpha z^{2} / H_\mathrm{g}^{2}$ and $\alpha  H_\mathrm{g}^{2} / r^{2}$. The scale height of the gas $H_\mathrm{g}(r)$ is \textit{a priori} a function of $r$. Two peculiarities of the system of equations should be identified prior to derivation in order to simplify the analytic treatment. First, the terms arising from advection and from the viscous stress tensor in the radial direction scale as $\alpha^{2}$ and can be safely neglected at this level of approximation. Secondly, the viscous stress tensor in the azimuthal direction contains terms in $\displaystyle \partial u_{\phi}/ \partial z^{2}$. This implies that terms of order $z^{4}$ should be preserved for the expression of $u_{\phi}$ and, therefore, in the gravitational potential, the radial pressure gradient and the orbital frequency $\Omega_{\rm k}$. However, keeping these terms gives rise to contributions of order $\alpha z^{2}H_\mathrm{g}^{2}/r^{4}$, which are negligible. Hence, it is both correct and more convenient to restrain the expansion of the gravitational potential to the leading order $z^{2}$. With $\Sigma_{\mathrm{g}}\propto r^{-p}$ and $T\propto r^{-m}$, we obtain from the balance of forces in the radial direction
\begin{equation}
u_{\phi} \simeq v_{\rm k,mid}\left[1 - \left( \frac{3 + m + 2p}{4}\right)\frac{H_\mathrm{g}^{2}}{r^{2}} -  \frac{m}{4} \frac{z^{2}}{r^{2}} \right] .
\end{equation}
This implies for the balance of forces in the azimuthal direction,
\begin{equation}
u_{r} \simeq \alpha  v_{\rm k,mid} \left[ \frac{6p + m - 3}{2} \frac{H^{2}}{r^{2}} +\frac{-9+ 5m}{2} \frac{z^{2}}{r^{2}} \right] .
\end{equation}
With the notations of \citet{takeuchi02a}, $\rho_{\mathrm{g}0}\propto r^{p_{\rm T}}$ and $T\propto r^{ q_{\rm T}}$, $p = - p_{\rm T} - \frac{q_{\rm T} + 3}{2}$ and $m = - q_{\rm T}$, we obtain
\begin{equation}
u_{\phi} \simeq v_{\rm k,mid}\left[1  + \frac{p_{\rm T} + q_{\rm T}}{2} \frac{H_\mathrm{g}^{2}}{r^{2}} + \frac{q_{\rm T}}{4} \frac{z^{2}}{r^{2}} \right] .
\label{eq:uphi_TL02}
\end{equation}
This implies for the balance of forces in the azimuthal direction,
\begin{equation}
u_{r} \simeq \alpha  v_{\rm k,mid} \left[ -\left( 3p_{\rm T} + 2q_{\rm T} + 6\right) \frac{H_\mathrm{g}^{2}}{r^{2}} - \frac{9+5q_{\rm T}}{2} \frac{z^{2}}{r^{2}} \right] ,
\end{equation}
which are the solutions derived by \citet{takeuchi02a}.

\newtext{
\section{Simplifying the linear expansion}
\label{app:approx}

In an inviscid disc, the linear expansion of the equations of motion (Eqs.~\ref{eq:1}--\ref{eq:4}) involves only the perturbed velocities of the gas and the dust with respect to the Keplerian orbit and not their derivatives. The system of equation can therefore be formally put under the form $\mathrm{\mathbf{D}}\mathbf{V} = \mathbf{F}$, where $\mathbf{V} = \left(u_{r},u_{\phi},v_{r},v_{\phi} \right)^{\top}$, $\mathbf{D}$ is a $4\times4$ invertible matrix that encompasses the drag coefficients and $\mathbf{F}$ is a constant driving force from the radial pressure gradient of the gas (from simplicity, we consider here only one dust phase, but the reasoning can be extended directly to any number of dust phases). The steady-state velocities are therefore given by $\mathbf{V} = \mathrm{\mathbf{D}}^{-1} \mathbf{F}$ \citep{nakagawa86a}. This formalism can safely be extended to an unstratified viscous disc when keeping only the leading term of order $\alpha  H_{\mathrm{g}}^{2} / r^{2}$ which drives the accretion flow, i.e. the viscous torque due to the Keplerian shear.
Rigorously, the analysis becomes much more complex in the case of a stratified disc. Formally, to the first order, the equations of motion at steady-state can be written under the form
\begin{equation}
\mathrm{\mathbf{D}}\mathbf{V} = \mathbf{F} + \mathcal{L}_{z}\left( \mathbf{V} \right) ,
\label{eq:full_exp_strat}
\end{equation}
where $\mathcal{L}_{z}$ denotes the linear operator corresponding to the viscosity associated to the shear in the vertical direction. As in Appendix~\ref{app:steady}, ansatz of the form $v_{0} + v_{1}\left(z/H_{\mathrm{g}} \right)^{2}$ should be used for $\textit{every}$ velocity (e.g. Eq.~\ref{eq:uphi_TL02}) and the steady-state solutions are obtained by inverting an $8\times 8$ system of coefficients that is not very tractable, but still physically meaningful.
Instead, we can approximate the rigorous solution by computing a steady-state where the drag terms are balanced by the viscous term in the dust-free flow. This is the approach followed by \citet{kanagawa17a} but generalised here for multiple grain sizes. To justify this approximation, we first note that the dust-free flow solution is the one derived in Sect.~\ref{app:steady}. Formally, the dust-free solution $\mathbf{\tilde{V}}$ satisfies 
\begin{equation}
\mathrm{\mathbf{\tilde{D}}}\mathbf{\tilde{V}} = \mathbf{F} + \mathcal{L}_{z}\left( \mathbf{\tilde{V}} \right) ,
\end{equation}
where $\mathrm{\mathbf{\tilde{D}}} = \mathrm{\mathbf{D}} \left( \mathrm{St} = 0 \right)$. Eqs.~\ref{eq:vradg}--\ref{eq:vthetag} consist in approximating $\mathcal{L}_{z}\left( \mathbf{V} \right)$ by $\mathcal{L}_{z}\left( \mathbf{\tilde{V}} \right)$. In this case, the approximated solution $\mathbf{V}'$ we obtain is 
\begin{equation}
\mathbf{V}' = \mathrm{\mathbf{D}}^{-1}\mathrm{\mathbf{\tilde{D}}} \mathbf{\tilde{V}} .
\label{eq:defvp}
\end{equation}
For the forthcoming derivation, we note at this stage that Eq.~\ref{eq:defvp} implies also that $\mathcal{L}_{z}\left(  \mathbf{V}' \right) = \mathrm{\mathbf{D}}^{-1}\mathrm{\mathbf{\tilde{D}}} \mathcal{L}_{z}\left( \mathbf{\tilde{V}} \right)$, since $\mathcal{L}_{z}$ is linear. Let us now denote $\mathbf{V} = \mathbf{V}' + \Delta \mathbf{V}$ and substitute this Ansatz in Eq.~\ref{eq:full_exp_strat}. Assuming that $\Delta \mathbf{V}$ is small and expanding the operator $\mathcal{L}_{z}$ to first order, we obtain
\begin{equation}
\left(\mathrm{\mathbf{D}} - \frac{\partial \mathcal{L}_{z}}{\partial \mathbf{V}} \right) \Delta \mathbf{V} = 0.
\label{eq:deltax}
\end{equation}
We have the estimate
\begin{equation}
\frac{\left\lVert \mathrm{\mathbf{D}} \right\rVert_{2}}{\left\lVert \displaystyle \frac{\partial \mathcal{L}_{z} }{\partial \mathbf{V}}    \right\rVert_{2}} \sim \sqrt{\frac{1/\epsilon t_{\rm s}}{\alpha \Omega_{\mathrm{k}}^{-1}}} ,
\label{eq:approx_odg}
\end{equation}
since $\mathcal{L}_{z}$ is an operator with two derivatives in the vertical direction. From Eq.~\ref{eq:approx_odg}, we see that Eq.~\ref{eq:deltax} reduces to $ \mathrm{\mathbf{D}} \Delta \mathbf{V}  \simeq 0$, i.e. $\Delta \mathbf{V} \simeq 0$ and $\mathbf{V}' \simeq \mathbf{V}$ if the condition 
\begin{equation}
\frac{\rm{St}}{\epsilon \alpha} > 1 
\label{eq:approx_cond}
\end{equation}
is fulfilled. This condition is satisfied in real discs, with the marginal exception of sub-micron grains. Since the total solid mass of these grains is negligible, this approximation also holds for multiple dust populations as well. 
Physically, the vertical component of the shear adds a small additional driving term that can directly affect the dust velocities and indirectly perturb the gas velocities via back-reaction.
We neglect this contribution since, in most cases, the driving is dominated by the radial pressure gradient of the gas (Eq.~\ref{eq:approx_cond}).

}

\label{lastpage}
\end{document}